\documentclass[11pt]{article}
\usepackage{fullpage}
\usepackage{amsmath, amsthm, amssymb, mathrsfs, graphicx, subfigure}
\usepackage{ifthen, url}
\usepackage[usenames]{xcolor}
\usepackage{enumerate}
\usepackage{hyperref}
\colorlet{blueblack}{blue!10!black!90}
\hypersetup{colorlinks,
    citecolor=blueblack,
    filecolor=blueblack,
    linkcolor=blueblack,
     urlcolor=blueblack}
\usepackage{ifdraft}
\usepackage{lscape}
\usepackage{cancel}
\usepackage{multirow} 
\usepackage{bigdelim} 
\usepackage[sort&compress]{natbib}
\bibpunct{(}{)}{;}{a}{}{,} 
\usepackage{booktabs}
\usepackage{titletoc}
\theoremstyle{plain}

\newtheorem{proposition}{Proposition}
\newtheorem{condition}{Condition}
\theoremstyle{remark}
\newtheorem{remark}{Remark}
\theoremstyle{definition}
\newtheorem{definition}{Definition}
\theoremstyle{lemma}
\newtheorem{lemma}{Lemma}

\theoremstyle{example}
\newtheorem{example}{Example}

\theoremstyle{definition}

\DeclareMathOperator*{\plim}{plim}

\makeatletter
\newcommand*{\defeq}{\mathrel{\rlap{%
                     \raisebox{0.3ex}{$\m@th\cdot$}}%
                     \raisebox{-0.3ex}{$\m@th\cdot$}}%
                     =}
\newcommand{\tbbX}{\widetilde{\mathbb{X}}}										
\newcommand{\tbbY}{\widetilde{\mathbb{Y}}}										
\newcommand{\tbbU}{\widetilde{\mathbb{U}}}

\makeatother

\title{{\Large On Model Selection Criteria for Climate Change Impact Studies}\footnote{\emph{Corresponding Author}: Dalia Ghanem, dghanem@ucdavis.edu. The authors are grateful to Felix Pretis, Zack Miller, Ariel Ortiz-Bobea and Glenn Rudebusch for helpful comments and suggestions. We also thank participants at the World Congress of the Econometric Society 2020 and the Climate Econometrics Virtual Seminar. Xiaomeng Cui acknowledges financial support from the National Natural Science Foundation of China (71903070) and the 111 Project of China (B18026).
}}
\author{Xiaomeng Cui\footnote{Jinan University, cuixiaomeng@jnu.edu.cn.}  \and Bulat Gafarov\footnote{UC Davis, bgafarov@ucdavis.edu} \and Dalia Ghanem\footnote{UC Davis, dghanem@ucdavis.edu.} \and Todd Kuffner\footnote{Washington University, St. Louis, kuffner@wustl.edu.}}\date{\today}
\usepackage{fancyhdr} 

\graphicspath{{./Figures/}}
\usepackage{titlesec}

\titleformat{\section}
  {\normalfont\fontsize{15}{15}\bfseries}{\thesection}{1em}{}
\titleformat{\subsection}
  {\normalfont\fontsize{12}{12}\bfseries}{\thesubsection}{1em}{}
\begin{document}
\maketitle

\begin{abstract}
\pagenumbering{gobble}
\noindent
Climate change impact studies inform policymakers on the estimated damages of future climate change on economic, health and other outcomes.  
In most studies, an annual outcome variable is observed, e.g. agricultural yield, along with a higher-frequency regressor, e.g. daily temperature. 
Applied researchers then face a problem of selecting a model to characterize the nonlinear relationship between the outcome and the high-frequency regressor to make a policy recommendation based on the model-implied damage function. 
We show that existing model selection criteria are only suitable for the policy objective if one of the models under consideration nests the true model. 
If all models are seen as imperfect approximations to the true nonlinear relationship, the model that performs well in the normal climate conditions is not guaranteed to perform well at the projected climate that is different from the historical norm.
We therefore propose a new criterion, the proximity-weighted mean-squared error (PWMSE), that directly targets precision of the damage function at the projected future climate. 
 To make this criterion feasible, we assign higher weights to prior years that can serve as weather analogs to the projected future climate when evaluating competing models using the PWMSE.
 We show that our approach selects the best approximate regression model that has the smallest weighted error of predicted impacts for a projected future climate.  
 A simulation study and an application revisiting the impact of climate change on agricultural production illustrate the empirical relevance of our theoretical analysis.   
\end{abstract}

\newpage
\pagenumbering{arabic}
\section{Introduction}

Using panel data, impacts of climate change have been extensively studied on aggregate economic productivity \citep{hsiang2010temperatures, dell2012temperature, burke2015global}, micro-level productivity and economic returns \citep{deryugina2017marginal, zhang2016temperature,addoum2020temperature,somanathan2021impact}, agricultural profits and crop production \citep{deschenes2007economic, SR2009, burke2016adaptation,aragon2021climate, cui2020beyond}, energy consumption \citep{li2019climate, auffhammer2017climate,wenz2017north}, migration and labor allocation \citep{feng2010linkages, mueller2014heat, jessoe2016climate,cattaneo2016migration}, human capital \citep{graff2018temperature,garg2020temperature,park2020heat}, health and mortality \citep{deschenes2011climate, barreca2016adapting, burke2018higher, heutel2021adaptation}, and conflicts \citep{hsiang2011civil, hsiang2013quantifying, harari2018conflict}. 

Researchers typically consider multiple models in their analysis of the relationship between the outcome and temperature, the so-called damage or response function.
This function not only informs policy design regarding climate adaptation in specific sectors and locations, but they also serve as empirical foundations for quantifying the social cost of carbon and influence decision-making on mitigating and adapting climate change at the regional and global scale \citep{Dell2014,Diaz:2017,Ricke:2018}.  
This paper formalizes the model selection problem in climate change impact studies given the policy objective, evaluates the suitability of existing criteria and proposes a new criterion that directly targets the policy objective by incorporating the information about the projected future climate.

In typical empirical climate change impact studies, for $i=1,2,\dots,n$, $t=1,2,\dots,T$, we observe an outcome $Y_{it}$ and a regressor $W_{ith}$, which is observed at a higher frequency $h=1,2,\dots,H$.   
Practitioners tend to present results for a set of models $\{\mathbf{M}_{\alpha}\}_{\alpha=1}^{A}$, where each model uses different summary statistics of the higher-frequency weather variable as a model of the response function, $\mu_{\alpha}(\mathcal{W}_{it})$, of the outcome to the high-frequency regressor time series, $\mathcal{W}_{it}\equiv \{W_{ith}\}_{h=1}^H$.\footnote{   Among the most commonly used summary statistics of temperature are the annual average \citep[e.g.,][]{dell2012temperature}, various degree day measures \citep[e.g.,][]{burke2016adaptation}, seasonal averages \citep[e.g.,][]{mendelsohn1994impact} as well as temperature bins \citep[e.g.,][]{deschenes2011climate}.  To capture nonlinearities in the annual average temperature, a quadratic function has also been employed \citep[e.g.,][]{burke2015global}. }  For a given $\alpha$, $\mathbf{M}_{\alpha}$ specifies a linear-in-parameter model,
\begin{align}
Y_{it}=&\mu_{\alpha}(\mathcal{W}_{it})+a_{i,\alpha}+u_{it,\alpha}.\label{Malpha} 
\end{align} 
Here, $\mu_\alpha(\mathcal{W}_{it})$ is known up to a finite-dimensional parameter, $a_{i,\alpha}$ is a fixed effect and $u_{it,\alpha}$ constitutes idiosyncratic shocks.\footnote{In practice, additional covariates, year fixed effects and flexible time trends are included.  To simplify our presentation, we do not include these additional features.  However, our analysis extends in a straightforward manner to accommodating them as we illustrate in our empirical applications.}

Our first contribution is to formalize the model selection problem and policy objective in the climate change impacts literature. To do so, we first provide a review of recent work published in leading economics journals with a focus on modeling the response function between an outcome and temperature. Building on the review, we characterize the class of models in the empirical literature, define nested, non-nested overlapping and strictly non-nested models as well as formalize the policy objective of climate projections. 
Suppose that the outcome is given by the true model, $\mathbf{M}_{\star}$,
\begin{align}Y_{it}&=\mu_{\star}(\mathcal{W}_{it})+a_i+u_{it},\end{align}
where $\mu_{\star}(\cdot)$ is the true response function. 
The policy objective is to forecast the impact of the projected change in climate in period $T$ for each location $i$, $\mu_{\star}(\mathcal{W}_{i,T+\tau}^f)-\mu_{\star}(\mathcal{W}_{iT})$, where  $\mathcal{W}_{i,T+\tau}^f$ is the projected climate in a future period $T+\tau$. 
With the formal expression of the policy objective, we proceed to define the ideal mean-squared error (MSE) target for this policy problem, specifically
\begin{align}E[(\mu_{\alpha}(\mathcal{W}_{i,T+\tau}^f)-\mu_{\alpha}(\mathcal{W}_{iT})-(\mu_{\star}(\mathcal{W}_{i,T+\tau}^f)-\mu_{\star}(\mathcal{W}_{iT})))^2],\label{eq:ccp_criterion_intro}\end{align}
which is the mean of squared errors in predicting the climate change impact using the model $\mathbf{M}_\alpha$.

We begin our analysis with an assessment of the suitability of the existing consistent model selection criteria, such as Monte Carlo cross-validation (MCCV) and Generalized Information Criteria (GICs), in our context. We extend results from the classical literature on MCCV and GICs \citep{Shao:1993,Shao:1997,Vuong:1989,Sin:1996} to the mixed-frequency panel data setting in the climate change impacts literature.
 We show that MCCV with a vanishing training-to-full sample ratio, BIC, and the criteria proposed in \citet{Sin:1996} are model selection consistent if one of the models under consideration nests the true model.\footnote{Let $\mathbf{M}_{\star}$ denote the true model and $\widehat{\mathbf{M}}_{C}$ denote the model selected by criterion $C$, a criterion is said to be model selection consistent if $P(\widehat{\mathbf{M}}_{C}=\mathbf{M}_{\star})\rightarrow 1$ as sample size grows. For additional discussion, see Sections \ref{subsec:existing_criteria} and \ref{app:gic}.} In contrast,  only the criteria proposed in \citet{Sin:1996} are consistent if all models are misspecified.\footnote{This property is referred to as pseudo-model selection consistency in \citet{Sin:1996}. We provide a definition and related discussions in Section \ref{app:gic}. An interesting byproduct of this analysis is that we find that in the context of mixed-frequency panel data BIC can be pseudo-inconsistent in nested model selection problems when all models under consideration are misspecified, whereas in standard nonlinear model selection problems prior literature establishes that BIC is pseudo-inconsistent in non-nested model selection problems \citep{Sin:1996,Hong:2012}. We formally discuss this issue in Online Appendix \ref{app:pseudo-true}.} Since consistent model selection criteria select the true model (or the most parsimonious model that nests it) with probability approaching one if such a model is under consideration, they would asymptotically achieve the ideal MSE in Eq. \eqref{eq:ccp_criterion_intro}. While it is plausible for empirical researchers to correctly specify the model for outcomes with a clear and well-studied physical relationship with weather, there is a wide range of economic outcomes of interest for which this would not be plausible. 
  
If none of the models under consideration nest the true model, consistent model selection criteria are not guaranteed to select models that perform well in terms of the ideal MSE target for the policy objective given in \eqref{eq:ccp_criterion_intro}. We therefore proceed to propose an alternative criterion suitable for the policy objective that treats all models as potentially misspecified.
Building on our formal definition of the policy objective, we propose a new criterion that accounts for the projected change in the weather distribution in the (infeasible) ideal MSE target in \eqref{eq:ccp_criterion_intro} and incorporates the information about the future climate in the evaluation. 
The criterion we propose is based on a proximity-weighted mean of the squared errors (PWMSE) in predicting the impact of the projected change in climate between year $T$ relative to prior years available in the data $T-r$ for $r=1,\dots,T-1$,
\begin{align}  PWMSE_{\alpha}&=
\sum_{r=1}^{T-1} E[( {\mu}_{\alpha}(\mathcal{W}_{i,T-r})- {\mu}_{\alpha} (\mathcal{W}_{iT})-(\mu_{\star}(\mathcal{W}_{i,T-r})-\mu_{\star}(\mathcal{W}_{iT})))^2\pi(\mathcal{W}_{i,T-r},\mathcal{W}_{i,T+\tau}^f)].\end{align}
The weighting function in the PWMSE is designed to give higher weight to prior years in the sample that are more similar to the projected future climate at $T+\tau$. 
The weighting function therefore ensures that the selected model performs well in terms of predicting the impacts of projected climate change from $\mathcal{W}_{iT}$ to $\mathcal{W}_{i,T+\tau}^f$ in line with the policy objective. To estimate the PWMSE, we propose a resampling procedure and show that the model that minimizes the estimated PWMSE minimizes the population PWMSE with probability approaching one as the sample size grows. We demonstrate numerically that the simulated ideal MSE target is minimized by the true model as well as models that nest it in our simulation examples. 

We conduct a simulation study to examine the finite-sample performance of the MCCV, GIC and PWMSE criteria. 
Our simulations demonstrate that, when the true model is under consideration, all model selection criteria select the true model with high probabilities. 
However, when all candidate models are misspecified, the PWMSE is more likely to select the model that minimizes the simulated ideal MSE target since it incorporates the information about the future forecasts in the evaluation.
Since the PWMSE requires practitioners to specify the weighting function, we also examine the finite-sample performance of the PWMSE with different choices for the weighting function and provide guidance to practitioners on this choice in Online Appendix \ref{app:pwmse}.

To illustrate the empirical relevance of our results, we compare the behavior of existing criteria and our proposed PWMSE in the context of an application revisiting the relationship between temperature and agricultural yield. 
In this application, while different criteria select different models, the associated damage functions are qualitatively similar --- rising temperature becomes detrimental at high temperatures. Consistent with our theoretical predictions, the \citet{Sin:1996} criteria  (SW) choose the most parsimonious among those models. Our proposed criteria of PWMSE, which incorporates the HadCM3-B1 climate projection in the weighting function, selects a less parsimonious model that provides flexible estimates of the nonlinear relationship consistent with the agronomic literature. Although the models selected by the SW and PWMSE criteria result in qualitatively similar damage functions, the quantitative differences between their predicted outcomes matter for policy arrangements on adaptation toward the specific future climate projected by HadCM3-B1. This application highlights the value of the PWMSE criterion in terms of ensuring that the selected model targets the policy objective for a given climate change projection.
 
\vspace{0.4cm}

\noindent \textbf{Implications for practice}. This paper has practical implications for the use of model selection criteria in climate change impact studies. We show that the existing model selection criteria, although useful, have important limitations in their application to climate change impact studies. The usefulness of consistent model selection criteria is limited to the case when the true model is nested in one of the models under consideration. While this is plausible in applications where the relationship between temperature and the outcome of interest is well-studied, such as for agricultural outcomes, it is restrictive for many other economic outcomes of interest. Therefore, the PWMSE serves as a useful addition to the existing criteria as it ensures improved performance for the policy objective of climate change projections. We emphasize, however, that rather than reporting the results of a particular model selection criterion, applied researchers should report the results of the different model selection criteria we consider, MCCV, GIC and PWMSE.  The reasoning behind this recommendation stems from the fact that if the true response function is nested in one or more of the models under consideration, then the different model selection criteria \emph{should} select models that have similar response functions, albeit different numbers of parameters, which is consistent with our theoretical and simulation analysis. The usefulness of reporting the different criteria in interpreting the empirical results is also clearly illustrated in our empirical application. We elaborate on these practical implications in Section \ref{sec:implications} and illustrate these insights in the context of our empirical application in Section \ref{sec:empirical}.

\vspace{0.4cm}
\noindent\textbf{Related literature}. This paper builds on the literature on the asymptotic properties of model selection criteria \citep[e.g.][]{Claeskens:2008,Arlot:2010}. We build on the literature on the asymptotic behavior of MCCVs in linear models \citep{Shao:1993, Shao:1997} as well as nonlinear model selection based on GICs \citep{Sin:1996,Hong:2012}. The latter strand of the literature builds on the seminal work in \citet{Vuong:1989} which shows that the convergence rate of the quasi-likelihood ratio depends on whether the models under consideration are nested or not.\footnote{Recent work in this literature develops approaches to uniformly valid testing and post-selection inference for non-nested models \citep{shi2015nondegenerate,schennach2017simple,liao2020nondegenerate}.}

The analysis of model selection consistency in the context of the mixed-frequency panel data models is of independent interest and relates to an important body of work on aggregation in mixed-frequency time series.  While the goal in this literature is starkly different from the climate change impacts literature, both share a common theme which is the need to aggregate regressors due to the mixed-frequency nature of the empirical setting. There is a large body of work in the mixed-frequency time series literature providing different aggregation schemes   \citep{Andreou:2008,GSV:2006,GSV:2007,chambers2016estimation,miller2016conditionally,miller2018simple}.\footnote{We point to some interesting connections between notions of misspecification bias in this paper and the literature on aggregation in cointegration models \citep{chambers2003asymptotic,chambers2011cointegration,chambers2007frequency,miller2014mixed,miller2016conditionally} in Online Appendix \ref{app:pseudo-true}.}
One promising direction for future work that complements existing work on specification testing \citep{AGK:2010,groenvik2018self,kvedaras2012testing,miller2018simple,liu2019choice} is to adopt the proposed model selection framework to examine the aggregation problem in mixed-frequency time series from the perspective of optimal  selection of an approximate model.

The PWMSE criterion proposed in this paper has some apparent similarity with analog forecasting procedures  (as proposed in \cite{lorenz1969atmospheric}) used in short-term climate forecasting. 
The analog forecasting uses past trajectories of the weather with initial conditions that approximately match the current conditions. Such approach is based on the fact that the climate systems follow deterministic differential equations that are time-invariant. So any two solutions for such systems with close initial conditions remain close to each other at least for a short time. The 
PWMSE criterion selects the model that performs best on average for past observations that resemble the future projected climate. 
Despite sharing this common feature, the goals of the two methods are different: the analog forecasting seeks to forecast the climate using past patterns, while PWMSE-minimizing model forecasts a nonlinear impact of the future weather that is taken as given.

The remainder of the paper is organized as follows. Section \ref{sec:model_selection} provides the policy motivation behind the present paper, summarizes current empirical practice, and formalizes the model selection problem and policy objective. Section \ref{sec:selection_criteria} evaluates the suitability of existing criteria for the policy objective of climate change projections and introduces the PWMSE criterion. Section \ref{sec:empirical} illustrates the empirical relevance of our theoretical analysis in the context of an empirical application revisiting the relationship between temperature and agricultural yields. 
\section{Why Model Selection in Climate Change Impact Studies?}\label{sec:model_selection}
Panel estimates of climate change impacts have served as critical inputs for policy making on climate change mitigation and adaptation. In this section, we first briefly introduce the rationale behind the existing empirical studies and review existing methods in recent work. We then formalize the model selection problem and the policy objective in this context, which also provides a foundation for understanding the behavior of existing criteria as well as the rationale behind our proposed PWMSE criteria in Section \ref{sec:selection_criteria}.

\subsection{Damage functions as policy parameters}

Designing cost-effective policy on climate change mitigation and adaptation requires a precise and thorough understanding on how the key climatic factor (e.g., temperature) affects human society. In responding to this mission, extensive efforts have been made to properly estimate the economic damage functions, which characterize the quantitative relationship between certain economic outcomes and climatic factors, especially temperature. 

\begin{figure}[htbp]
\begin{tabular}{cc}
A. Monthly averages  & B. One-knot linear spline \\~\\
\includegraphics[width=.48\linewidth]{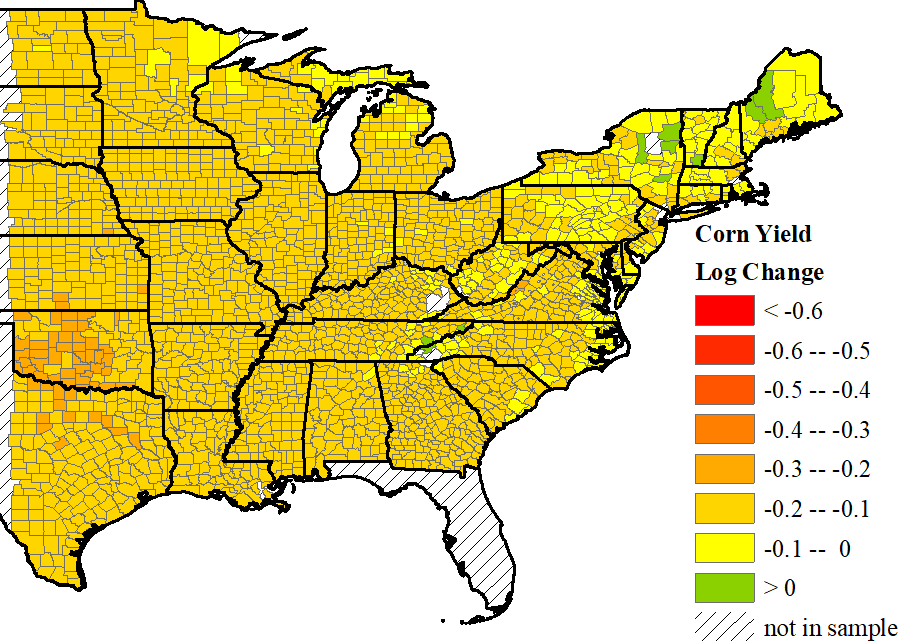}&
\includegraphics[width=.48\linewidth]{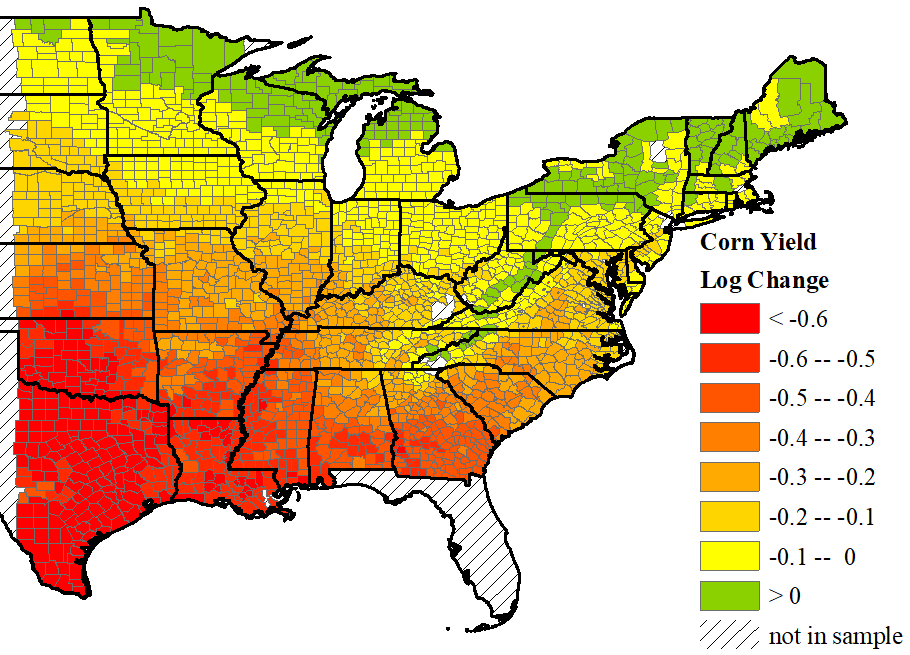}\\
\end{tabular}
\caption{Yield Impacts Projected under Future Climate}\label{fig:motivate_maps}
{\footnotesize {\it Notes:} The county-level log changes in corn yields are obtained by applying different empirically estimated damage functions to the temperature of 2050 projected under HadCM3-B1. The damage functions are estimated based on monthly average temperatures in panel A and linear spline with one endogenously determined knot in panel B.}
\end{figure}

The estimated damage functions serve two purposes. From a micro perspective, they help improve the design of regulatory and adaptation policies that target specific contexts. Using agricultural production as an example, Figure \ref{fig:motivate_maps} shows sharply contrasted warming implications on corn yields in the United States under two distinct but reasonable damage functions. Both panels reflect predicted warming-induced yield losses by 2050 under future temperatures projected by the HadCM3-B1 model. Panel A relies on a damage function empirically estimated based on a specification of monthly average temperatures, while Panel B relies on that of linear spline with an endogenously determined knot.\footnote{We describe the precise steps for obtaining these estimates and projections in Section \ref{sec:empirical}.} It is clear that the chosen damage function would directly influence policy arrangements on where and how corn production needs to be adapted to mitigate potential losses under future warming.

The policy relevance of damage functions is not restricted to policies targeting specific outcomes and sectors.
From a macro perspective, the cost-benefit evaluation of a host of domestic policies as well as international agreements is directly influenced by the estimated social cost of carbon (SCC). Improving this estimate hinges on those damage functions as critical inputs \citep{NAS:2017}. This point is epitomized in \cite{Moore:2017} as they show that the SCC is doubled by simply updating the damage functions of agriculture with the most recent empirical estimates in the FUND model.\footnote{The FUND model is a widely used integrated assessment model (IAM) developed by Richard Tol and David Anthoff. See more details at: {\url{http://www.fund-model.org}}.}

\subsection{Current Empirical Practice}\label{sec:empirical_practice}

The empirical studies for pinning down the damage functions typically rely on panel data with large $n$ and relatively small $T$. For example, the empiricist may observe some economic outcomes at the county level for more than a thousand counties, and the outcomes are observed annually over a period of a few decades. This practice was first initiated in the seminal work of \cite{deschenes2007economic} on identifying temperature impacts on agriculture, it then became popular and widely adopted in understanding climatic impacts on many dimensions of human society. 

In this empirical literature, researchers commonly use a panel fixed effects model where, conceptually, the location fixed effects control for impacts of the time-invariant factors like county characteristics.\footnote{To the best of our knowledge, there are no studies in this literature relying on random effects models.}
With this setup, identifying the damage function would mostly rely on year-to-year within-county variation in weather that is arguably exogenous.\footnote{The literature has acknowledged that this year-to-year weather variation is different from long-run climate change, but utilizing the arguably exogenous weather variation for pinning down damage functions is still regarded as highly useful for policy purposes \citep{Auffhammer:2018}.} This empirical approach rests on the validity of the strict exogeneity assumption. To focus our attention on the model selection problem in this literature, we therefore maintain this assumption here.

A critical component in these econometric examinations is to estimate the particularly damaging effects associated with high temperatures, commonly referred to as the nonlinear temperature effects. With raw temperature data observed at a higher frequency than that for the panel regressions, identifying this nonlinearity would require the empiricist to first summarize the high-frequency regressor data in certain ways such that the variables with summarized temperature information can support estimating nonlinear temperature effects. 

\begin{table}[htbp]
  \centering
  \fontsize{8}{8}\selectfont
  \caption{Summary of Temperature Models in Leading Economics Journals since 2015}
    \begin{tabular}{p{6cm}lccccccc}\hline
    Article & Journal & \multicolumn{7}{c}{Temperature Models Considered} \\ \cline{3-9}
          &       & mean  & max   & degree & bins  & poly- & splines & heat  \\
          &       & temp. & temp. & days & & nomials &   &   index\\ \hline
          \\
    \cite{Levinson:2016} & \textit{AER} &       &       & \checkmark     &       &       &       &  \\
    \cite{barreca2016adapting} & \textit{JPE} &       &       &       & \checkmark    &       &       &  \\
    \cite{somanathan2021impact} & \textit{JPE} &       &       & \checkmark    & \checkmark    & \checkmark    &       &  \\
    \cite{Busse:2015} & \textit{QJE} &       & \checkmark    &       & \checkmark    &       &       &  \\
    \cite{Heyes:2019} & \textit{AEJ-App} & \checkmark    &       &       & \checkmark    &       &       & \checkmark\\
    \cite{Colmer:2021} & \textit{AEJ-App} & \checkmark    &       & \checkmark    & \checkmark    &       &       &  \\
    \cite{Lopalo:forthcoming} & \textit{AEJ-App} &       &       &       & \checkmark    &       &       &  \\
    \cite{burke2016adaptation} & \textit{AEJ-Pol} &       &       & \checkmark    &       &       &       &  \\
    \cite{park2020heat} & \textit{AEJ-Pol} & \checkmark    &       &       & \checkmark    &       &       &  \\
    \cite{aragon2021climate} & \textit{AEJ-Pol} &       &       & \checkmark    & \checkmark    &       &       &  \\
    \cite{Cohen:forthcoming} & \textit{AEJ-Pol} &       &       &       & \checkmark    &       &       &  \\
    \cite{Liu:forthcoming} & \textit{AEJ-Pol} & \checkmark    &       &       &       &       &       &  \\
    \cite{Anderson:2017} & \textit{EJ} & \checkmark    &       &       &       &       &       &  \\
    \cite{jessoe2016climate} & \textit{EJ} &       &       & \checkmark    & \checkmark    &       &       &  \\
    \cite{Jagnani:2021} & \textit{EJ} &       &       & \checkmark    & \checkmark    &       &       &  \\
    \cite{Adhvaryu:2020} & \textit{REStat} &       &       &       &       & \checkmark    & \checkmark    &  \\
    \cite{heutel2021adaptation} & \textit{REStat} &       &       &       & \checkmark    &       &       &  \\
    \cite{Novan:forthcoming} & \textit{REStat} & \checkmark    &       &       &       &       & \checkmark    &  \\ \hline
    \\
    \multicolumn{9}{l}{\parbox[t]{16cm}{\footnotesize {\it Notes}: This survey covers empirical studies on the Top 5 journals, AEJs, EJ, and REStat since 2015. We focus on studies for which identifying the response function to temperatures is of critical importance. The temperature models are broadly categorized. Within each model category, the specific formulation of the model still varies across different articles. Formal definitions of these models are provided in Online Appendix \ref{sec:model_def}.}}
    \end{tabular}
  \label{tab:lit_sum}
\end{table}

There is no consensus or best practice on how to implement this process. Empiricists usually make decisions on how they summarize high-frequency temperature information to eventually obtain the estimates on the nonlinearity. In Table \ref{tab:lit_sum}, we summarize temperature models considered in recent empirical studies in leading economics journals, including the ``Top 5'' journals, American Economic Journals ({\it AEJ}), Economic Journal ({\it EJ}), and Review of Economics and Statistics ({\it REStat}). We limit our attention to the studies for which identifying the response function to temperature fluctuations is of critical importance for the research question.\footnote{Under this criterion, this survey does not include articles where response to temperature is only a first-stage or temperatures are simply considered as control variables.}

The chosen models vary across different studies, and none of the listed empirical studies adopt a selection rule for determining which model should be used to inform policy.\footnote{While not included this review, we note that the influential study in \citet{SR2009}, which we revisit in Section \ref{sec:empirical},  relies on Monte Carlo cross-validation to select between the models under consideration.} Even though many studies favor the use of degree days and bins, the specific ways of constructing these variables as well as the critical thresholds involved are determined by the empiricists.\footnote{We note one exception in \cite{burke2016adaptation} where the authors rely on in-sample fits (i.e., $R^2$) to determine the whole-number cutoff degree for separating growing and heat degree days.} It is important to note, however, that the revealed nonlinearity (or the absence of it) obtained from the econometric estimation is shaped or at least indirectly influenced by the decision made on how to summarize the high-frequency regressor data, as we have illustrated in Figure \ref{fig:motivate_maps}. If empiricists improperly summarize the information and adopt a poorly specified model, they could potentially reach biased conclusions on climate change impacts.

In the following subsection, we make precise definitions and formalize the model selection problem and policy objective for these empirical studies on identifying climate change impacts.

\subsection{Formalizing the Model Selection Problem and Policy Objective}\label{subsec:CC_MS}
\subsubsection{Class of Models in Empirical Practice}

In the setting of climate change impact studies, researchers observe an annual outcome $Y_{it}$ and a vector of regressors observed at a higher frequency, such as daily or hourly, $\mathcal{W}_{it}=\{W_{ith}\}_{h=1}^H$.  In this mixed-frequency panel data setting, the models considered in the literature as presented in Section \ref{sec:empirical_practice} are fixed effects models that differ in terms of the response functions used to model the relationship between the outcome, $Y_{it}$, and the high-frequency regressor time series, $\mathcal{W}_{it}$.  For a given model $\mathbf{M}_{\alpha}$, the outcome equation is thus given by the following\begin{align}Y_{it}&=\mu_{\alpha}(\mathcal{W}_{it})+a_{i,\alpha}+u_{it,\alpha}.
\end{align}
While control variables and additional fixed effects can be included, we omit them in our analysis to simplify the exposition and to focus our analysis on the main issue we consider here which is the potential misspecification of the response function. 

The class of response functions considered in the empirical literature can be formally represented as follows,
\begin{align}\mu_{\alpha}(\mathcal{W}_{it})&=X_{it,\alpha}'\beta_{\alpha},\\
X_{it,\alpha}&=\psi_{\alpha}\left(\sum_{h=1}^Hf_{\alpha}(W_{ith})^\prime\phi_{\alpha,h}\right),\end{align}
where $f_{\alpha}(\cdot)$ is a $d_{f_\alpha}\times 1$ vector of linear and/or nonlinear transformations of the high-frequency regressor, $W_{ith}$, and $\phi_{\alpha,h}$ is a $d_{f_\alpha}\times 1$ vector of $h$-specific weights. Once the weighted transformations of $W_{ith}$ are aggregated over $h$, the $d_{\psi_\alpha}\times 1$ transformation $\psi_\alpha(\cdot)$ is applied to them. For the purposes of this paper, we assume that $d_{f_{\alpha}}$, $d_{\phi_{\alpha}}$ and $d_{\psi_{\alpha}}$ are finite, such that $k_\alpha=dim(X_{it,\alpha})<\infty$. 

While the class of models considered in the literature preserves linearity in the parameters and separability between observables and unobservables, it can allow for nonlinearities in the high-frequency regressor. To show this, we provide a few examples to illustrate the different roles played by $f_{\alpha}$, $\phi_{\alpha,h}$ and $\psi_{\alpha}$. The annual mean model, where $X_{it,\alpha}=\bar{W}_{it}=\sum_{h=1}^HW_{ith}/H$, can be obtained by letting  $f_{\alpha}(W_{ith})=W_{ith}$, $\phi_{\alpha,h}=\frac{1}{H}$, and $\psi_\alpha(\bar{W}_{it})=\bar{W}_{it}$, where $\bar{W}_{it}=\sum_{h=1}^HW_{ith}/H$. The quadratic in annual mean model, where $X_{it,\alpha}=(\bar{W}_{it},\bar{W}_{it}^2)^\prime$, relies on the same $f_{\alpha}$ and $\phi_{\alpha,h}$ as the annual mean model, but uses a second-order polynomial function $\psi_\alpha$, such that $\psi_\alpha({W}_{ith})=({W}_{ith},{W}_{ith}^2)^\prime$. To obtain the quarterly mean model, we set $f_{\alpha}(W_{ith}) = W_{ith} (1,1,1,1)^\prime$, $$\phi_{\alpha,h}=(1\{h\in Q_1\}/|Q_1|,1\{h\in Q_2\}/|Q_2|,1\{h\in Q_3\}/|Q_3|, 1\{h\in Q_4\}/|Q_4|)^\prime,$$ and $\psi_{\alpha}$ is the identity function. For $j=1,\dots,4$, $Q_j$ denotes the subset of indices $\{1,\dots,H\}$ that are in the $j^{th}$ quarter and $|\mathcal{A}|$ denotes the cardinality for a set $\mathcal{A}$. Finally, the model with temperature bins, where $dim(W_{ith})=1$, can be obtained by setting $$f_{\alpha}(W_{ith})=\left(1\{W_{ith}\in[l_1,u_1)\},1\{W_{ith}\in[l_2,u_2)\},\dots, 1\{W_{ith}\in[l_{d_{f_{\alpha}}},u_{d_{f_{\alpha}}}]\}\right)^\prime,$$
$\phi_{\alpha,h}=(1,1,\dots,1)^\prime$, $\psi_{\alpha}$ is the identity function and $\{[l_j,u_j)\}_{j=1}^{d_{f_\alpha}}$ is a set of intervals.
In Online Appendix \ref{sec:model_def}, we also provide formal definitions for other commonly used models in the literature, such as degree days, polynomials, splines, etc.

\subsubsection{Defining Nested and Non-nested Models}
Since all models considered rely on different aggregated transformations of the same underlying high-frequency regressor, the models are likely to be overlapping as defined below.  However, we would like to differentiate between different cases of overlapping models.  Assume without loss of generality $k_{\alpha}<k_{\gamma}$.  Let $\omega=\{\omega_{h}\}_{h=1}^H$ denote a realization of $\mathcal{W}_{it}$.  For a fixed realization $\omega$, the realizations of $X_{it,\alpha}$ and $X_{it,\gamma}$ are given by $x_{\omega,\alpha}=\psi_{\alpha}\left(\sum_{h=1}^Hf_{\alpha}(\omega_h)\phi_{\gamma,h}\right)$ and $x_{\omega,\gamma}=\psi_{\gamma}\left(\sum_{h=1}^Hf_{\gamma}(\omega_h)\phi_{\gamma,h}\right)$, respectively.  Let $\mathcal{B}_{\alpha}$ denote the parameter space of $\beta_{\alpha}$ and $\beta_{\alpha}^{k}$ the $k^{th}$ element of $\beta_{\alpha}$.  

We next provide formal definitions for when two models, $\mathbf{M}_{\alpha}$ and $\mathbf{M}_{\gamma}$, are nested, non-nested overlapping or strictly non-nested. In the following, let $\mathbf{0}_{k_{\alpha}}$ denote a $k_\alpha$-dimensional column vector of zeros.
\begin{definition}
\begin{enumerate}[(i)]
\item  $\mathbf{M}_{\alpha}$ is nested in $\mathbf{M}_{\gamma}$ 
 iff $x_{\omega,\alpha}'\beta_{\alpha}=R_{\alpha,\gamma}x_{\omega,\gamma}'\beta_{\gamma}$ for all $\omega$ and $\beta_{\alpha}\in\mathcal{B}_{\alpha}$, where $\beta_{\gamma}\in\mathcal{B}_{\gamma}$ and $R_{\alpha,\gamma}$ is a $k_{\alpha}\times k_{\gamma}$ non-random matrix with full row rank,
\item $\mathbf{M}_{\alpha}$ and $\mathbf{M}_{\gamma}$ are non-nested, overlapping iff $\mathbf{M}_{\gamma}$ does not nest $\mathbf{M}_{\alpha}$, but $x_{\omega,\alpha}'\beta_\alpha=x_{\omega,\gamma}'\beta_{\gamma}$ for all $\omega$ and some $\beta_{\alpha}\in\mathcal{B}_{\alpha}\setminus \{\mathbf{0}_{k_{\alpha}}\}$ and $\beta_{\gamma}\in\mathcal{B}_{\gamma}\setminus \{\mathbf{0}_{k_{\gamma}}\}$,
\item  $\mathbf{M}_{\alpha}$ and $\mathbf{M}_{\gamma}$ are strictly non-nested iff they are not nested and ${x}_{\omega,\alpha}'\beta_{\alpha}\neq x_{\omega,\gamma}'\beta_{\gamma}$ for all $\omega$, $\beta_{\alpha}\in\mathcal{B}_{\alpha}$ and $\beta_{\gamma}\in\mathcal{B}_{\gamma}$.\\
\end{enumerate}
\end{definition}
\noindent Note that according to (i), a model contains another if the regressors in the latter can be expressed as a linear combination of the regressors in the former.  This is different from the typical linear regression framework where a model contains another if the regressors in the latter are a subset of the regressors in the former, i.e. the elements in $R_{\alpha,\gamma}$ can only be zero or one.  We illustrate the above definitions with the following example. 
\begin{example}\label{ex:AQinAQ} (Annual Mean, Quarterly Mean, Quadratic in Annual Mean   and   Temperature Bin Models)\\
Let $\mathbf{M}_{\alpha}$ denote the annual mean model, with outcome equation
\begin{align}
Y_{it}&=X_{it,\alpha}'\beta_{\alpha}+a_{i,\alpha}+u_{it,\alpha} \; ,
\end{align}
where $X_{it,\alpha}=\bar{W}_{it}\equiv \sum_{h=1}^H W_{ith}/H$.  The quarterly mean model uses instead the quarterly means of $\mathcal{W}_{it}$ as regressors. In the quarterly mean model,  $X_{it,\gamma}=(\sum_{h\in Q_{1}}W_{ith}/|Q_{1}|,\dots,\sum_{h\in Q_{4}}W_{ith}/|Q_{4}|).$  Then $\mathbf{M}_{\gamma}$ prescribes the outcome equation
\begin{align}
Y_{it}&=X_{it,\gamma}'\beta_{\gamma}+a_{i,\gamma}+u_{it,\gamma} .
\end{align}
Note that for all $\beta_{\alpha}\in\mathcal{B}_{\alpha}$ $X_{it,\alpha}'\beta_{\alpha}=R_{\alpha,\gamma}X_{it,\gamma}'\beta_{\gamma}$, where 
\begin{align}
R_{\alpha,\gamma}&=\frac{1}{H}\left(\begin{array}{cccc}|Q_{1}|,&|Q_{2}|,&|Q_{3}|,&|Q_{4}|\end{array}\right), \nonumber\\
\beta_{\gamma}&=\beta_{\alpha}(1,1,1,1)'.\nonumber
\end{align}
Hence, $\mathbf{M}_{\alpha}$ is nested in $\mathbf{M}_{\gamma}$.

The quadratic in annual mean model, $\mathbf{M}_{\delta}$, uses $X_{it,\delta}=(\bar{W}_{it},\bar{W}_{it}^{2})'$ as regressors.  Even though the quadratic in annual mean and the quarterly mean models are not nested, if $\beta_{\delta}^{2}=0$ and $\beta_{\gamma}^{k}=\beta_{\gamma}^{k'}$ for $k\neq k'$, with $k, k'\in\{1,2,3,4\}$, then both models yield the annual mean model given $\mathcal{W}_{it}$.  Hence, they are overlapping, non-nested.

To complete the example, note that annual mean model is both non-nested and non-overlapping with the temperature bin model $\mathbf{M}_{\rho}$ with regressor $    X_{it,\rho}=\frac{1}{H}  \sum_{h=1}^H 1\{ W_{ith}\geq 0\}/H $  as long as $W_{ith}$ has continuous support. 
Indeed, one can construct multiple examples of elementary events $\omega$  such that the number of the days with positive values $W_{ith}$ being equal, but the annual average being different and vice versa. Thus the only overlapping submodel of  $\mathbf{M}_{\alpha}$ and  $\mathbf{M}_{\rho}$ is an empty model.

\label{eg:models}
\end{example}

\subsubsection{Policy Objective: Climate Change Projections}

Consider a set of locations $i=1,\dots,n$, for which we observe an annual outcome and daily temperature time series over $T$ time periods, $\{Y_{it},\mathcal{W}_{it}\}_{t=1}^{T}$.\footnote{Note that our setting can allow for a vector of high-frequency regressors, but to fix ideas and to remain consistent with the policy question, we motivate this section with $\mathcal{W}_{it}$ as the temperture time series.} We would like to forecast the impact of climate change on the outcome between period $T$ and $T+\tau$. The location is projected to experience the daily weather time series in $T+\tau$ periods ahead, $\mathcal{W}^f_{i,T+\tau}$. In this setting, $\{\mathcal{W}^f_{i,T+\tau}\}_{i=1}^{n}$ is obtained from a forecast model of a specific projection for a given climate change scenario.\footnote{We note that uncertainties exist in forecasting future climate and the forecasts may vary across different projections even for the same climate change scenario. We restrict our attention to the case that the policy maker is interested in the impact of climate forecasted by a specific projection.}

Our goal is to predict the impact of this change in climate on an outcome of interest. One of the unique features of this policy objective is that the goal of the prediction exercise here is a causal parameter.  As a result, to make progress here, we have to define a true model that generates the outcome. For the purposes of this paper, we assume that the outcome, $Y_{it}$, is given by the following model, $\mathbf{M}_{\star}$,
\begin{align}Y_{it}&=\mu_{\star}(\mathcal{W}_{it})+a_i+u_{it}.\end{align}
Similar to the models imposed in the literature, $\mathbf{M}_{\star}$ imposes separability between unobservables and observables. As a result, the potential source of misspecification of a given model $\mathbf{M}_\alpha$ would stem from the misspecification of the true response function, $\mu_{\star}(\cdot)$. 

Assuming strict exogeneity, $E[u_{it}|\mathcal{W}_{i1},\dots,\mathcal{W}_{i,T+\tau}=\mathcal{W}_{i,T+\tau}^f,a_i]=0$, which imposes the mean independence of $u_{it}$ not only of $\{\mathcal{W}_{it}\}_{t=1}^T$ but also $\mathcal{W}_{i,T+\tau}=W_{i,T+\tau}^f$, we can formally define the object of interest for each location $i$
\begin{align}E[Y_{i,T+\tau}|\mathcal{W}_{i,T+\tau}=\mathcal{W}^f_{i,T+\tau},a_i]-E[Y_{iT}|\mathcal{W}_{iT},a_i]&=\mu_{\star}(\mathcal{W}^f_{i,T+\tau})-\mu_{\star}(\mathcal{W}_{iT}).\end{align}
Let us define $ Supp \mathcal{W}_t$ as the support of $\mathcal{W}_{it}$. Our object of interest is a functional of $\omega_1\in Supp\mathcal{W}_T$ and $\omega_2\in Supp \mathcal{W}_{T+\tau}^f$
\begin{align}\mu_{\star}(\omega_2)-\mu_{\star}(\omega_1).\end{align}
We are therefore interested in estimating the function with the highest precision in a specific set of values that correspond to the support of $\mathcal{W}_{i,T+\tau}^f$ and $\mathcal{W}_{iT}$, respectively.

In practice, researchers and policymakers do not know the true functional form of $\mu_{\star}$, they rely on a set of models $\{\mathbf{M}_{\alpha}\}_{\alpha=1}^A$ to estimate the impact of projected climate change.  The question that arises here is what criterion policymakers should use to choose between the different models and their implied projections. For instance, in Figure \ref{fig:motivate_maps}, how should a policymaker decide between the two models considered which give substantially different climate change projections and therefore have very different policy implications?  We tackle this question in the following section, evaluating the usefulness of existing criteria and proposing a new criterion tailored to the policy objective in climate change impact studies. We then illustrate the empirical relevance of our theoretical analysis in the context of the empirical application that corresponds to Figure \ref{fig:motivate_maps}.

\section{Model Selection Criteria for Climate Change Impact Studies}\label{sec:selection_criteria}

Given the policy objective of climate change projections, an ideal criterion to choose between a set of models under consideration would be to minimize the following mean-squared error (MSE) criterion,
\begin{align}&E[({\mu}_{\alpha}(\mathcal{W}^f_{i,T+\tau})-{\mu}_{\alpha}(\mathcal{W}_{iT})-(\mu_{\star}(\mathcal{W}^f_{i,T+\tau})-\mu_{\star}(\mathcal{W}_{iT})))^2].\label{eq:ccp_criterion}
\end{align}
This MSE criterion is based on the error of predicting the impact of the projected climate change from $\mathcal{W}_{iT}$ to $\mathcal{W}_{i,T+\tau}^f$. A challenge here, as in other forecasting problems, is of course that this ideal target is infeasible. Given the use of MCCV in practice, we first examine the asymptotic properties of this criterion as well as GICs (Section \ref{subsec:existing_criteria}). Since our theoretical analysis of these existing criteria underscores the importance of at least one of the models in the set $\{\mathbf{M}_{\alpha}\}_{\alpha=1}^{A}$ nesting $\mathbf{M}_{\star}$ for these criteria to minimize the ideal MSE target in \eqref{eq:ccp_criterion}, we proceed to propose a class of alternative criteria that minimize feasible counterparts of the ideal target whether or not the models under consideration are correctly specified (Section \ref{subsec:CCP-MSE}).

In the following, we consider asymptotics that let $n\rightarrow\infty$ while holding $T$ fixed. Given the observational nature of the panel data under consideration in this setting, we do not restrict the conditional mean of $a_i$ given $\{\mathcal{W}_{it}\}_{t=1}^T$, but maintain the strict exogoneity assumption, $E[u_{it}|\mathcal{W}_{i1},\dots,\mathcal{W}_{iT},a_i]=0$. As a result, we focus our attention on fixed effects (FE) estimators that rely on the within-group transformation.\footnote{Our results can be extended to the first-difference estimator in a straightforward manner.} For a random variable $V_{it}$, $\widetilde{V}_{it}=V_{it}-\sum_{t=1}^T V_{it}/T$ (in particular, $\tilde{\mu}_{\alpha}(\mathcal{W}_{it}) = {\mu}_{\alpha}(\mathcal{W}_{it})-  \sum_{t=1}^T{\mu}_{\alpha}(\mathcal{W}_{it})/T$) and $\widetilde{V}_{i}=(\widetilde{V}_{i1},\dots,\widetilde{V}_{iT})$.

\subsection{Asymptotic Properties of Existing Model Selection Criteria}\label{subsec:existing_criteria}

In this section, we consider two classes of model selection criteria, MCCV and GIC, and discuss the conditions under which these criteria are suitable for our policy objective. To maintain a simple exposition in this section, we relegate all formal conditions and statements to Online Appendix \ref{app:existing_criteria} and provide a summary of the implications of these results for our setting here.

MCCV is a very popular method in practice, because it directly measures out-of-sample prediction error and seems ``model-free''.  It has been used in \citet{SR2009} to justify their model selection choice.  To formally introduce it, let $Y_{i}=(Y_{i1},\dots,Y_{iT})$ and $X_{i}=(X_{i1},\dots,X_{iT})$.  Given observations $\{Y_{i},X_{i}\}_{i=1}^{n}$, to compute the MCCV mean squared error, we randomly draw a collection $\mathcal{R}$ of $b$ subsets of $\{1,\dots,n\}$ with size $n_{v}$ (test sample size). For a given choice of $n_v$, the model selected by MCCV minimizes the following out-of-sample MSE criteria among a set of models $\{\mathbf{M}_{\alpha}\}_{\alpha=1}^A$,
\begin{align}
\hat{\Gamma}_{\alpha,nT}^{MCCV}&=\frac{1}{n_{v}Tb}\sum_{s\in\mathcal{R}}\|\mathbb{Y}_{s}-\hat{\mathbb{Y}}_{\alpha,s^{c}}\|^{2}.
\end{align}
Here $\mathbb{Y}_{s}=(Y_{i}')_{i\in s}$ is an $nT\times 1$ vector that vertically stacks $Y_{i}'$ for all $i\in s$ and $\hat{\mathbb{Y}}_{\alpha,s^{c}}=\tilde{\mathbb{X}}_{s,\alpha}\hat{\beta}_{\alpha}^{s^{c}}$, where $\tilde{\mathbb{X}}_{s,\alpha}$ denotes the within-demeaned version of $\mathbb{X}_{s,\alpha}=(X_{i,\alpha}')_{i\in s}$  and $\hat{\beta}_{\alpha}^{s^{c}}$ is the estimator of the parameter vector of $\mathbf{M}_{\alpha}$ using the training data set $\{Y_{i},\mathcal{W}_{i}\}_{i\in s^{c}}$, where $s$ denotes a set in $\mathcal{R}$ and $s^{c}$ denotes the complement of $s$, i.e. the remaining $b-1$ subsets in the collection $\mathcal{R}$ after removing subset $s$. 

GICs have been shown to be closely related to MCCV criteria in the classical variable selection problem with cross-sectional data in \citet{Shao:1997}. Unlike MCCV, however, GICs can be estimated using the original sample without any resampling. Given a penalty level $\lambda_{nT}$, the model selected by GIC among a set of models $\{\mathbf{M}_{\alpha}\}_{\alpha=1}^A$ minimizes the following criterion
\begin{align}
GIC_{\alpha,\lambda_{nT}}&=\hat{\ell}_{nT}^{\alpha}-\lambda_{nT}k_{\alpha},
\end{align}
where $\hat{\ell}_{nT}^\alpha\equiv -n(T-1)\log(\hat{\sigma}_{\alpha}^{2}(\hat{\beta}_{\alpha})),$
$\hat{\sigma}^2(\beta_\alpha)=\sum_{i=1}^n\sum_{t=1}^T(\tilde{Y}_{it}-\tilde{X}_{it}'\beta_\alpha)^2/(n(T-1))$ and $$\hat{\beta}_{\alpha}=\arg\max_{\beta\in\mathcal{B}}\ell_{nT}^{\alpha}(\beta_{\alpha},\hat{\sigma}^{2}(\beta_{\alpha})).$$ For a formal derivation of the concentrated likelihood $\ell_{nT}^\alpha(\beta_{\alpha},\hat{\sigma}^{2}(\beta_{\alpha}))$, see Appendix \ref{app:gic}.

When examining the asymptotic behavior of MCCV and GICs, there are two properties that have been examined in the literature, model selection efficiency (optimality) as well as model selection consistency \citep[e.g.,][]{Shao:1993,Shao:1997,Claeskens:2008}.\footnote{See \citet{Yang:2005}, for an interesting analysis that poses the question of whether these two objectives can be combined and points to a clear trade-off between the two objectives.}

\begin{figure}[htbp]\caption{Illustrating the Difference between Minimizing the MSE of Impacts and Levels}\label{fig:MSE_levelvsdiff}
\centering
\includegraphics[width=\linewidth]{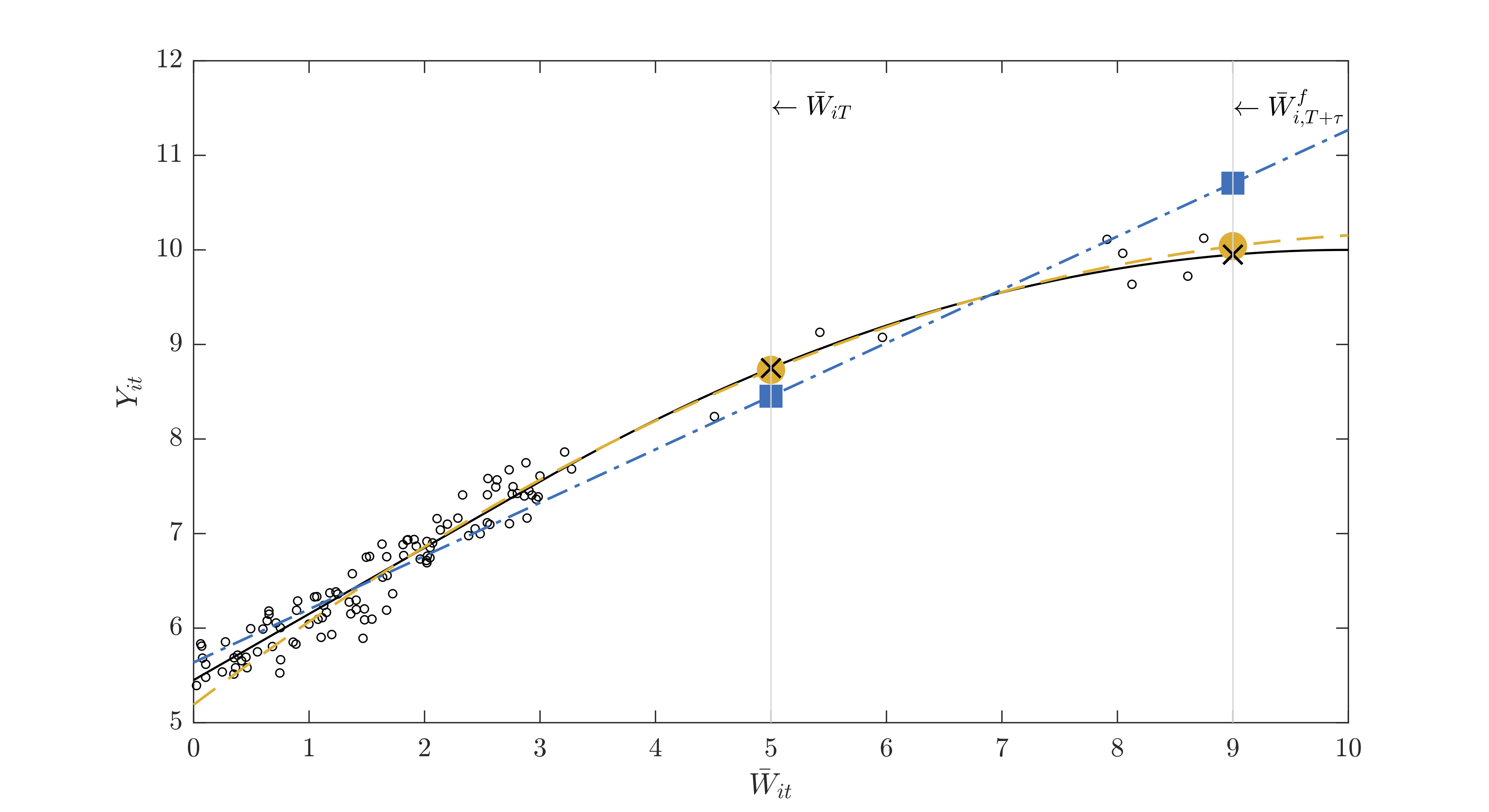}\\
\parbox{14cm}{\footnotesize \emph{Notes}:
The black solid line corresponds to the true model $\mu_{\star}(\mathcal{W}_{it})$ (nonlinear in $\bar W_{it}$);
blue dot-dashed line corresponds to the best approximating model that minimizes the  in-sample MSE (linear in $\bar W_{it}$),
the gold dashed line corresponds to the best approximating model that minimizes the impact MSE (quadratic in $\bar W_{it}$).
The small black circles correspond to a random sample of $(Y_{it},\bar W_{it})$ for a given $i$.  }
\end{figure}

A model selection criterion is said to be efficient \citep{Shao:1997,Claeskens:2008} if the ratio of the mean squared error (MSE) of the model selected by these criteria and the theoretical minimizer among those under consideration will converge in probability to one. Formally, a criterion $C$ is said to be model selection efficient in our context relying on FE estimators, if the following ratio converges to one
\begin{align}&\frac{L_{nT}(\widehat{\mathbf{M}}_C)}{L_{nT}(\mathbf{M}_{m})}\overset{p}{\rightarrow} 1,\end{align}
where $\widehat{\mathbf{M}}_C$ is the model selected by the criterion $C$, $\mathbf{M}_m$ is the theoretical minimizer of the loss function, $E_{\widehat{\mathbf{M}}_C}$ integrates over all random quantities except for $\widehat{\mathbf{M}}_C$,
\begin{align}
&L_{nT}(\widehat{\mathbf{M}}_C)=\sum_{i=1}^n\sum_{t=1}^TE_{\widehat{\mathbf{M}}_C}[(\tilde{Y}_{it}-\hat{\tilde{\mu}}_{\widehat{\mathbf{M}}_C}(\mathcal{W}_{it}))^2],\nonumber\\
&L_{nT}(\mathbf{M}^m)=\sum_{i=1}^n\sum_{t=1}^TE[(\tilde{Y}_{it}-\tilde{\mu}_{\mathbf{M}_m}(\mathcal{W}_{it}))^2].\nonumber
\end{align}

Among MCCV and GIC criteria, leave-one-out cross-validation (MCCV with $n_v=1$) and AIC (GIC with $\lambda_{nT}=2$) are model selection efficient \citep{Shao:1997}. While such criteria select models that minimize the MSE of the within-group transformed outcome,
they are not guaranteed to perform well for the ideal target for climate change projections in \eqref{eq:ccp_criterion}. The MSE of the within-group transformed outcome and the ideal target in \eqref{eq:ccp_criterion} can in fact result in substantially different best approximate models, even in settings where the true response function $\mu_\star$ depends only on a scalar $\bar{W}_{it}$. To illustrate this point, consider the prediction of the impact of projected climate change for a unit $i$ illustrated in Figure \ref{fig:MSE_levelvsdiff}. In this graphical illustration, both linear  and quadratic in $\bar W_{it}$ models are misspecified, but the quadratic model predicts the impact of the projected change from $\bar{W}_{iT}$ to $\bar{W}^f_{i,T+\tau}$ better. To see this, compare the gold circles with the blue squares, noting that the black crosses correspond to the  true impact. In this diagram, most observations are far away from the projected climate $\bar W^f_{i,T+\tau}$ and lie in the domain where the true impact function is linear. As a result, minimizing the MSE in levels results in a linear-in-$\bar W_{it}$ model, which has a larger bias at the projected climate. This example provides an intuition for why model selection efficiency relying on the MSE in levels, and therefore AIC and leave-one-out cross-validation, are not guaranteed to minimize our ideal target.

Consistent model selection can be suitable for the policy objective of climate change projections under certain conditions. Specifically, if one of the models under consideration nests the true model, a consistent model selection criterion $C$ chooses $\mathbf{M}_{\star}$ with probability approaching one asymptotically, $P(\widehat{\mathbf{M}}_C=\mathbf{M}_\star)\rightarrow 1$ as $n\rightarrow\infty$. As a result, consistent model selection criteria minimize the ideal target in \eqref{eq:ccp_criterion}
and particularly attain zero by selecting $\mathbf{M}_{\star}$ with probability one asymptotically.  As a result, we provide conditions on the tuning parameter choices for the MCCV and GIC that ensure consistent model selection in Online Appendix \ref{app:existing_criteria}. These results extend \citet{Shao:1993} and \citet{Sin:1996} to the mixed-frequency panel data setting we examine. We specifically show that MCCV with vanishing training-to-full sample ratios ($n_c/n\rightarrow 0$) and the GIC criteria proposed in \citet{Sin:1996} (e.g. GIC with $\lambda_{nT}=\log(\log(nT))$ as well as BIC (GIC with $\lambda_{nT}=\log(nT)$) are model selection consistent and will choose the true model or the most parsimonious model that nests it with probability approaching one as $n\rightarrow\infty$, \emph{assuming} that such a model is contained in the set under consideration. Otherwise, the criteria proposed in \citet{Sin:1996} are pseudo-consistent, i.e. consistent under misspecification, in the sense that they minimize the Kullback-Leibler divergence. The problem, however, is that minimizing the Kullback-Leibler divergence does not guarantee that we minimize the ideal target in \eqref{eq:ccp_criterion}.

In sum, this section underscores that the suitablity of consistent model selection criteria for the policy objective of climate change projections is limited to cases where at least one model under consideration nests the true model. While this is possible in applications where the scientific literature can characterize the physical relationship between temperature and the outcome, the literature on climate change impacts examines a variety of economic outcomes for which the relationship with temperature is difficult to fully characterize. When all models are misspecified, some of the existing criteria minimize the MSE of the within-group transformed outcome, whereas others minimize the Kullback-Leibler divergence. Neither of these two targets ensures that the selected model would perform well in terms of the ideal target in \eqref{eq:ccp_criterion}. As a result, we next propose the PWMSE that seeks to mimic the ideal target.

\subsection{Targeting the MSE of Projected Climate Change Impacts}\label{subsec:CCP-MSE}
In this section, we propose a new MSE criterion that targets the prediction of climate change impacts directly for a given climate projection $\mathcal{W}_{i,T+\tau}^f$, which we refer to as proximity-weighted MSE (PWMSE). We then propose an MCCV procedure to estimate the PWMSE and show that the model minimizing the estimated PWMSE is asymptotically minimizing its population analogue.

To provide an intuition for this criterion, let us re-examine Figure \ref{fig:MSE_levelvsdiff}. Recall that in this example, minimizing the MSE of the levels of the outcome would not lead us to choose the best approximate model at $W_{i,T+\tau}^f$. Now consider an alternative approach where instead of relying on the within-group transformation, we use within-differences between $T$ and prior years $T-r$, but only use prior years with weather ``close'' to $\bar W^f_{i,T+\tau}$ for model selection. This approach would guarantee that we choose the better approximate model, quadratic in $\bar W_{it}$. This insight guides the formulation of the PWMSE. Note that under strict exogeneity and additional regularity conditions,  minimizing the ideal MSE target in \eqref{eq:ccp_criterion} is equivalent to minimizing
\begin{align}
      \bar E[({\mu}_{\alpha}(\mathcal{W}^f_{i,T+\tau})-{\mu}_{\alpha}(\mathcal{W}_{i,T})-(Y_{i,T+\tau}  -Y_{iT}))^2]+ \bar E[(  u_{i,T+\tau}^f-u_{iT})^2],\label{eq:ccp_criterion_decomposition}
\end{align}
 where $u_{i,T+\tau}^f=\mu_{\star}(\mathcal{W}_{i,T+\tau})-\mu_{\star}(\mathcal{W}_{i,T+\tau}^f)+u_{i,T+\tau}$ and $\bar E$ is limit of the cross-sectional average of expectations that accounts for heterogeneity in distributions over space (see Condition \ref{cond:PWMSE}.5 below for a formal definition). 
 Since the second term in \eqref{eq:ccp_criterion_decomposition} does not depend on $\textbf{M}_{\alpha}$, we can select the model based on the first term. 
 
 In practice, the problem, as with other forecasting settings, is that we do not observe $Y_{i,T+\tau}$.  The standard approach in forecasting would be to mimic the $\tau$-step-ahead forecasting in the following way, 
\begin{align} 
\frac{1}{T-\tau}\sum_{t=\tau+1}^{T} \bar E[({\mu}_{\alpha}(\mathcal{W}_{it})-{\mu}_{\alpha}(\mathcal{W}_{i,t-\tau})-(Y_{it}-Y_{i,t-\tau}))^2].\label{eq:MSE_criterion_tau}\end{align}
This approach is not suitable in our setting since climate change implies that the distribution of weather is strongly persistent and possibly non-stationary.
As a result, this criterion targets precision of the functional approximation of $\mu_{\star}(w_2)-\mu_{\star}(w_1)$ under the past as opposed to the projected future climate.
For example, if in the past the climate was on average more moderate, while the projected climate is relatively hot, the above criterion will give relatively higher weight to moderate climates.
Moreover, there is no guarantee that the projected changes in the climate will follow a linear trend that would make changes between $T-\tau$ and $T$ comparable to changes between  $T$ and $T+\tau$. Finally, this criterion can be infeasible in this setting, since $\tau/T$ tends to be close to or greater than 1 in practice. 

We suggest the following criterion that relies on a weighting function that gives higher weights to historical years with climates that are more similar to the projected climate $\mathcal{W}_{i,T+\tau}^f$ and thereby mimics the ideal target more closely,
\begin{align}  
PWMSE_{\alpha}&=\sum_{r=1}^{T-1} \bar E[( {\mu}_{\alpha}(\mathcal{W}_{i,T-r})- {\mu}_{\alpha} (\mathcal{W}_{iT})-(\mu_{\star}(\mathcal{W}_{i,T-r})-\mu_{\star}(\mathcal{W}_{iT})))^2\pi(\mathcal{W}_{i,T-r},\mathcal{W}_{i,T+\tau}^f)],\label{eq:fccp_criterion}\end{align}
where   $\pi(\mathcal{W}_{j,T-r},\mathcal{W}_{i,T+\tau}^f)$ is a weighting function that gives higher weight to prior years with similar weather to the projected weather. The PWMSE can provide a good approximation to the ideal target in \eqref{eq:ccp_criterion}, if for each $i$ there exists a year in the sample that resembles the corresponding projected climate. 
The two criteria trivially coincide  if $T=2$ and   $\mathcal{W}_{i,1}=\mathcal{W}^f_{i,T+\tau}$ for all $i$. Intuitively, the weighting function $\pi(\cdot,\mathcal{W}^f_{i,T+\tau})$ ensures that the selected model provides improved prediction of climate change impacts for years with climates that resemble the corresponding projected climate, $\mathcal{W}^f_{i,T+\tau}$.

In practice, the weighting function is a user-specified component in the PWMSE criterion.
The function should be non-negative, bounded, and decrease with the distance between $\mathcal{W}_{i,T-r}$ and the target $\mathcal{W}_{i,T+\tau}^f$.
A natural choice for the weighting function is the following:
$$\pi(\mathcal{W}_{i,T-r},\mathcal{W}_{i,T+\tau}^f) =e^{-\frac{\|\mathcal{W}_{i,T-r}-\mathcal{W}_{i,T+\tau}^f\|_p^2}{h}},$$ 
which decays exponentially as the distance between $\mathcal{W}_{i,T-r}$ and $\mathcal{W}_{i,T+\tau}^f$ grows as measured by the norm, $\|.\|_p$. 

The choice of the norm should be guided by what aspects of the projected climate the researcher or policymaker would like to mimic. In the simulation section, we consider $L^2$ norms of the difference in annual and monthly mean temperature with a moderate level of the tuning parameter $h$. We also examine the impact of other different norms as well as the choice of the tuning parameter in additional simulations in Online Appendix \ref{sec:PWMSE_sims}, and we provide guidance for practitioners on the choice of norms and the tuning parameter $h$ in Online Appendix \ref{sec:guide_weight}. 

In order to estimate the PWMSE in practice, we propose the following MCCV procedure,
 \begin{enumerate}

\item For each $s$ randomly split  individuals $i=1,\dots, n$ into a training sample $\mathcal{I}^c_s$ of size $n_c$ and a testing sample  $\mathcal{I}^v_s$ of size $n_v$. Use training sample $\mathcal{I}^c_s$ to estimate $\hat{\mu}^s_{\alpha}(\cdot)$ and use the testing sample $n_v$ to compute the out-of-sample MSE,

\begin{align}  
\widehat{PWMSE}_{\alpha}^s=\sum_{r=1}^{T-1}\frac{1}{n_v} \sum_{i\in \mathcal{I}^v_s}  (\hat{\mu}^{s}_{\alpha}(\mathcal{W}_{iT})-\hat{\mu}^{s}_{\alpha}(\mathcal{W}_{i,T-r})-(Y_{iT}-Y_{i,T-r}))^2\pi(\mathcal{W}_{i,T-r},\mathcal{W}_{i,T+\tau}^f),\label{eq:MSE_criterion_sumtaustar}\end{align}
where $\pi(\mathcal{W}_{i,T-r},\mathcal{W}_{i,T+\tau}^f) =e^{-\frac{\|\mathcal{W}_{i,T-r}-\mathcal{W}_{i,T+\tau}\|_p^2}{h}}$.

  \item Repeat Step 1 $S$ times and compute the average (for variance reduction purposes), 
  \begin{align}
  \widehat{PWMSE}_{\alpha}&=\frac{1}{S}\sum_{s=1}^{S}\widehat{PWMSE}_{\alpha}^s.\end{align}  
\end{enumerate}

Next, we show that the model that minimizes $\widehat{PWMSE}_{\alpha}$ minimizes $PWMSE_\alpha$ with probability one as $n\rightarrow\infty$. Before we proceed to the proposition, we introduce the required conditions.

\begin{condition}[Consistency of FE]\label{cond:PWMSE} The following conditions hold:
\begin{enumerate}[1.]
\item $E[u_{it}|\mathcal{W}_{i1},\dots,\mathcal{W}_{iT},\mathcal{W}^f_{i,T+\tau},a_i]=0$ for all $i$ and $t$.
 \item  $(u_{i1},\dots,u_{iT},\mathcal{W}_{i1},\dots,\mathcal{W}_{iT},\mathcal{W}^f_{i,T+\tau})$ are (stochastically) independent  across $i$. 
  \item There exist constants   ${\varepsilon}>0$ and $M<  \infty$ such that for all  $i=1,\dots,N$, $t =1,\dots,T$, and $\alpha \in  \{\star,1,\dots, A\}$ the following moment conditions hold   
 \begin{equation}
      E [ {u}_{it}^{2+\varepsilon}]<M,  \quad E[({X}_{it,\alpha}^\prime  {X}_{it,\alpha})^{1+\varepsilon} ] <M,\label{eq:momentCondtions}
 \end{equation}
where ${X}_{it,\star}=\mu_{\star}(\mathcal{W}_{it})$. 
\item $ \frac{1}{T n} \sum_{i=1}^{n}\sum_{t=1}^{T}   E [\widetilde{X}_{it,\alpha}\widetilde{X}_{it,\alpha}']$ is uniformly positive definite for all $\alpha=1,\dots, A$ and $n$.
\item The following limits exist for all  $i=1,\dots,N$, $t,t'=1,\dots,T$,  $\alpha,\alpha' \in  \{\star,1,\dots, A\}$, and $h>0$ 
\begin{align}
  \bar E [ {u}_{it} {u}_{it'} \pi_{it}]&\equiv   \lim_{n\to\infty} \frac{1}{n}\sum_{i=1}^{n}  E  [{u}_{it} {u}_{it'} \pi_{it}] , \\
      \bar E  [\tilde{X}_{it,\alpha}\tilde{X}_{it',\alpha'}'\pi_{it}]&\equiv\lim_{n\to\infty} \frac{1}{n}\sum_{i=1}^{n}  E [\tilde{X}_{it,\alpha}\tilde{X}_{it',\alpha'}'\pi_{it}],   
\end{align}
where $\pi_{it}  = e^{-\frac{\|\mathcal{W}_{i,t}-\mathcal{W}_{i,T+\tau}\|_p^2}{h}}$.
\end{enumerate}
\end{condition}

Condition  \ref{cond:PWMSE} provides the set of sufficient conditions for the consistency of the $\widehat{PWMSE}_{\alpha}$ to be established in Proposition \ref{prop:CCP-MSE}. It includes the conditions required for the almost sure consistency of the FE estimator for misspesified models  in the case of independent, heterogeneously distributed observations. 
They are  analogous to sufficient conditions for the consistency of the OLS estimator in \citet[][Exercise 3.14]{white2014asymptotic} but have two subtle differences, Eq. \eqref{eq:momentCondtions} and  Condition \ref{cond:PWMSE}.5, which we explain below.

Condition \ref{cond:PWMSE}.1  is an augmented version of the strict exogeneity assumption that includes the projected climate $\mathcal{W}_{i,T+\tau}^f$. Augementation is required to make sure that the weights $\pi_{it}$ are not correlated with the scores $  {u}_{it} \tilde{X}_{it,\alpha}$  when establishing consistency of $\widehat{PWMSE}_{\alpha}$.\footnote{    In principle, one can imagine that the economic variables $Y_{it}$ in a location $i$ may have effect on the weather projections for that location $ W^f_{i,T+r}$ through the  anthropogenic climate feedback, which would invalidate Condition  1.1. However, this feedback is typically modeled only through the total global human influence and does not include direct impact of the local economic activity, which can be seen from the descriptions on general circulation models \citep{Teixeira2014}. As a result, any dependence of the mean of  $u_{it}$ on future projections $ W^f_{i,T+r}$ can be assumed negligible.}
  
 The cross-sectional independence assumption (Condition \ref{cond:PWMSE}.2) is imposed on both regressors and residuals. It is one of the sufficient conditions for the consistency of $\hat{\beta}_\alpha$ and the application of the uniform law of large numbers to the PWMSE. Remark \ref{rem:dependentData} below discusses that one can accommodate arbitrary dependence between the regressors $X_{it,\alpha}$ across $i$ in the presence of heterogeneity in the distribution under the assumption of bounded support. Independence of $u_{it}$ across $i$ (unconditional or conditional on the regressors) can be replaced by a weak-dependence assumption across $i$. This would, however, invalidate certain standard inference procedures, including cluster-robust standard errors for FE estimators. Finally, we emphasize that Condition \ref{cond:PWMSE}.2 allows for arbitrary serial correlation and does not restrict the dependence of $a_i$ across $i$. Therefore, the outcome may be spatially correlated under Condition \ref{cond:PWMSE}.2 through fixed effects $a_i$.

 Condition \ref{cond:PWMSE}.3 is a sufficient condition for the applicability of Markov's law of large numbers that results in the almost sure consistency of $\hat{\beta}_\alpha$ and   $\widehat{PWMSE}_{\alpha}$ for all $\alpha$. It is a rather mild moment restriction that is commonly assumed.
The difference with the standard conditions for consistency (Exercise 3.14 in \cite{white2014asymptotic}) is the moment restriction on $ {u}_{it} {u}_{it'} $ given in  \eqref{eq:momentCondtions}. It is required for the consistency of the PWMSE criterion.

Similar to the conventional OLS setup, Condition \ref{cond:PWMSE}.4 (no multicollinearity in approximate models) is required for the consistency of the FE estimator $\hat{\beta}_\alpha$.

 Finally,  Condition \ref{cond:PWMSE}.5 ensures the existence of a well-defined limit of the FE estimator in case of misspecified models (it is redundant for correctly specified models). 
 This condition also requires the existence of limits for the weighted averages of the means which is an important step to establish consistency of the PWMSE criterion. 
In the cases  of spatially distributed data, this condition would be satisfied if the corresponding mean functions considered as a function of continuous coordinates are Lebesgue integrable, in which case the limits on the right-hand side expression coincides with integrals over the geographic domain with weights equal to the spatial  intensity of the observations.\footnote{It is a direct corollary of Theorem 5.24 and Corollary 5.41 in \cite{kallenberg2017random}, as long as the spatial point process of weather measurement locations is stationary and ergodic.} 

\begin{remark}\label{rem:dependentData}
Note that the independence of $\mathcal{W}_{it}$ across $i$ is essentially without loss of generality. 
One can apply the theory in presence of arbitrary dependence of  weather measurements  across $i$ as long as ${X}_{it,\alpha}$ have bounded support, which is a realistic assumption in the case of the geophysical variables. 
In this case, one can interpret the independence of $u_{it}$ across $i$  as  independence  conditional on $(\mathcal{W}_{i1},\dots,\mathcal{W}_{iT},\mathcal{W}^f_{i,T+\tau})$   and moments \eqref{eq:momentCondtions} in conditional expectation sense with the same conditioning.
 Bounded support  for  ${X}_{it,\alpha}$ would immediately imply a conditional expectation version of \eqref{eq:momentCondtions}. 
 One should keep in mind, however, that the regression coefficients $\beta_\alpha$ and the notion of the best approximate model would depend on a particular draw of the spatial weather distribution. \qed
 \end{remark}

\begin{proposition}\label{prop:CCP-MSE} Consider a finite class of models $\mathbb{M}=\{\mathbf{M}_{\alpha}\}_{\alpha=1}^{A}$.
Suppose that Condition \ref{cond:PWMSE}  holds,  $n_c\to\infty $ and $n_v\to\infty $, as $n\rightarrow\infty$ holding $T$ fixed, and the split into training and validation samples is uniform and random.   
Then, the model that minimizes $ \widehat{PWMSE}_{\alpha}$  also minimizes  $ {PWMSE}_{\alpha} $
in the class of models with probability  1 as the sample size grows. 
\end{proposition}
The proof is provided in Appendix \ref{sec:proof_prop_3}.  If the model with minimal error ${PWMSE}_{\alpha}$ is unique,  Proposition~\ref{prop:CCP-MSE} implies that $ \widehat{PWMSE}_{\alpha}$ selects that model as the training and validation sample sizes grow. Otherwise, $ \widehat{PWMSE}_{\alpha}$ is guaranteed to select one of the models that minimize ${PWMSE}_{\alpha}$ with probability 1 as the training and validation sample sizes grow.

The feasible criterion ${PWMSE}_{\alpha}$ is only an approximate version of the ideal criterion \eqref{eq:ccp_criterion_decomposition} --- if all of the  historical data do not have any resemblance to the future climate, the criterion will fail at selecting a model that can perform well in terms of the ideal target.
In such an unfortunate occasion, however,  none of the existing alternative methods would perform better since they are not using any information about the  distribution of the future weather, unless one of the models under consideration nests the true model.  
That is, in the absence of historical observations analogous to the future climate, the only practical solution would be to make sure that one of the models under consideration is correctly specified and rely on a consistent model selection procedure.\footnote{Alternatively, one can impose the random effects assumption on $a_i$ and consider comparing model damage function predictions across locations and time (instead of time only). 
This approach can be used if the forecasts of the weather are very different from the historical data. For example, a climate projection may predict that London  becomes as hot as Cairo is now.
If there is no historical analog of Cairo-type weather for London, then under the random effect assumption one can use Cairo's observations as analog for future London climate in PWMSE computation. We leave this extension for future work. }

\subsection{Simulation Study}\label{sec:simulations_PWMSE}

We design a simulation study to illustrate the finite-sample performance of the MCCV, GIC and PWMSE criteria. We consider three DGPs of an outcome-temperature relationship: (i) the annual mean model ($A$), (ii) the quadratic in annual mean model ($QinA$), and (iii) the quarterly mean model ($Q$), and we evaluate the performance of each model selection criterion for selecting among a broader set of models.

The following functions generate the outcome $Y_{it}$ for the three DGPs we consider which correspond to the examples we provide in Example \ref{ex:AQinAQ}:
\begin{itemize}
\item Annual Mean ($A$): $Y_{it}=\bar{W}_{it}+a_{i,\alpha}+u_{it,\alpha}$,
\item Quadratic in Annual Mean ($QinA$): $Y_{it}=0.2\bar{W}_{it}-0.05\bar{W}_{it}^{2}+a_{i,\delta}+u_{it,\delta}$,
\item Quarterly Mean ($Q$): $Y_{it}=-0.25\bar{W}_{it}^{{Q}_{1}}+0.75\bar{W}_{it}^{Q_{3}}+a_{i,\gamma}+u_{it,\gamma}$,
\end{itemize}

For each of the DGPs, we use a random sample of counties from the National Climatic Data Center (NCDC) temperature dataset for the years 1968-1977 as $\mathcal{W}_{it}$ for $i=1,\dots,n$ and $t=1,\dots,T$, where $T=10$.  For each simulation replication, we generate $a_{i}|\mathcal{W}_{i1},\mathcal{W}_{i2},\dots,\mathcal{W}_{iT}\overset{i.i.d.}{\sim} N(0.5\bar{W}_{i},1)$, where $\bar{W}_{i}=\sum_{t=1}^{T}\sum_{\tau=1}^{H}W_{it\tau}/(TH)$.  The idiosyncratic shocks $u_{it}$ are generated as a bivariate mixture normal that is heteroskedastic and serially correlated as follows.  Let $u_{i}=(u_{i1},\dots,u_{iT})=\epsilon_{i}^{1}+\epsilon_{i}^{2}$, where $\epsilon_{i}^{1}|\mathcal{W}_{i1},\dots,\mathcal{W}_{iT},a_{i}\overset{i.i.d.}{\sim} N(-0.5,\Sigma_{1})$ and $\epsilon_{i}^{2}|\mathcal{W}_{i1},\dots,\mathcal{W}_{iT},a_{i}\overset{i.i.d.}{\sim}N(0.5,\Sigma_{2})$, with

{\fontsize{7}{7}\selectfont
\begin{align}\Sigma_{1}=\left(\begin{array}{cccccccccc}
   1     & 0.5   & 0.1   & 0     & 0     & 0     & 0     & 0     & 0     & 0 \\
    0.5   & 1     & 0.5   & 0.1   & 0     & 0     & 0     & 0     & 0     & 0 \\
    0.1   & 0.5   & 1     & 0.5   & 0.1   & 0     & 0     & 0     & 0     & 0 \\
    0     & 0.1   & 0.5   & 1     & 0.5   & 0.1   & 0     & 0     & 0     & 0 \\
    0     & 0     & 0.1   & 0.5   & 1     & 0.5   & 0.1   & 0     & 0     & 0 \\
    0     & 0     & 0     & 0.1   & 0.5   & 1     & 0.5   & 0.1   & 0     & 0 \\
    0     & 0     & 0     & 0     & 0.1   & 0.5   & 1     & 0.5   & 0.1   & 0 \\
    0     & 0     & 0     & 0     & 0     & 0.1   & 0.5   & 1     & 0.5   & 0.1 \\
    0     & 0     & 0     & 0     & 0     & 0     & 0.1   & 0.5   & 1     & 0.5 \\
    0     & 0     & 0     & 0     & 0     & 0     & 0     & 0.1   & 0.5   & 1 \\
\end{array}
\right),&~~\Sigma_{2}=\left(\begin{array}{cccccccccc}
    1     & 0.5   & 0.1   & 0     & 0     & 0     & 0     & 0     & 0     & 0 \\
    0.5   & 0.75  & 0.5   & 0.1   & 0     & 0     & 0     & 0     & 0     & 0 \\
    0.1   & 0.5   & 1     & 0.5   & 0.1   & 0     & 0     & 0     & 0     & 0 \\
    0     & 0.1   & 0.5   & 0.75  & 0.5   & 0.1   & 0     & 0     & 0     & 0 \\
    0     & 0     & 0.1   & 0.5   & 1     & 0.5   & 0.1   & 0     & 0     & 0 \\
    0     & 0     & 0     & 0.1   & 0.5   & 0.75  & 0.5   & 0.1   & 0     & 0 \\
    0     & 0     & 0     & 0     & 0.1   & 0.5   & 1     & 0.5   & 0.1   & 0 \\
    0     & 0     & 0     & 0     & 0     & 0.1   & 0.5   & 0.75  & 0.5   & 0.1 \\
    0     & 0     & 0     & 0     & 0     & 0     & 0.1   & 0.5   & 1     & 0.5 \\
    0     & 0     & 0     & 0     & 0     & 0     & 0     & 0.1   & 0.5   & 0.75 \\
\end{array}
\right).\nonumber
\end{align}
}

For each DGP, we consider different subsets of the following set of models:
\begin{itemize}
\item Annual Mean ($A$): 
$Y_{it}=\beta_{1}\bar{W}_{it}+a_{i}+u_{it}$,
\item Bi-annual Mean ($B$): 
$Y_{it}=\sum_{k=1}^{2} \beta_{k}\bar{W}^{B_k}_{it}+a_{i}+u_{it}$,
\item Quarterly Mean ($Q$):
$Y_{it}=\sum_{k=1}^{4} \beta_{k}\bar{W}^{Q_k}_{it}+a_{i}+u_{it}$,
\item Monthly Mean ($M$):
$Y_{it}=\sum_{k=1}^{12} \beta_{k}\bar{W}^{M_k}_{it}+a_{i}+u_{it}$,
\item Quadratic in Annual Mean ($QinA$):
$Y_{it}=\beta_{1}\bar{W}_{it}+\beta_{2}\bar{W}^2_{it}+a_{i}+u_{it}$,
\item 10$^\circ$F Bins ($Bins$):
$Y_{it}=\sum_{k=1}^{9} \beta_{k}{Bin^k}_{it}+a_{i}+u_{it}$.
\end{itemize}

Since the ideal target in \eqref{eq:ccp_criterion} involves future climate, we use the year of 1997 as the future period, twenty years after the last period of the 1968-1977 sample (i.e., $\tau=20$). In addition, we also consider other future climates by treating 1992 and 2002 as alternative future periods (i.e., $\tau=15,25$), respectively.
Data for these three years come from the same county-level NCDC temperature dataset. 

Given the importance of the pseudo-true parameter values as well as our ideal target evaluated at these values in our theoretical analysis, we simulate these quantities for models $A$, $B$, $Q$, $M$, $QinA$, and $Bins$ using $2,000$ simulation replications using the sample of all counties in our dataset ($n=3,074$) to ensure that our simulated quantities are as close as possible to their population analogues.  

Table \ref{tab:pseudo_true_param} presents simulated pseudo-true parameter values $\bar{\beta}_{\alpha}$.\footnote{We compute these values by taking the simulation mean of the estimated coefficients using the entire NCDC sample for each model across 2,000 simulations.} We also simulate the ideal target for our sample $\frac{1}{n}\sum_{i=1}^n(\bar{\mu}_{\alpha}(\mathcal{W}^f_{i,T+\tau})-\bar{\mu}_{\alpha}(\mathcal{W}_{iT})-(\mu_{\star}(\mathcal{W}^f_{i,T+\tau})-\mu_{\star}(\mathcal{W}_{iT})))^2$ for $\tau=20$, where $\bar{\mu}_{\alpha}(\mathcal{W}_{it})=X_{it,\alpha}'\bar{\beta}_{\alpha}$.
The results indicate that the ideal target is minimized when the model being considered is the DGP (true model). However, we note that the simulated ideal target is very close to that of the true model for models that nest it. In Appendix Table \ref{tab:ideal_target_plus}, we also report the ideal targets calculated for $\tau=15$ and $\tau=25$, respectively. The results are qualitatively similar.

We use the same random sample of $n$ counties from the full NCDC sample of 3,074 counties and use the temperature data for these counties between 1968-77 as our high-frequency regressor $\{\mathcal{W}_i\}_{i=1}^n$ in all the simulation designs. The outcome variable is generated using the DGP in question. All regression models are implemented on the generated data and the eight model selection criteria presented below are calculated for each model:

\begin{itemize}
\item MCCV ($n_{c}/n$)
\begin{enumerate}[-]
\item $n_{c}/n=p=0.75$ (MCCV with fixed training-to-full sample ratios, hereinafter MCCV-$p$),
\item $n_{c}=n^{-1/4}$ (MCCV with vanishing training-to-full sample ratios, hereinafter MCCV-Shao); 
\end{enumerate}
\item $GIC_{\alpha,\lambda_{nT}}=-n(T-1)\log(\hat{\sigma}_{\alpha}^{2})-\lambda_{nT}k_{\alpha}$, where $\hat{\sigma}^{2}=\sum_{i=1}^{n}\sum_{t=1}^{T}(\tilde{y}_{it}-\tilde{x}_{it,\alpha}'\hat{\beta}_{\alpha})^{2}/(nT)$,
\begin{enumerate}[-]
\item $\lambda_{nT}=2$ (AIC),
\item $\lambda_{nT}=\log(nT)$ (BIC),
\item $\lambda_{nT}=\sqrt{nT\log(\log(nT))}$ (SW$_1$),
\item $\lambda_{nT}=\sqrt{nT\log(nT)}$ (SW$_2$);
\end{enumerate}
\item PWMSE with weights specified as
\begin{enumerate}[-]
    \item $L^2$ norm of monthly differences with the tuning parameter $h=10$ (PWMSE$_1$),
    \item $L^2$ norm of yearly differences with the tuning parameter $h=10$ (PWMSE$_2$).\footnote{For formal definitions of these norms, see Section \ref{sec:PWMSE_sims}.}
\end{enumerate}
\end{itemize}
The simulation probabilities (proportions) of selecting a particular model using each of the criteria are computed using 500 simulation replications.\footnote{For the Monte Carlo cross-validation procedure of PWMSE as described in Online Appendix \ref{subsec:CCP-MSE}, we use $n_c/n=0.75$ and $S=100$ within each simulation.}

Figure \ref{fig:sim_model_select} presents the simulation results for $n=3,000$. The figure is arranged in three rows with each corresponding to a specific DGP (i.e., $A$, $QinA$, $Q$). Within each row, we consider two groups of comparisons: (i) across nested models (i.e., $A$, $B$, $Q$, $M$), and (ii) across possibly non-nested models (i.e., $A$, $QinA$, $Q$, $Bin$). Within the nested/non-nested models, we compare across all the four models as well as all combinations of every three out of the four models, which allows us to make systematic evaluations when the true model is included or excluded in the set of candidate models. Within each panel, each bar corresponds to a specific model selection criterion, and the height of a colored bar indicates the simulation proportion of a model being selected.

The results illustrate that, when the true DGP is in the set of models under consideration, all model selection criteria select the true model with high simulation probabilities. The MCCV with vanishing training-to-full sample ratios, BIC and the SW criteria select it with simulation probability approaching one, while AIC, MCCV with fixed training-to-full sample ratios, and the two PWMSE criteria show a non-zero probability to choose the less parsimonious model that nests the true DGP in certain circumstances. 

When all models under consideration are misspecified, however, different criteria tend to select different models. This difference is most evident in the results for DGP$=QinA$. In particular, the SW criteria tend to select the most parsimonious model, while the PWMSE is more likely to select the model that corresponds to the smallest value of the ideal target. This can be seen when considering the set of models $\{A, B, Q\}$. Referring to Table \ref{tab:pseudo_true_param}, under DGP$=QinA$, $A$ is the most parsimonious model but $B$ is the misspecified model that corresponds to the smallest value of the ideal target.\footnote{Note that BIC exhibits pseudo-inconsistency in this context, selecting $B$ and $Q$, unlike the pseudo-consistent SW criteria which select $A$ with probability equal to 1. We provide a formal explanation of the behavior of BIC in nested, misspecified model selection problems in Online Appendix \ref{app:pseudo-true}.}

We note that here we only report results considering the future period of $\tau=20$ and our PWMSE formations are based on two weight specifications. We present a set of extended simulation results under different future climates as well as using PWMSE with weights formed based on different combinations of the norms and tuning parameters. We make detailed discussions of these additional results in Online Appendix \ref{sec:PWMSE_sims}.

\subsection{Implications for Practice}\label{sec:implications}
The theoretical and simulation results have several implications for the use of model selection criteria in climate change impact studies. When the true model is considered in a set of models, all criteria we have discussed above tend to select the true model with high probabilities. In particular, MCCV with vanishing training-to-full sample ratios, BIC, and the SW criteria select it with probability approaching one as sample size grows. This illustrates the usefulness of the existing criteria in the context where the scientific literature can guide the applied researchers on the underlying relationship between climatic factors and the economic outcome of interest.

When all models being considered are misspecified, which is likely for many outcomes of interest in practice, different model selection criteria may favor different models. This issue is particularly concerning when the policy objective is to understand climatic impacts under a specific future climate projection. The existing model selection criteria do not incorporate the projected future climate into model selection and therefore would choose the same model regardless of the climate projection in question. By contrast, the PWMSE chooses the best approximate model for a given climate projection. Consistent with our theoretical predictions, the advantage from the PWMSE relative to existing criteria is greater when all models under consideration are misspecified. Given the importance of the choice of the weighting function when using the PWMSE in practice, we provide detailed guidance in Online Appendix \ref{sec:guide_weight} on the choice of the norm and tuning parameter.

Finally, we recommend that practitioners report the results of the different criteria we consider, including the PWMSE. As we illustrate in the next section, comparing the models selected by the different criteria can aid practitioners in interpreting their empirical results.

\section{Empirical Application: Temperature and Crop Yields}\label{sec:empirical}

In this section, we illustrate the model selection criteria we examine in a classic context of the empirical literature --- identifying nonlinear temperature effects on crop yields.
The agronomic studies have documented that the accumulation of heat is only beneficial to crop growth over certain ranges of the temperature and becomes detrimental otherwise \citep{ritchie1991temperature}. Previous statistical analyses also find evidence of nonlinearity in crop yield response to temperature under different estimation specifications \citep[e.g.,][]{SR2009,burke2016adaptation,gammans2017}. However, the qualitative similarity of nonlinearity does not diminish the importance of exploring the quantitative difference between alternative specifications, especially considering that nuances in the estimation results could be substantially magnified when it comes to projecting future climate impacts.

In this empirical application, we consider different specifications of temperature variables in the following model,
\begin{align} \log(Y_{it}) = \mu_{\alpha}(\mathcal{T}_{it}) + \theta_1 P_{it} + \theta_2 P_{it}^2 + \delta_{1,s} t + \delta_{2,s} t^2 + a_{i} + \epsilon_{it}, \nonumber\end{align}
where $Y_{it}$ represents corn yield (bushels/acre) in county $i$ in crop year $t$. The response function $\mu_{\alpha}(\mathcal{T}_{it})=X_{it,\alpha}'\beta_{\alpha}$ is a linear function of regressors constructed from the daily temperature time series over the growing season $\mathcal{T}_{it}\equiv \{T_{ith}\}_{h=1}^H$.\footnote{Previous studies have shown the importance of considering within-season temperature variation in modeling the response of crop yields since the seminal work in \cite{schlenker2006impact} and \cite{SR2009}.}  
 $P_{it}$ represents growing-season total precipitation, $\delta_{1,s}$ and $\delta_{2,s}$ characterize state-level quadratic trends, $a_i$ represents county fixed effect and $\epsilon_{it}$ is the error term.
 
We consider the following set of temperature specifications: (a) reference model with no temperature variables, (b) monthly average temperatures, (c) 1$^\circ$C daily temperature bins, (d) 3$^\circ$C step function, (e) degree days in the fashion of \cite{schlenker2006impact} (SHF degree days, hereafter), (f) piecewise linear function with one knot, and (g) piecewise linear function with two knots. Models (a)-(f) are model candidates considered in \cite{SR2009}, and model (g) is a more flexible variant of (f).\footnote{Model (c) uses bins constructed based on daily average temperatures, while (d) further utilizes diurnal variation in temperature by employing a sinusoidal interpolation between daily maximum and minimum temperatures before forming relevant bins.} All the models above only consider growing-season temperatures. The two piecewise linear specifications rely on knots selected by minimizing MSE.\footnote{We present the smallest ten MSEs in Appendix Table \ref{tab:knots_mse}.} Although the specifications considered here are not exhaustive, we believe they are sufficiently rich to illustrate the differentiated performances among different model selection criteria.

We obtain county-level corn yield data covering 1950-2015 from USDA Quick Stats. The source of historical weather information is the PRISM dataset, which provides spatially gridded daily data at 4km-by-4km resolution. We follow the data managing procedure in \cite{SR2009} and obtain county-level daily temperature and precipitation over 1950-2015. Based on the merged county-level data, we first conduct estimation using an unbalanced panel of all available observations. This unbalanced sample contains 2,278 counties with a total of 120,995 observations. The estimation results, as reported in Appendix Table \ref{tab:yld_est_unbalanced}, are in line with previous findings.\footnote{We also report the SNR for each model, and these ratios are mostly close to 0.40. When calculating, we consider all weather variables as the signal component, and we obtain the SNRs by projecting out all the time trends and fixed effects.} We also implement the estimation based on a balanced panel of 679 counties in the core region of the corn belt, with a total of 44,814 observations. The estimation results, presented in Appendix Table \ref{tab:yld_est_balanced}, are qualitatively similar. 

We illustrate the different projection results based on these empirical estimates in the mid-century. Specifically, we recalculate all relevant temperature variables for the 2050 climate at the county level using projected climate data from the HadCM3-B1 model. We then calculate the predicted changes in these temperature variables at the county level by differencing their 2050 values with their 2015 counterparts, and form predicted changes in log yields by applying different sets of the empirical estimates. The predicted changes in log yields based on different models are mapped in Figure \ref{fig:ag_projections}.

As for model selection, we first apply the existing model selection criteria, including the two MCCV and the four GICs. Table \ref{tab:ms_temp_yield} shows that, evaluated on the unbalanced panel, AIC and BIC select the 3$^\circ$C step function, the two cross-validation procedures and SW$_1$ select the two-knot piecewise function (knots at 24$^\circ$C and 26$^\circ$C), and SW$_2$ selects the one-knot piecewise linear function (knot at 29$^\circ$C). When evaluated on the balanced panel, both AIC and BIC still select the 3$^\circ$C step function, while the two SW criteria and the two MCCV procedures all choose the one-knot piecewise function for which the MSE-minimizing knot is at 30$^\circ$C. We visualize these selected nonlinear temperature response functions in Figure \ref{fig:yld_fn}. Although the nonlinear patterns are qualitatively similar for these selected models, those piecewise models selected by MCCV and SW criteria are more parsimonious while the step function selected by AIC and BIC is more complex. 

As we discussed earlier, none of the criteria above are specifically tailored for the policy objective of better understanding potential future impacts of climate change. Therefore, we apply our proposed PWMSE criteria that incorporate projected future climatic conditions to the seven candidate models with $S=1000$. Since the finest temporal resolution of our HadCM3-B1 projection data is at the monthly level, we consider the norms based on monthly and annual differences when forming weights. In Table \ref{tab:pwmse_yields}, we present the PWMSE for each model evaluated with different weights for the unbalanced panel and the balanced panel, respectively. Specifically, we consider the case with no weight specified, and the cases with weights based on the $L^\infty$ and $L^2$ norms of annual and monthly temperatures (i.e., M1, M2, Y1, and Y2, respectively), scaled by $h=1,10,100$, respectively. 
For the unbalanced panel, the PWMSEs are minimized by the model of 3$^\circ$C step function except for the cases of $h=1$ where the selected model is degree days in the fashion of \cite{schlenker2006impact} (i.e., SHF degree days).\footnote{The SHF degree days include three variables: the growing degree days (GDD), the square of GDD, and the square-root of heat degree days (HDD). GDD accumulates degree-by-days in the range of 8-32$^\circ$C and HDD accumulates degree-by-days above 34$^\circ$C.} But we note that setting $h=1$ in this unbalanced panel exercise may have assigned much higher weights to some of the historical observations that correspond to less frequent growing counties on the peripheral regions of the corn belt. For the balanced panel, the model of 3$^\circ$C step function delivers the lowest PWMSE regardless of how we specify the weights.

Compared with the models selected by the SW criteria, the PWMSE criterion selects models that result in a qualitatively similar damage function, as we illustrated visually in Figure \ref{fig:yld_fn}. We also note that the piecewise functions selected by the SW criteria deliver PWMSEs that are only slightly larger than those of the 3$^\circ$C step function. These similarities may reflect that both models are fairly close approximations of the underlying DGP, especially considering that they deliver damage functions that are consistent qualitatively with agronomic predictions. 

The models selected by the SW criteria are more parsimonious compared with those selected by the PWMSE criterion. This result is consistent with our theoretical discussion and simulation illustration. Nevertheless, we highlight that only the PWMSE criterion takes the projected future climate  into account directly through specifying the weights, while the SW criteria only depend on the historical observations. Our results indicate that the larger model of the 3$^\circ$C step function provides higher flexibility that potentially improves the projection-specific prediction for the policy objective of climate change projection. This tailored model selection has quantitative implications since the predicted outcomes under HadCM3-B1, although similar in their spatial gradient, have noticeable differences in their extents of yield reductions across the piecewise and 3$^\circ$C step functions, as illustrated in Appendix Figure \ref{fig:kdensity_hadcm3}. These quantitative differences matter for the policy arrangements of investment activities and adaptation efforts that target specific areas toward a given future climate projection.

\section{Conclusion}
This paper formalizes the model selection problem as well as the policy target faced by applied researchers and policymakers interested in examining climate change impacts on outcomes of interest.  Building on this crucial first step, the paper first provides conditions under which existing criteria, specifically MCCV and GICs, are appropriate for the policy objective. We show that consistent model selection criteria are suitable if at least one of the models under consideration nests the truth. Since this requirement is restrictive for economic outcomes for which the relationship with temperature is not well understood, we propose a proximity-weighted MSE criterion that targets the MSE of projected climate change impacts directly. We illustrate that these criteria choose models that minimize the ideal target with high probability in a simulation analysis. We demonstrate the empirical relevance of our theoretical analysis in the context of an application on climate change projection of agricultural yields.

While this paper constitutes an important first step toward principled model selection in this policy-relevant empirical context, there are several interesting directions for future research. In light of recent work on exogeneity in climate econometrics \citep{Pretis:2021}, developing methods that relax the strict exogeneity assumption is an important direction for future work.  More flexible procedures to estimate the response functions would also be a good substitute to the model selection approach taken in this literature.  Furthermore, allowing for possible nonlinearities between regressors and fixed effects is another important departure from the setup in this paper.  Finally, providing valid post-selection inference for the aforementioned methods constitutes a priority for future work.
\appendix

\linespread{1}

\section{Proof of Proposition \ref{prop:CCP-MSE}}\label{sec:proof_prop_3}

The proof proceeds in 5 steps.

\noindent\textbf{Step 1 (Auxiliary Limits)}.
Consider any subsample  $\mathcal{I}$ out of $n$ cross-sectional units of growing size $n_\mathcal{I}$. Under Conditions \ref{cond:PWMSE}.2 and \ref{cond:PWMSE}.3, one can apply Propositions 3.9 and Corollary 3.9 (Markov's law of large numbers) from \cite{white2014asymptotic} to show that  the following sums converge to 0 almost surely,
\begin{align}
   &  \frac{1}{n_\mathcal{I}}\sum_{i\in \mathcal{I}}  ({u}_{it} {u}_{it'}    \pi_{ir} - E({u}_{it} {u}_{it'}  \pi_{ir}))\overset{a.s.}{\to}0,\\
   &  \frac{1}{n_\mathcal{I}}\sum_{i\in \mathcal{I}}  ({u}_{it} \tilde{X}_{it',\alpha'}   \pi_{ir} - E({u}_{it} \tilde{X}_{it',\alpha'} \pi_{ir})) \overset{a.s.}{\to}0,\\
  &\frac{1}{n_\mathcal{I}} \sum_{i\in \mathcal{I}}    \big(\tilde{X}_{it,\alpha}\tilde{X}_{it',\alpha'}'\pi_{ir}) 
   - E \tilde{X}_{it,\alpha}\tilde{X}_{it',\alpha'}'\pi_{ir})  \big)\overset{a.s.}{\to}0, 
\end{align}
where  $\pi_{ir}=\pi(\mathcal{W}_{i,T-r},\mathcal{W}^f_{i,T+\tau})=e^{-\frac{\|\mathcal{W}_{i,T-r}-\mathcal{W}_{i,T+\tau}\|_p^2}{h}}$ for each $i,r$ (indices domains are  $t,t',r=1,\dots,T$; $\alpha,\alpha' \in  \{\star,1,\dots, A\}$ and $h>0$).

\noindent\textbf{Step 2 (Consistency of FE estimator)}. Consider the FE estimator for a subsample with split index $s$ for the model $\alpha$ that determines $\hat{\mu}^{s}_{\alpha}(\cdot)$,
\begin{align}
     \hat\beta^{s}_\alpha= \left(\frac{1}{T n_c} \sum_{t=1}^{T} \sum_{i\in \mathcal{I}^c_s} [\widetilde{X}_{it,\alpha}\widetilde{X}_{it,\alpha}']\right)^{-1}\frac{1}{T n_c} \sum_{t=1}^{T}  \sum_{i\in \mathcal{I}^c_s} [\widetilde{X}_{it,\alpha}(\widetilde{u}_{it}+ \widetilde{X}_{it,\star}  )].
\end{align}
(As before, the true regressors $X_{it,\star} $  can be chosen equal to   $ \mu_{\star}(\mathcal{W}_{it})$ even if the true model does not admit finite linear index representation.)

Under Condition~\ref{cond:PWMSE}, the corresponding population analog $\beta_{\alpha} $ exists.
Since the split is taken at random, the limit coincides with the full sample limit,
\begin{align}
     \beta_{\alpha}=\left( \sum_{t=1}^{T} \bar{E}[\widetilde{X}_{i,\alpha}\widetilde{X}_{i,\alpha}']\right)^{-1} \sum_{t=1}^{T} \bar{E}[\widetilde{X}_{i,\alpha}\widetilde{X}_{i,\star}'].
\end{align}
By Step 1,  the law of large numbers  implies  $\hat\beta^{s}_\alpha\overset{a.s.}{\to} \beta_{\alpha}$ as $n_c\to \infty$ for each $s$ and $\alpha$ under consideration.
As a result, one can define the approximate damage function
as  $\mu_\alpha(\mathcal{W}_{it}) = {X}_{it}^\prime \beta_{\alpha}$.

\noindent\textbf{Step 3 (Uniform LLN for PWMSE).} 
 Consider  the PWMSE estimate based on a single sample split $s$,
  \begin{align}
  \widehat{PWMSE}_{\alpha}^s=\sum_{r=1}^{T-1}\frac{1}{n_v} \sum_{i\in \mathcal{I}^v_s}  (\hat{\mu}^{s}_{\alpha}(\mathcal{W}_{iT})-\hat{\mu}^{s}_{\alpha}(\mathcal{W}_{i,T-r})-(Y_{iT}-Y_{i,T-r}))^2\pi(\mathcal{W}_{i,T-r},\mathcal{W}^f_{i,T+\tau})
  \end{align}
and  its component for an individual time period $r$,
\begin{align}
  \frac{1}{n_v} \sum_{i\in \mathcal{I}^v_s}  (\hat{\mu}^{s}_{\alpha}(\mathcal{W}_{iT})-\hat{\mu}^{s}_{\alpha}(\mathcal{W}_{i,T-r})-(Y_{iT}-Y_{i,T-r}))^2\pi(\mathcal{W}_{i,T-r}, \mathcal{W}^f_{i,T+\tau}).\label{eq:PWMSE_rcomponent}
\end{align}

Let $\mathcal{B}_{\alpha}$ be any compact subset of $\mathbb{R}^{k_{\alpha}}$ that contains $\beta_{\alpha}$ as an interior point.
By Step 2,  $ \hat\beta^s_\alpha\in \mathcal{B}_\alpha$ with probability 1 as $n_c$ grows.

For each $\alpha=1,\dots,A$, consider an empirical process indexed by a vector $b_\alpha\in \mathcal{B}_\alpha$.
\begin{align}
 \Xi^{r,s}_{\alpha,n_v}(b_\alpha)= ( E^s_{n_v} -\bar{E}) \big( ({X}_{iT,\alpha}-{X}_{i,T-r,\alpha})^\prime b_\alpha -(Y_{iT}-Y_{i,T-r})\big)^2\pi_{ir}.
\end{align}
where operator notation $(E^s_{n_v} -\bar{E})f_i= E^s_{n_v}f_i -\bar{E} f_i $ is used to define an empirical process indexed by $f$.
This process can be rewritten as 
\begin{align}
 \Xi^{r,s}_{\alpha,n_v}(b_\alpha)&= b^\prime_\alpha  ( E^s_{n_v} -\bar{E}) \big( ({X}_{iT,\alpha}-{X}_{i,T-r,\alpha})({X}_{iT,\alpha}-{X}_{i,T-r,\alpha})^\prime   \pi_{ir}\big)  b_\alpha ,\\
 & -2( E^s_{n_v} -\bar{E}) \big((u_{iT}-u_{i,T-r}  + {X}_{iT,\star}-{X}_{i,T-r,\star}) ({X}_{iT,\alpha}-{X}_{i,T-r,\alpha})^\prime   \pi_{ir}\big)  b_\alpha  ,\\
 & + ( E^s_{n_v} -\bar{E}) \big(  (u_{iT}-u_{i,T-r}  + {X}_{iT,\star}-{X}_{i,T-r,\star})^2 \pi_{ir}\big) 
\end{align}
By Step 1, the LLN applies for each of the coefficient matrices above. 
Since the index set $\mathcal{B}_\alpha$ for the empirical process is a compact set in a finite-dimensional Euclidean space,   
\begin{equation}
  \max_{b_\alpha\in   \mathcal{B}_{\alpha}} |\Xi^{r,s}_{\alpha,n_v}(b)|\overset{a.s.}{\to} 0\quad as \quad n_v\to\infty.
\end{equation}
It implies   for each  $\alpha=1,...,A$ as $ n_v\to\infty $ 
  \begin{align}
   \frac{1}{S}\sum_{s=1}^{S}  \sum_{r=1}^{T-1}\frac{1}{n_v} \sum_{i\in \mathcal{I}^v_s}  (\hat{\mu}^{s}_{\alpha}(\mathcal{W}_{iT})-\hat{\mu}^{s}_{\alpha}(\mathcal{W}_{i,T-r})-(Y_{iT}-Y_{i,T-r}))^2\pi_{ir}\nonumber \\
 -\frac{1}{S}\sum_{s=1}^{S}  \sum_{r=1}^{T-1} \bar E \big[ (\hat{\mu}^{s}_{\alpha}(\mathcal{W}_{iT})-\hat{\mu}^{s}_{\alpha}(\mathcal{W}_{i,T-r})-(Y_{iT}-Y_{i,T-r}))^2\pi_{ir}| \hat\beta^{s}_{\alpha} \big] \overset{a.s.}{\to} 0,\label{eq:UniformLLN}
  \end{align}
where
\begin{align}
  &\bar E \big[ (\hat{\mu}_{\alpha}(\mathcal{W}_{iT})-\hat{\mu}_{\alpha}(\mathcal{W}_{i,T-r})-(Y_{iT}-Y_{i,T-r}))^2\pi_{ir}| \hat\beta^{s}_{\alpha} \big]\\
  &= ( \hat\beta^{ s}_{\alpha} )^\prime \bar{E} \big( ({X}_{iT,\alpha}-{X}_{i,T-r,\alpha})({X}_{iT,\alpha}-{X}_{i,T-r,\alpha})^\prime   \pi_{ir}\big)   \hat\beta^{s}_{\alpha},\\
 & -2   \bar{E}  \big((u_{iT}-u_{i,T-r}  + {X}_{iT,\star}-{X}_{i,T-r,\star}) ({X}_{iT,\alpha}-{X}_{i,T-r,\alpha})^\prime   \pi_{ir}\big)   \hat\beta^{s}_{\alpha}  ,\\
 & +    \bar{E}  \big(  (u_{iT}-u_{i,T-r}  + {X}_{iT,\star}-{X}_{i,T-r,\star})^2 \pi_{ir}\big) 
\end{align}

\noindent\textbf{Step 4 (Conditional Expectation of PWMSE).}
 This component can be rewritten as
  \begin{align}
  &\bar E [ (\hat{\mu}^{s}_{\alpha}(\mathcal{W}_{iT})-\hat{\mu}^{s}_{\alpha}(\mathcal{W}_{i,T-r})-(\mu_\star (\mathcal{W}_{iT})-\mu_\star (\mathcal{W}_{i,T-r})))^2\pi_{ir} | \hat\beta^{s}_{\alpha}]\label{eq:pr3_rterm1}\\
   & + \bar E [\big(\hat{\mu}^{s}_{\alpha}(\mathcal{W}_{iT})-\hat{\mu}^{s}_{\alpha}(\mathcal{W}_{i,T-r})-(\mu_\star (\mathcal{W}_{iT})-\mu_\star (\mathcal{W}_{i,T-r}))\big)(u_{iT}-u_{i,T-r}) \pi_{ir} | \hat\beta^{s}_{\alpha}]\label{eq:pr3_rterm2}\\
  & + \bar E (u_{iT}-u_{i,T-r})^2\pi_{ir} \label{eq:pr3_rterm3}.
  \end{align}
The   term    \eqref{eq:pr3_rterm1}  can be further decomposed as follows,
\begin{align}
  &\bar E  [\big(\hat{\mu}^{s}_{\alpha}(\mathcal{W}_{iT})-\hat{\mu}^{s}_{\alpha}(\mathcal{W}_{i,T-r})-(\mu_\star (\mathcal{W}_{iT})-\mu_\star (\mathcal{W}_{i,T-r}))\big)^2\pi_{ir}  | \hat\beta^{s}_{\alpha}]\\
 &=\bar E  [ \big(\hat{\mu}^{s}_{\alpha}(\mathcal{W}_{iT})-\hat{\mu}^{s}_{\alpha}(\mathcal{W}_{i,T-r})-( {\mu}_{\alpha}(\mathcal{W}_{iT})- {\mu}_{\alpha}(\mathcal{W}_{i,T-r})\big)^2\pi_{ir} | \hat\beta^{s}_{\alpha}] \\
  &+\bar E \big( {\mu}_{\alpha}(\mathcal{W}_{iT})- {\mu}_{\alpha}(\mathcal{W}_{i,T-r})-(\mu_\star (\mathcal{W}_{iT})-\mu_\star (\mathcal{W}_{i,T-r}))\big)^2\pi_{ir}\\
    &+ 2\bar E [  \big( {\mu}_{\alpha}(\mathcal{W}_{iT})- {\mu}_{\alpha}(\mathcal{W}_{i,T-r})-(\mu_\star (\mathcal{W}_{iT})-\mu_\star (\mathcal{W}_{i,T-r}))\big)\nonumber\\
   & \times (\hat{\mu}^{s}_{\alpha}(\mathcal{W}_{iT})-\hat{\mu}^{s}_{\alpha}(\mathcal{W}_{i,T-r})-( {\mu}_{\alpha}(\mathcal{W}_{iT})- {\mu}_{\alpha}(\mathcal{W}_{i,T-r}))) \pi_{ir}| \hat\beta^{s}_{\alpha}]. \label{eq:pr3_crossterm}
\end{align}

The conditional expectation of the cross-product term  \eqref{eq:pr3_crossterm} is given by 
\begin{align}
&\bar E\Big[\big( {\mu}_{\alpha}(\mathcal{W}_{iT})- {\mu}_{\alpha}(\mathcal{W}_{i,T-r})-(\mu_\star (\mathcal{W}_{iT})-\mu_\star (\mathcal{W}_{i,T-r}))\big)\\
&\times(\hat{\mu}^{s}_{\alpha}(\mathcal{W}_{iT})-\hat{\mu}^{s}_{\alpha}(\mathcal{W}_{i,T-r}) -  {\mu}^{s}_{\alpha}(\mathcal{W}_{iT})+ {\mu}^{s}_{\alpha}(\mathcal{W}_{i,T-r})) |\hat\beta^{s}  \Big]\\
&= \bar E \big[\big( {\mu}_{\alpha}(\mathcal{W}_{iT})- {\mu}_{\alpha}(\mathcal{W}_{i,T-r})-(\mu_\star (\mathcal{W}_{iT})-\mu_\star (\mathcal{W}_{i,T-r}))\big)\nonumber  \\
&\times(\tilde{X}_{iT,\alpha}-\tilde{X}_{i,T-r,\alpha})^\prime\big] ( \hat\beta^{s}_\alpha-\beta_\alpha). 
 \end{align}
where the last equality follows from the independence of the training and validation samples.
In addition, under strict exogeneity (Condition \ref{cond:PWMSE}.1),   we obtain the following,
\begin{align}
  \bar E\left[\big(\hat{\mu}^{s}_{\alpha}(\mathcal{W}_{iT})-\hat{\mu}^{s}_{\alpha}(\mathcal{W}_{i,T-r})-(\mu_\star (\mathcal{W}_{iT})-\mu_\star (\mathcal{W}_{i,T-r}))\big)(u_{iT}-u_{i,T-r}) \pi_{ir} \right]=& \\
   \bar E\left[\big({\mu}_{\alpha}(\mathcal{W}_{iT})-{\mu}_{\alpha}(\mathcal{W}_{i,T-r})-(\mu_\star (\mathcal{W}_{iT})-\mu_\star (\mathcal{W}_{i,T-r}))\big)(u_{iT}-u_{i,T-r}) \pi_{ir} \right]=& 0.
\end{align}
Summarizing the previous steps,
\begin{align}
    &\bar E [(\hat{\mu}^{s}_{\alpha}(\mathcal{W}_{iT})-\hat{\mu}^{s}_{\alpha}(\mathcal{W}_{i,T-r})-(Y_{iT}-Y_{i,T-r}))^2\pi_{ir} |\hat\beta^{s} _\alpha]\\
    =& \bar E\big[ ( {\mu}_{\alpha}(\mathcal{W}_{i,T-r})- {\mu}_{\alpha} (\mathcal{W}_{iT})-(\mu_{\star}(\mathcal{W}_{i,T-r})-\mu_{\star}(\mathcal{W}_{iT}))\big)^2\pi_{ir}\big] \\
    +&  ( \hat\beta^{s}_{\alpha}-\beta_\alpha)^\prime \Omega_{r,\alpha}     ( \hat\beta^{s}_{\alpha}-\beta_{\alpha})  \label{eq:omegaTerm}\\
    +& \bar E \big[\big( {\mu}_{\alpha}(\mathcal{W}_{iT})- {\mu}_{\alpha}(\mathcal{W}_{i,T-r})-(\mu_\star (\mathcal{W}_{iT})-\mu_\star (\mathcal{W}_{i,T-r}))\big)  \\
&\times(\tilde{X}_{iT,\alpha}-\tilde{X}_{i,T-r,\alpha})^\prime\big] ( \hat\beta^{s}_{\alpha}-\beta_{\alpha})\\
    +& \bar E (u_{iT}-u_{i,T-r})^2\pi_{ir} ,
\end{align}
where $\Omega_{r,\alpha} =\bar E (\tilde{X}_{iT,\alpha}-\tilde{X}_{i,T-r,\alpha})(\tilde{X}_{iT,\alpha}-\tilde{X}_{i,T-r,\alpha})^\prime \pi_{ir} $.
 The last term in the displayed formula above does not depend on $\alpha$ so it will not affect the model selection.

\noindent\textbf{Step 5 (Using continuous mapping theorem)}. As a result of equation  \eqref{eq:UniformLLN} and  Step 4, the model $\alpha$  that minimizes $\widehat{PWMSE}_{\alpha}$   as $n_v\to\infty$  also minimizes   with probability 1 the following expression
\begin{align}
  {PWMSE}_{\alpha}  +
   \frac{1}{S}\sum_{s=1}^{S} ( \hat\beta^s _{\alpha}-\beta_{\alpha})^\prime \Omega_\alpha  ( \hat\beta^s _{\alpha}-\beta_{\alpha}) +  \frac{1}{S}\sum_{s=1}^{S} \Theta_{\alpha} ( \hat\beta^{s}_{\alpha}-\beta_{\alpha}), \label{eq:asy_PWMSE_proof}
\end{align}
where $\Omega_\alpha = \sum_{r=1}^{T-1} \Omega_{r,\alpha} $ and $$\Theta_{\alpha}=  \sum_{r=1}^{T-1} \bar E \big[\big( {\mu}_{\alpha}(\mathcal{W}_{iT})- {\mu}_{\alpha}(\mathcal{W}_{i,T-r})-(\mu_\star (\mathcal{W}_{iT})-\mu_\star (\mathcal{W}_{i,T-r}))\big)  \times(\tilde{X}_{iT,\alpha}-\tilde{X}_{i,T-r,\alpha})^\prime\big] $$ by the continuous mapping theorem.
Since by Step 2    $   \hat\beta^s _{\alpha} \overset{a.s.}{\to} \beta_{\alpha}$ ,    the model selected by $\widehat{PWMSE}_{\alpha}$ also minimizes   ${PWMSE}_{\alpha}$ with probability  1 as the sample size grows. 
\qed

\bibliography{cggk_2022}

\begin{thebibliography}{81}
\providecommand{\natexlab}[1]{#1}
\providecommand{\url}[1]{\texttt{#1}}
\expandafter\ifx\csname urlstyle\endcsname\relax
  \providecommand{\doi}[1]{doi: #1}\else
  \providecommand{\doi}{doi: \begingroup \urlstyle{rm}\Url}\fi

\bibitem[Addoum et~al.(2020)Addoum, Ng, and Ortiz-Bobea]{addoum2020temperature}
J.~M. Addoum, D.~T. Ng, and A.~Ortiz-Bobea.
\newblock Temperature shocks and establishment sales.
\newblock \emph{The Review of Financial Studies}, 33\penalty0 (3):\penalty0
  1331--1366, 2020.

\bibitem[Adhvaryu et~al.(2020)Adhvaryu, Kala, and Nyshadham]{Adhvaryu:2020}
A.~Adhvaryu, N.~Kala, and A.~Nyshadham.
\newblock The light and the heat: Productivity co-benefits of energy-saving
  technology.
\newblock \emph{Review of Economics and Statistics}, 102\penalty0 (4):\penalty0
  779--792, 2020.

\bibitem[Anderson et~al.(2017)Anderson, Johnson, and Koyama]{Anderson:2017}
R.~W. Anderson, N.~D. Johnson, and M.~Koyama.
\newblock Jewish persecutions and weather shocks: 1100--1800.
\newblock \emph{The Economic Journal}, 127\penalty0 (602):\penalty0 924--958,
  2017.

\bibitem[Andreou and Ghysels(2006)]{Andreou:2008}
E.~Andreou and E.~Ghysels.
\newblock Sampling frequency and window length trade-offs in data-driven
  volatility estimation: appraising the accuracy of asymptotic approximations.
\newblock In D.~Terrell and T.~B. Fomby, editors, \emph{{Econometric Analysis
  of Financial and Economic Time Series}}, {Advances in Econometrics, Volume 20
  Part 1}, pages 155--181. Emerald Group Publishing Limited, 2006.

\bibitem[Andreou et~al.(2010)Andreou, Ghysels, and Kourtellos]{AGK:2010}
E.~Andreou, E.~Ghysels, and A.~Kourtellos.
\newblock Regression models with mixed sampling frequencies.
\newblock \emph{J. Econometrics}, 158\penalty0 (2):\penalty0 246--261, 2010.
\newblock ISSN 0304-4076.
\newblock \doi{10.1016/j.jeconom.2010.01.004}.
\newblock URL \url{http://dx.doi.org/10.1016/j.jeconom.2010.01.004}.

\bibitem[Arag{\'o}n et~al.(2021)Arag{\'o}n, Oteiza, and Rud]{aragon2021climate}
F.~M. Arag{\'o}n, F.~Oteiza, and J.~P. Rud.
\newblock Climate change and agriculture: Subsistence farmers' response to
  extreme heat.
\newblock \emph{American Economic Journal: Economic Policy}, 13\penalty0
  (1):\penalty0 1--35, 2021.

\bibitem[Arellano(2003)]{arellano2003}
M.~Arellano.
\newblock \emph{{Panel Data Econometrics}}.
\newblock Oxford: Oxford University Press, 2003.

\bibitem[Arlot and Celisse(2010)]{Arlot:2010}
S.~Arlot and A.~Celisse.
\newblock A survey of cross-validation procedures for model selection.
\newblock \emph{Statistics Surveys}, 4:\penalty0 40--79, 2010.
\newblock \doi{10.1214/09-SS054}.

\bibitem[Auffhammer(2018)]{Auffhammer:2018}
M.~Auffhammer.
\newblock Quantifying economic damages from climate change.
\newblock \emph{Journal of Economic Perspectives}, 32\penalty0 (4):\penalty0
  33--52, 2018.

\bibitem[Auffhammer et~al.(2017)Auffhammer, Baylis, and
  Hausman]{auffhammer2017climate}
M.~Auffhammer, P.~Baylis, and C.~H. Hausman.
\newblock Climate change is projected to have severe impacts on the frequency
  and intensity of peak electricity demand across the united states.
\newblock \emph{Proceedings of the National Academy of Sciences}, 114\penalty0
  (8):\penalty0 1886--1891, 2017.

\bibitem[Barreca et~al.(2016)Barreca, Clay, Deschenes, Greenstone, and
  Shapiro]{barreca2016adapting}
A.~Barreca, K.~Clay, O.~Deschenes, M.~Greenstone, and J.~S. Shapiro.
\newblock Adapting to climate change: The remarkable decline in the {US}
  temperature-mortality relationship over the twentieth century.
\newblock \emph{Journal of Political Economy}, 124\penalty0 (1):\penalty0
  105--159, 2016.

\bibitem[Burke and Emerick(2016)]{burke2016adaptation}
M.~Burke and K.~Emerick.
\newblock Adaptation to climate change: Evidence from {US} agriculture.
\newblock \emph{American Economic Journal: Economic Policy}, 8\penalty0
  (3):\penalty0 106--40, 2016.

\bibitem[Burke et~al.(2015)Burke, Hsiang, and Miguel]{burke2015global}
M.~Burke, S.~M. Hsiang, and E.~Miguel.
\newblock Global non-linear effect of temperature on economic production.
\newblock \emph{Nature}, 527\penalty0 (7577):\penalty0 235--239, 2015.

\bibitem[Burke et~al.(2018)Burke, Gonz{\'a}lez, Baylis, Heft-Neal, Baysan,
  Basu, and Hsiang]{burke2018higher}
M.~Burke, F.~Gonz{\'a}lez, P.~Baylis, S.~Heft-Neal, C.~Baysan, S.~Basu, and
  S.~Hsiang.
\newblock Higher temperatures increase suicide rates in the united states and
  mexico.
\newblock \emph{Nature climate change}, 8\penalty0 (8):\penalty0 723--729,
  2018.

\bibitem[Busse et~al.(2015)Busse, Pope, Pope, and Silva-Risso]{Busse:2015}
M.~R. Busse, D.~G. Pope, J.~C. Pope, and J.~Silva-Risso.
\newblock The psychological effect of weather on car purchases.
\newblock \emph{The Quarterly Journal of Economics}, 130\penalty0 (1):\penalty0
  371--414, 2015.

\bibitem[Cattaneo and Peri(2016)]{cattaneo2016migration}
C.~Cattaneo and G.~Peri.
\newblock The migration response to increasing temperatures.
\newblock \emph{Journal of Development Economics}, 122:\penalty0 127--146,
  2016.

\bibitem[Chambers(2003)]{chambers2003asymptotic}
M.~J. Chambers.
\newblock The asymptotic efficiency of cointegration estimators under temporal
  aggregation.
\newblock \emph{Econometric Theory}, 19\penalty0 (1):\penalty0 49--77, 2003.

\bibitem[Chambers(2011)]{chambers2011cointegration}
M.~J. Chambers.
\newblock Cointegration and sampling frequency.
\newblock \emph{The Econometrics Journal}, 14\penalty0 (2):\penalty0 156--185,
  2011.

\bibitem[Chambers(2016)]{chambers2016estimation}
M.~J. Chambers.
\newblock The estimation of continuous time models with mixed frequency data.
\newblock \emph{Journal of Econometrics}, 193\penalty0 (2):\penalty0 390--404,
  2016.

\bibitem[Chambers and McCrorie(2007)]{chambers2007frequency}
M.~J. Chambers and J.~R. McCrorie.
\newblock Frequency domain estimation of temporally aggregated gaussian
  cointegrated systems.
\newblock \emph{Journal of Econometrics}, 136\penalty0 (1):\penalty0 1--29,
  2007.

\bibitem[Claeskens and Hjort(2008)]{Claeskens:2008}
G.~Claeskens and N.~L. Hjort.
\newblock \emph{Model Selection and Model Averaging}.
\newblock Cambridge Series in Statistical and Probabilistic Mathematics.
  Cambridge University Press, 2008.
\newblock \doi{10.1017/CBO9780511790485}.

\bibitem[Cohen and Dechezlepr{\^e}tre(Forthcoming)]{Cohen:forthcoming}
F.~Cohen and A.~Dechezlepr{\^e}tre.
\newblock Mortality, temperature, and public health provision: evidence from
  mexico.
\newblock \emph{American Economic Journal: Economic Policy}, Forthcoming.

\bibitem[Colmer(2021)]{Colmer:2021}
J.~Colmer.
\newblock Temperature, labor reallocation, and industrial production: Evidence
  from india.
\newblock \emph{American Economic Journal: Applied Economics}, 13\penalty0
  (4):\penalty0 101--24, 2021.

\bibitem[Cui(2020)]{cui2020beyond}
X.~Cui.
\newblock Beyond yield response: Weather shocks and crop abandonment.
\newblock \emph{Journal of the Association of Environmental and Resource
  Economists}, 7\penalty0 (5):\penalty0 901--932, 2020.

\bibitem[Dell et~al.(2012)Dell, Jones, and Olken]{dell2012temperature}
M.~Dell, B.~F. Jones, and B.~A. Olken.
\newblock Temperature shocks and economic growth: Evidence from the last half
  century.
\newblock \emph{American Economic Journal: Macroeconomics}, 4\penalty0
  (3):\penalty0 66--95, 2012.

\bibitem[Dell et~al.(2014)Dell, Jones, and Olken]{Dell2014}
M.~Dell, B.~F. Jones, and B.~A. Olken.
\newblock What do we learn from the weather? the new climate-economy
  literature.
\newblock \emph{Journal of Economic Literature}, 52:\penalty0 740--798, 2014.

\bibitem[Deryugina and Hsiang(2017)]{deryugina2017marginal}
T.~Deryugina and S.~Hsiang.
\newblock The marginal product of climate.
\newblock Technical report, National Bureau of Economic Research, 2017.

\bibitem[Desch\^{e}nes and Greenstone(2007)]{deschenes2007economic}
O.~Desch\^{e}nes and M.~Greenstone.
\newblock The economic impacts of climate change: evidence from agricultural
  output and random fluctuations in weather.
\newblock \emph{American Economic Review}, 97\penalty0 (1):\penalty0 354--385,
  2007.

\bibitem[Desch\^{e}nes and Greenstone(2011)]{deschenes2011climate}
O.~Desch\^{e}nes and M.~Greenstone.
\newblock {Climate change, mortality, and adaptation: Evidence from annual
  fluctuations in weather in the US}.
\newblock \emph{American Economic Journal: Applied Economics}, 3\penalty0
  (4):\penalty0 152--185, 2011.

\bibitem[Diaz and Moore(2017)]{Diaz:2017}
D.~Diaz and F.~Moore.
\newblock Quantifying the economic risks of climate change.
\newblock \emph{Nature Climate Change}, 7\penalty0 (11):\penalty0 774--782,
  2017.

\bibitem[Feng et~al.(2010)Feng, Krueger, and Oppenheimer]{feng2010linkages}
S.~Feng, A.~B. Krueger, and M.~Oppenheimer.
\newblock {Linkages among climate change, crop yields and Mexico--US
  cross-border migration}.
\newblock \emph{Proceedings of the National Academy of Sciences}, 107\penalty0
  (32):\penalty0 14257--14262, 2010.

\bibitem[Gammans et~al.(2017)Gammans, M\'{e}rel, and Ortiz-Bobea]{gammans2017}
M.~Gammans, P.~M\'{e}rel, and A.~Ortiz-Bobea.
\newblock {Negative impacts of climate change on cereal yields: statistical
  evidence from France}.
\newblock \emph{Environmental Research Letters}, 12\penalty0 (5):\penalty0
  054007, 2017.

\bibitem[Garg et~al.(2020)Garg, Jagnani, and Taraz]{garg2020temperature}
T.~Garg, M.~Jagnani, and V.~Taraz.
\newblock Temperature and human capital in india.
\newblock \emph{Journal of the Association of Environmental and Resource
  Economists}, 7\penalty0 (6):\penalty0 1113--1150, 2020.

\bibitem[Ghysels et~al.(2006)Ghysels, Santa-Clara, and Valkanov]{GSV:2006}
E.~Ghysels, P.~Santa-Clara, and R.~Valkanov.
\newblock Predicting volatility: getting the most out of return data sampled at
  different frequencies.
\newblock \emph{J. Econometrics}, 131\penalty0 (1-2):\penalty0 59--95, 2006.
\newblock ISSN 0304-4076.
\newblock \doi{10.1016/j.jeconom.2005.01.004}.

\bibitem[Ghysels et~al.(2007)Ghysels, Sinko, and Valkanov]{GSV:2007}
E.~Ghysels, A.~Sinko, and R.~Valkanov.
\newblock {MIDAS} regressions: further results and new directions.
\newblock \emph{Econometric Rev.}, 26\penalty0 (1):\penalty0 53--90, 2007.
\newblock ISSN 0747-4938.
\newblock \doi{10.1080/07474930600972467}.

\bibitem[Graff~Zivin et~al.(2018)Graff~Zivin, Hsiang, and
  Neidell]{graff2018temperature}
J.~Graff~Zivin, S.~M. Hsiang, and M.~Neidell.
\newblock Temperature and human capital in the short and long run.
\newblock \emph{Journal of the Association of Environmental and Resource
  Economists}, 5\penalty0 (1):\penalty0 77--105, 2018.

\bibitem[Groenvik and Rho(2018)]{groenvik2018self}
H.~Groenvik and Y.~Rho.
\newblock A self-normalizing approach to the specification test of
  mixed-frequency models.
\newblock \emph{Communications in Statistics-Theory and Methods}, 47\penalty0
  (8):\penalty0 1913--1922, 2018.

\bibitem[Harari and Ferrara(2018)]{harari2018conflict}
M.~Harari and E.~L. Ferrara.
\newblock Conflict, climate, and cells: a disaggregated analysis.
\newblock \emph{Review of Economics and Statistics}, 100\penalty0 (4):\penalty0
  594--608, 2018.

\bibitem[Heutel et~al.(2021)Heutel, Miller, and Molitor]{heutel2021adaptation}
G.~Heutel, N.~H. Miller, and D.~Molitor.
\newblock Adaptation and the mortality effects of temperature across us climate
  regions.
\newblock \emph{Review of Economics and Statistics}, 103\penalty0 (4):\penalty0
  740--753, 2021.

\bibitem[Heyes and Saberian(2019)]{Heyes:2019}
A.~Heyes and S.~Saberian.
\newblock Temperature and decisions: evidence from 207,000 court cases.
\newblock \emph{American Economic Journal: Applied Economics}, 11\penalty0
  (2):\penalty0 238--65, 2019.

\bibitem[Hong and Preston(2012)]{Hong:2012}
H.~Hong and B.~Preston.
\newblock Bayesian averaging, prediction and nonnested model selection.
\newblock \emph{Journal of Econometrics}, 167\penalty0 (2):\penalty0 358 --
  369, 2012.
\newblock ISSN 0304-4076.
\newblock \doi{https://doi.org/10.1016/j.jeconom.2011.09.021}.
\newblock Fourth Symposium on Econometric Theory and Applications (SETA).

\bibitem[Hsiang(2010)]{hsiang2010temperatures}
S.~M. Hsiang.
\newblock {Temperatures and cyclones strongly associated with economic
  production in the Caribbean and Central America}.
\newblock \emph{Proceedings of the National Academy of Sciences}, 107\penalty0
  (35):\penalty0 15367--15372, 2010.

\bibitem[Hsiang et~al.(2011)Hsiang, Meng, and Cane]{hsiang2011civil}
S.~M. Hsiang, K.~C. Meng, and M.~A. Cane.
\newblock Civil conflicts are associated with the global climate.
\newblock \emph{Nature}, 476\penalty0 (7361):\penalty0 438--441, 2011.

\bibitem[Hsiang et~al.(2013)Hsiang, Burke, and Miguel]{hsiang2013quantifying}
S.~M. Hsiang, M.~Burke, and E.~Miguel.
\newblock Quantifying the influence of climate on human conflict.
\newblock \emph{Science}, 341\penalty0 (6151):\penalty0 1235367, 2013.

\bibitem[Jagnani et~al.(2021)Jagnani, Barrett, Liu, and You]{Jagnani:2021}
M.~Jagnani, C.~B. Barrett, Y.~Liu, and L.~You.
\newblock Within-season producer response to warmer temperatures: Defensive
  investments by kenyan farmers.
\newblock \emph{The Economic Journal}, 131\penalty0 (633):\penalty0 392--419,
  2021.

\bibitem[Jessoe et~al.(2018)Jessoe, Manning, and Taylor]{jessoe2016climate}
K.~Jessoe, D.~T. Manning, and J.~E. Taylor.
\newblock Climate change and labour allocation in rural {M}exico: Evidence from
  annual fluctuations in weather.
\newblock \emph{The Economic Journal}, 128\penalty0 (608):\penalty0 230--261,
  2018.

\bibitem[Kallenberg et~al.(2017)]{kallenberg2017random}
O.~Kallenberg et~al.
\newblock \emph{Random measures, theory and applications}, volume~1.
\newblock Springer, 2017.

\bibitem[Kvedaras and Zemlys(2012)]{kvedaras2012testing}
V.~Kvedaras and V.~Zemlys.
\newblock Testing the functional constraints on parameters in regressions with
  variables of different frequency.
\newblock \emph{Economics Letters}, 116\penalty0 (2):\penalty0 250--254, 2012.

\bibitem[Levinson(2016)]{Levinson:2016}
A.~Levinson.
\newblock How much energy do building energy codes save? evidence from
  california houses.
\newblock \emph{American Economic Review}, 106\penalty0 (10):\penalty0
  2867--94, 2016.

\bibitem[Li et~al.(2019)Li, Pizer, and Wu]{li2019climate}
Y.~Li, W.~A. Pizer, and L.~Wu.
\newblock Climate change and residential electricity consumption in the yangtze
  river delta, china.
\newblock \emph{Proceedings of the National Academy of Sciences}, 116\penalty0
  (2):\penalty0 472--477, 2019.

\bibitem[Liao and Shi(2020)]{liao2020nondegenerate}
Z.~Liao and X.~Shi.
\newblock A nondegenerate vuong test and post selection confidence intervals
  for semi/nonparametric models.
\newblock \emph{Quantitative Economics}, 11\penalty0 (3):\penalty0 983--1017,
  2020.

\bibitem[Liu et~al.(Forthcoming)Liu, Shamdasani, and Taraz]{Liu:forthcoming}
M.~Y. Liu, Y.~Shamdasani, and V.~Taraz.
\newblock Climate change and labor reallocation: Evidence from six decades of
  the indian census.
\newblock \emph{American Economic Journal: Economic Policy}, Forthcoming.

\bibitem[Liu and Rho(2019)]{liu2019choice}
Y.~Liu and Y.~Rho.
\newblock On the choice of instruments in mixed frequency specification tests.
\newblock \emph{Communications in Statistics-Theory and Methods}, 48\penalty0
  (24):\penalty0 6098--6118, 2019.

\bibitem[LoPalo(Forthcoming)]{Lopalo:forthcoming}
M.~LoPalo.
\newblock Temperature, worker productivity, and adaptation: evidence from
  survey data production.
\newblock \emph{American Economic Journal: Applied Economics}, Forthcoming.

\bibitem[Lorenz(1969)]{lorenz1969atmospheric}
E.~N. Lorenz.
\newblock Atmospheric predictability as revealed by naturally occurring
  analogues.
\newblock \emph{Journal of Atmospheric Sciences}, 26\penalty0 (4):\penalty0
  636--646, 1969.

\bibitem[Mendelsohn et~al.(1994)Mendelsohn, Nordhaus, and
  Shaw]{mendelsohn1994impact}
R.~Mendelsohn, W.~D. Nordhaus, and D.~Shaw.
\newblock The impact of global warming on agriculture: a {R}icardian analysis.
\newblock \emph{American Economic Review}, pages 753--771, 1994.

\bibitem[Miller(2014)]{miller2014mixed}
J.~I. Miller.
\newblock Mixed-frequency cointegrating regressions with parsimonious
  distributed lag structures.
\newblock \emph{Journal of Financial Econometrics}, 12\penalty0 (3):\penalty0
  584--614, 2014.

\bibitem[Miller(2016)]{miller2016conditionally}
J.~I. Miller.
\newblock Conditionally efficient estimation of long-run relationships using
  mixed-frequency time series.
\newblock \emph{Econometric Reviews}, 35\penalty0 (6):\penalty0 1142--1171,
  2016.

\bibitem[Miller(2018)]{miller2018simple}
J.~I. Miller.
\newblock Simple robust tests for the specification of high-frequency
  predictors of a low-frequency series.
\newblock \emph{Econometrics and Statistics}, 5:\penalty0 45--66, 2018.

\bibitem[Moore et~al.(2017)Moore, Baldos, Hertel, and Diaz]{Moore:2017}
F.~C. Moore, U.~Baldos, T.~Hertel, and D.~Diaz.
\newblock New science of climate change impacts on agriculture implies higher
  social cost of carbon.
\newblock \emph{Nature Communications}, 8\penalty0 (1):\penalty0 1--9, 2017.

\bibitem[Mueller et~al.(2014)Mueller, Gray, and Kosec]{mueller2014heat}
V.~Mueller, C.~Gray, and K.~Kosec.
\newblock Heat stress increases long-term human migration in rural pakistan.
\newblock \emph{Nature climate change}, 4\penalty0 (3):\penalty0 182--185,
  2014.

\bibitem[{National Academies of Sciences, Engineering, and
  Medicine}(2017)]{NAS:2017}
{National Academies of Sciences, Engineering, and Medicine}.
\newblock \emph{Valuing climate damages: updating estimation of the social cost
  of carbon dioxide}.
\newblock National Academies Press, 2017.

\bibitem[Novan et~al.(Forthcoming)Novan, Smith, and Zhou]{Novan:forthcoming}
K.~Novan, A.~Smith, and T.~Zhou.
\newblock Residential building codes do save energy: Evidence from hourly
  smart-meter data.
\newblock \emph{The Review of Economics and Statistics}, pages 1--45,
  Forthcoming.

\bibitem[Park et~al.(2020)Park, Goodman, Hurwitz, and Smith]{park2020heat}
R.~J. Park, J.~Goodman, M.~Hurwitz, and J.~Smith.
\newblock Heat and learning.
\newblock \emph{American Economic Journal: Economic Policy}, 12\penalty0
  (2):\penalty0 306--39, 2020.

\bibitem[Pretis(2021)]{Pretis:2021}
F.~Pretis.
\newblock Exogeneity in climate econometrics.
\newblock \emph{Energy Economics}, 96:\penalty0 105122, 2021.
\newblock ISSN 0140-9883.
\newblock \doi{https://doi.org/10.1016/j.eneco.2021.105122}.
\newblock URL
  \url{https://www.sciencedirect.com/science/article/pii/S014098832100027X}.

\bibitem[Ricke et~al.(2018)Ricke, Drouet, Caldeira, and Tavoni]{Ricke:2018}
K.~Ricke, L.~Drouet, K.~Caldeira, and M.~Tavoni.
\newblock Country-level social cost of carbon.
\newblock \emph{Nature Climate Change}, 8\penalty0 (10):\penalty0 895--900,
  2018.

\bibitem[Ritchie and Nesmith(1991)]{ritchie1991temperature}
J.~T. Ritchie and D.~S. Nesmith.
\newblock Temperature and crop development.
\newblock In J.~Hanks and J.~T. Ritchie, editors, \emph{Modeling Plant and Soil
  Systems, Agronomy 31}, pages 5--29. American Society of Agronomy, Crop
  Science Society of America, Soil Science Society of America, 1991.

\bibitem[Schennach and Wilhelm(2017)]{schennach2017simple}
S.~M. Schennach and D.~Wilhelm.
\newblock A simple parametric model selection test.
\newblock \emph{Journal of the American Statistical Association}, 112\penalty0
  (520):\penalty0 1663--1674, 2017.

\bibitem[Schlenker and Roberts(2009)]{SR2009}
W.~Schlenker and M.~J. Roberts.
\newblock {Nonlinear temperature effects indicate severe damages to U.S. crop
  yields under climate change}.
\newblock \emph{Proceedings of the National Academy of Sciences}, 106, 2009.

\bibitem[Schlenker et~al.(2006)Schlenker, Hanemann, and
  Fisher]{schlenker2006impact}
W.~Schlenker, W.~M. Hanemann, and A.~C. Fisher.
\newblock The impact of global warming on us agriculture: an econometric
  analysis of optimal growing conditions.
\newblock \emph{Review of Economics and statistics}, 88\penalty0 (1):\penalty0
  113--125, 2006.

\bibitem[Shao(1993)]{Shao:1993}
J.~Shao.
\newblock Linear model selection by cross-validation.
\newblock \emph{Journal of the American Statistical Association}, 88\penalty0
  (422):\penalty0 486--494, 1993.
\newblock ISSN 0162-1459.

\bibitem[Shao(1997)]{Shao:1997}
J.~Shao.
\newblock An asymptotic theory for linear model selection.
\newblock \emph{Statistica Sinica}, 7\penalty0 (2):\penalty0 221--264, 1997.
\newblock ISSN 1017-0405.
\newblock With comments and a rejoinder by the author.

\bibitem[Shi(2015)]{shi2015nondegenerate}
X.~Shi.
\newblock A nondegenerate vuong test.
\newblock \emph{Quantitative Economics}, 6\penalty0 (1):\penalty0 85--121,
  2015.

\bibitem[Sin and White(1996)]{Sin:1996}
C.-Y. Sin and H.~White.
\newblock Information criteria for selecting possibly misspecified parametric
  models.
\newblock \emph{Journal of Econometrics}, 71\penalty0 (1):\penalty0 207 -- 225,
  1996.
\newblock ISSN 0304-4076.
\newblock \doi{https://doi.org/10.1016/0304-4076(94)01701-8}.

\bibitem[Somanathan et~al.(2021)Somanathan, Somanathan, Sudarshan, and
  Tewari]{somanathan2021impact}
E.~Somanathan, R.~Somanathan, A.~Sudarshan, and M.~Tewari.
\newblock The impact of temperature on productivity and labor supply: Evidence
  from indian manufacturing.
\newblock \emph{Journal of Political Economy}, 129\penalty0 (6):\penalty0
  1797--1827, 2021.

\bibitem[Teixeira et~al.(2014)Teixeira, Taylor, Persson, and
  Matheou]{Teixeira2014}
J.~Teixeira, M.~Taylor, A.~Persson, and G.~Matheou.
\newblock Atmospheric general circulation models.
\newblock In E.~G. Njoku, editor, \emph{Encyclopedia of Remote Sensing}, pages
  35--37. Springer New York, New York, NY, 2014.

\bibitem[Vuong(1989)]{Vuong:1989}
Q.~H. Vuong.
\newblock Likelihood ratio tests for model selection and non-nested hypotheses.
\newblock \emph{Econometrica}, 57\penalty0 (2):\penalty0 307--333, 1989.

\bibitem[Wenz et~al.(2017)Wenz, Levermann, and Auffhammer]{wenz2017north}
L.~Wenz, A.~Levermann, and M.~Auffhammer.
\newblock North--south polarization of european electricity consumption under
  future warming.
\newblock \emph{Proceedings of the National Academy of Sciences}, 114\penalty0
  (38):\penalty0 E7910--E7918, 2017.

\bibitem[White(2014)]{white2014asymptotic}
H.~White.
\newblock \emph{Asymptotic theory for econometricians}.
\newblock Academic press, 2014.

\bibitem[Yang(2005)]{Yang:2005}
Y.~Yang.
\newblock Can the strengths of aic and bic be shared? a conflict between model
  indentification and regression estimation.
\newblock \emph{Biometrika}, 92\penalty0 (4):\penalty0 937--950, 2005.
\newblock ISSN 00063444.
\newblock URL \url{http://www.jstor.org/stable/20441246}.

\bibitem[Zhang et~al.(2018)Zhang, Zhang, Deschenes, and
  Meng]{zhang2016temperature}
P.~Zhang, J.~Zhang, O.~Deschenes, and K.~Meng.
\newblock {Temperature effects on productivity and factor reallocation:
  Evidence from a half million Chinese manufacturing plants}.
\newblock \emph{Journal of Environmental Economics and Management},
  88:\penalty0 1--17, 2018.

\end{thebibliography}

\begin{table}[htbp]
  \centering
  \caption{Simulation Mean of Model Coefficients and Ideal Targets}\label{tab:pseudo_true_param}
  \fontsize{10}{10}\selectfont
    \begin{tabular}{llcccccccc} \hline
 {DGP:}&&\multicolumn{2}{c}{$A$}&&\multicolumn{2}{c}{$QinA$}&&\multicolumn{2}{c}{$Q$}\\
		\cline{3-4}\cline{6-7}\cline{9-10}
	$\mathbf{M}_{\alpha}$&	$X_{it,\alpha}^{k}$&$\bar{\beta}_{\alpha}^{k}$&\multicolumn{1}{c}{Ideal target}&&$\bar{\beta}_{\alpha}^{k}$&\multicolumn{1}{c}{Ideal target}&&$\bar{\beta}_{\alpha}^{k}$&\multicolumn{1}{c}{Ideal target}\\ \hline
	\\   
    A     & A     & 1.000 & \rdelim\}{1}{50pt}[0.25$\times$10$^{-6}$] &       & -5.130 & \rdelim\}{1}{50pt}[1.32] &       & 0.109 & \rdelim\}{1}{50pt}[5.60] \\
          &       &       &       &       &       &       &       &       &  \\
    B     & B1    & 0.496 & \rdelim\}{2}{50pt}[0.37$\times$10$^{-6}$] &       & -2.497 &\rdelim\}{2}{50pt}[1.25] &       & -0.382 & \rdelim\}{2}{50pt}[4.68] \\
          & B2    & 0.504 &       &       & -2.627 &       &       & 0.434 &  \\
          &       &       &       &       &       &       &       &       &  \\
    Q     & Q1    & 0.246 & \rdelim\}{4}{50pt}[1.03$\times$10$^{-6}$] &       & -1.244 & \rdelim\}{4}{50pt}[1.26] &       & -0.250 & \rdelim\}{4}{50pt}[1.03$\times$10$^{-6}$] \\
          & Q2    & 0.249 &       &       & -1.247 &       &       & 0.000 &  \\
          & Q3    & 0.252 &       &       & -1.304 &       &       & 0.750 &  \\
          & Q4    & 0.252 &       &       & -1.318 &       &       & 0.000 &  \\
          &       &       &       &       &       &       &       &       &  \\
    M     & M1    & 0.085 &\rdelim\}{12}{50pt}[1.19$\times$10$^{-6}$] &       & -0.451 & \rdelim\}{12}{50pt}[1.40] &       & -0.086 & \rdelim\}{12}{50pt}[1.19$\times$10$^{-6}$] \\
          & M2    & 0.077 &       &       & -0.380 &       &       & -0.078 &  \\
          & M3    & 0.085 &       &       & -0.429 &       &       & -0.086 &  \\
          & M4    & 0.082 &       &       & -0.361 &       &       & 0.000 &  \\
          & M5    & 0.085 &       &       & -0.472 &       &       & 0.000 &  \\
          & M6    & 0.082 &       &       & -0.439 &       &       & 0.000 &  \\
          & M7    & 0.085 &       &       & -0.396 &       &       & 0.253 &  \\
          & M8    & 0.085 &       &       & -0.381 &       &       & 0.253 &  \\
          & M9    & 0.082 &       &       & -0.487 &       &       & 0.245 &  \\
          & M10   & 0.085 &       &       & -0.377 &       &       & 0.000 &  \\
          & M11   & 0.082 &       &       & -0.481 &       &       & 0.000 &  \\
          & M12   & 0.085 &       &       & -0.430 &       &       & 0.000 &  \\
          &       &       &       &       &       &       &       &       &  \\
    QinA  & A     & 0.998 & \rdelim\}{2}{50pt}[0.28$\times$10$^{-6}$] &       & 0.198 & \rdelim\}{2}{50pt}[0.28$\times$10$^{-6}$] &       & -0.519 & \rdelim\}{2}{50pt}[5.47] \\
          & A$^2$  & 0.000 &       &       & -0.050 &       &       & 0.006 &  \\
          &       &       &       &       &       &       &       &       &  \\
    Bin   & Bin1  & -0.152 & \rdelim\}{9}{50pt}[0.07] &       & 0.723 & \rdelim\}{9}{50pt}[4.07] &       & 0.076 & \rdelim\}{9}{50pt}[0.99] \\
          & Bin2  & -0.103 &       &       & 0.513 &       &       & 0.075 &  \\
          & Bin3  & -0.083 &       &       & 0.414 &       &       & 0.048 &  \\
          & Bin4  & -0.055 &       &       & 0.288 &       &       & 0.023 &  \\
          & Bin5  & -0.028 &       &       & 0.157 &       &       & 0.013 &  \\
          & Bin6  & 0.026 &       &       & -0.146 &       &       & -0.008 &  \\
          & Bin7  & 0.054 &       &       & -0.290 &       &       & 0.051 &  \\
          & Bin8  & 0.076 &       &       & -0.414 &       &       & 0.109 &  \\
          & Bin9  & 0.120 &       &       & -0.734 &       &       & 0.211 &  \\ \hline
          \multicolumn{10}{l}{\parbox[t]{14cm}{\footnotesize {\it Notes:} The table presents simulated pseudo-true parameter values, $\bar{\beta}_\alpha^{k}$, which are computed as the simulation mean for each estimated element of the parameter vector in the models considered across 2,000 simulation replications for each DGP ($A$, $QinA$ and $Q$). In this design, $n=3,074$ (the total number of counties in the dataset) and $T=10$. We use $\bar{\beta}_\alpha^{k}$ to calculate the ideal target $\frac{1}{n}\sum_{i=1}^n(\bar{\mu}_{\alpha}(\mathcal{W}^f_{i,T+\tau})-\bar{\mu}_{\alpha}(\mathcal{W}_{iT})-(\mu_{\star}(\mathcal{W}^f_{i,T+\tau})-\mu_{\star}(\mathcal{W}_{iT})))^2$, where $\bar{\mu}_{\alpha}(\mathcal{W}_{it}) = X_{it}' \bar{\beta}_\alpha^k$ and $\tau=20$.}}          
    \end{tabular}
\end{table}

\clearpage
\begin{figure}[htbp]
{\fontsize{9}{5}\selectfont
\noindent
\begin{tabular}{cc}
\multicolumn{2}{c}{\bf DGP=$A$}\\
\includegraphics[width=.5\linewidth]{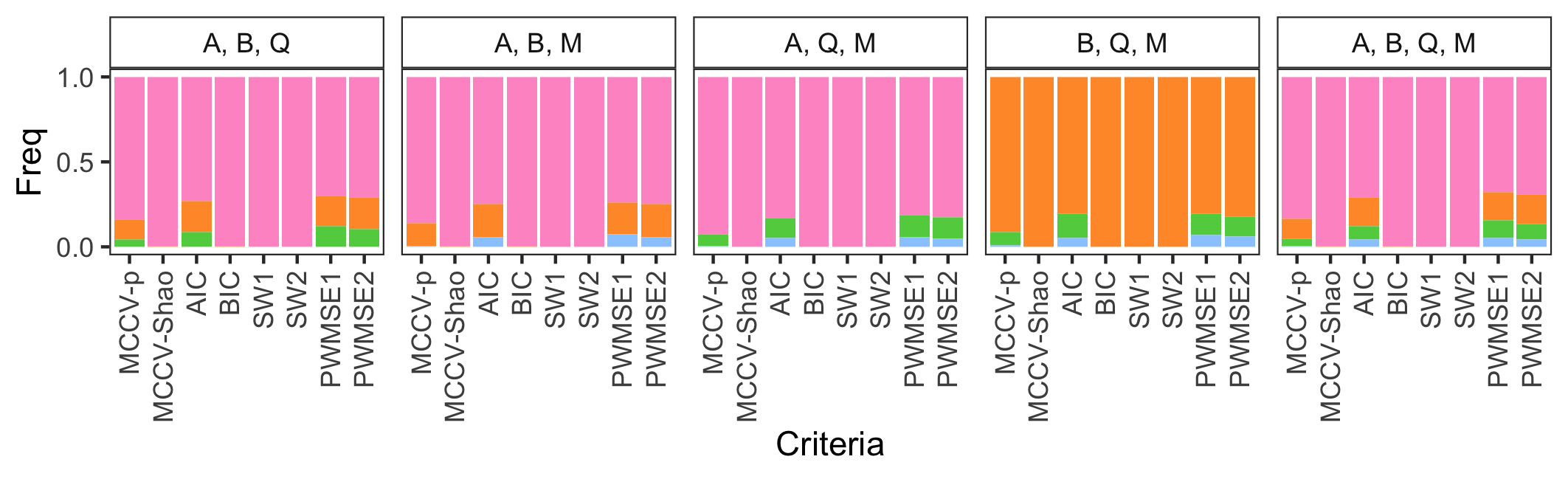}&
\includegraphics[width=.5\linewidth]{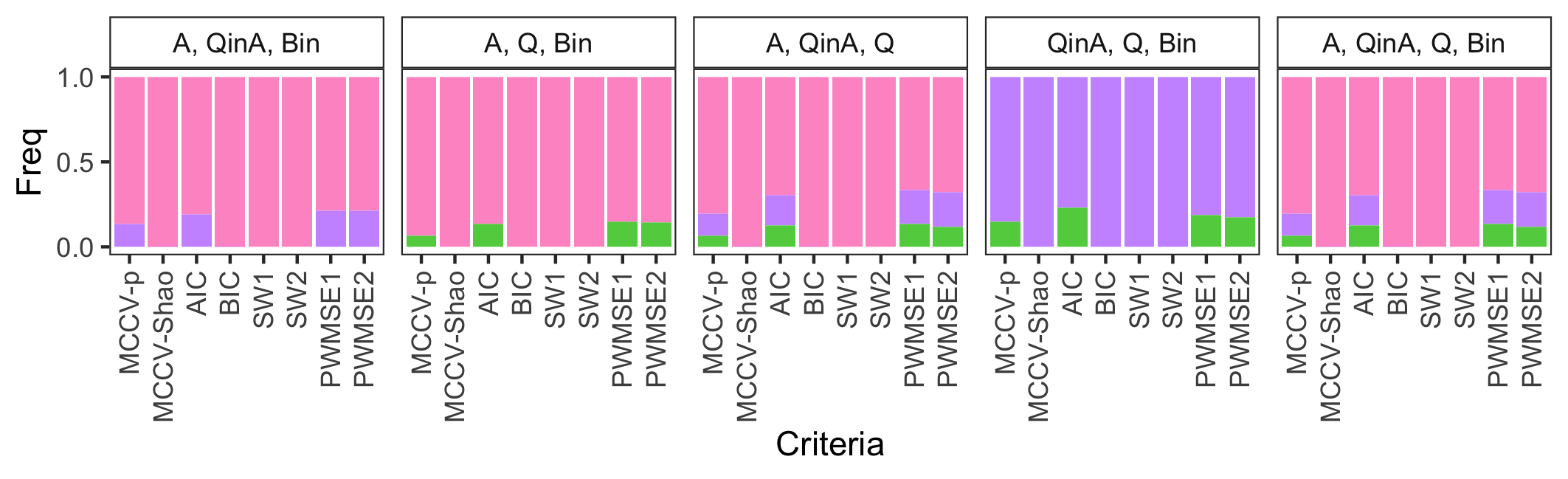}\\
\multicolumn{2}{c}{\bf DGP=$QinA$}\\
\includegraphics[width=.5\linewidth]{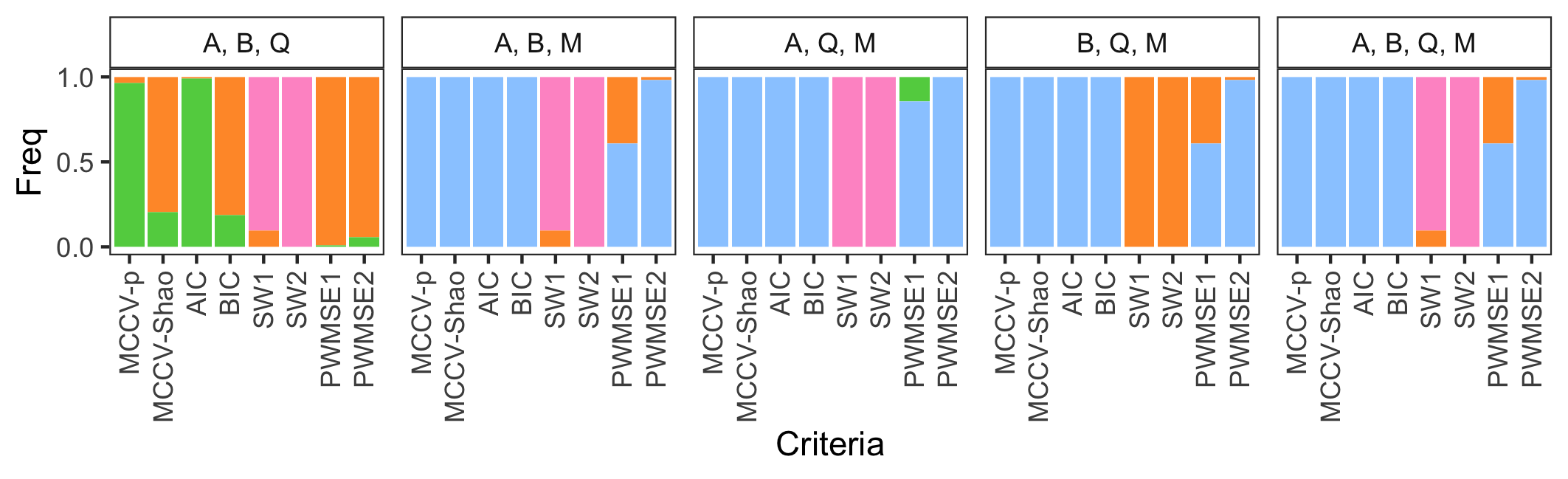}&
\includegraphics[width=.5\linewidth]{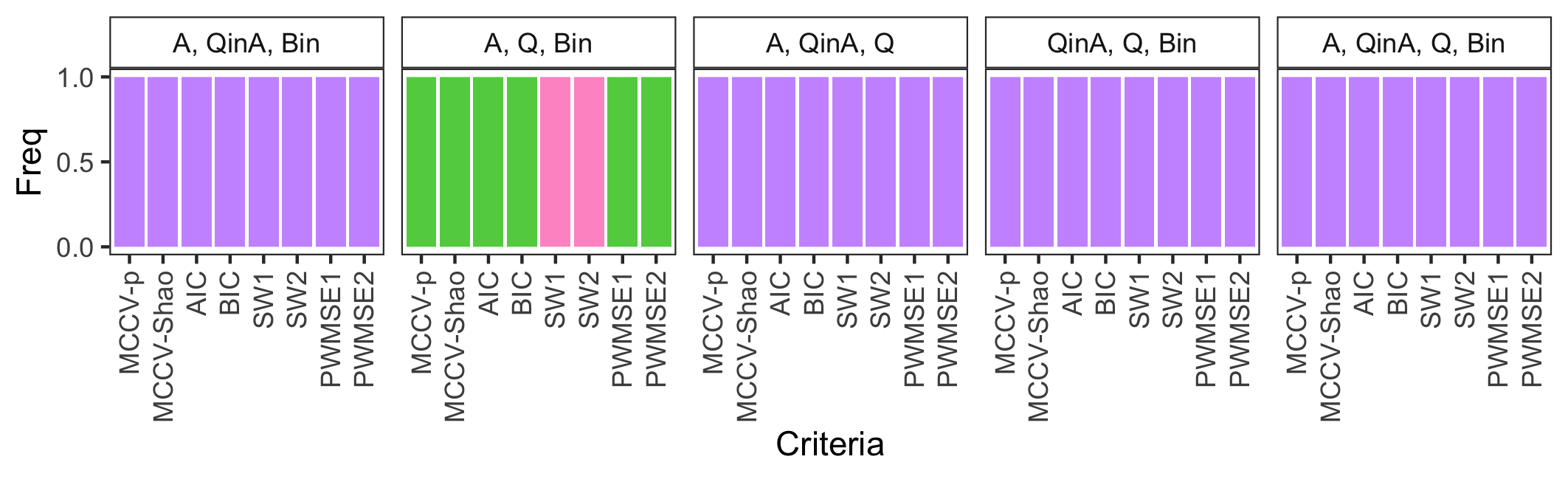}\\
\multicolumn{2}{c}{\bf DGP=$Q$}\\
\includegraphics[width=.5\linewidth]{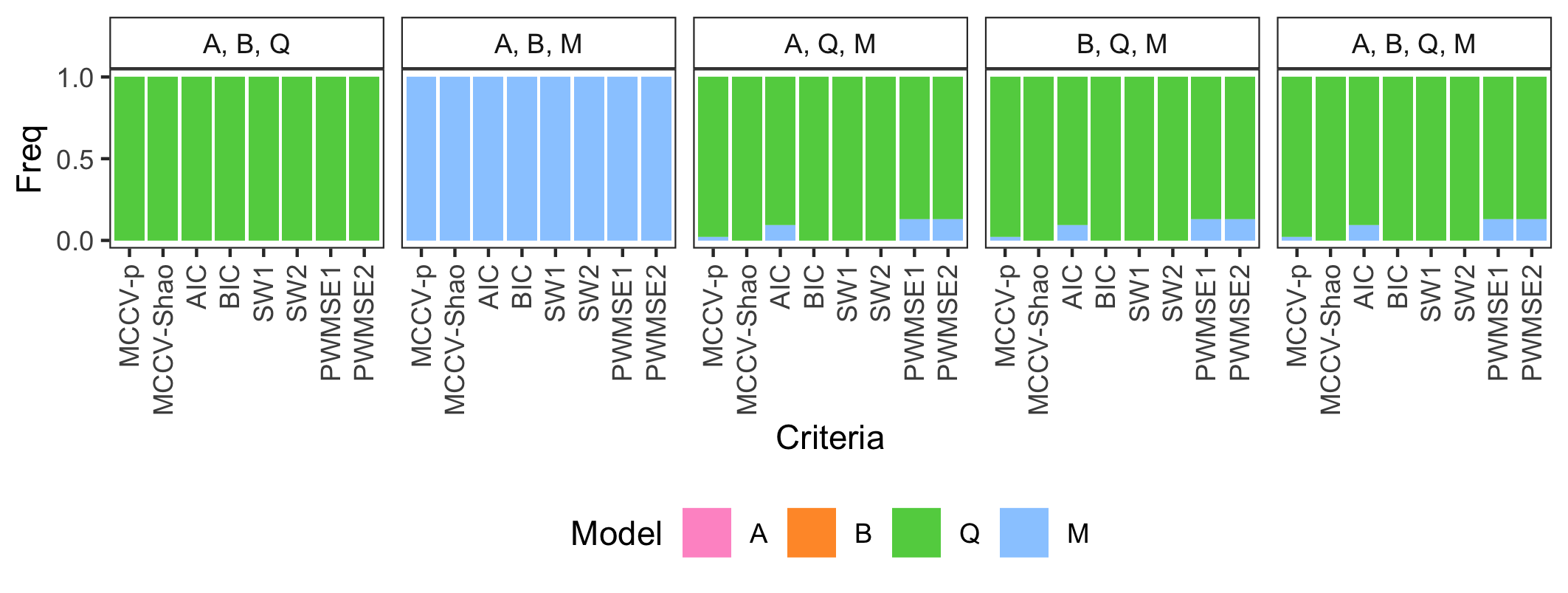}&
\includegraphics[width=.5\linewidth]{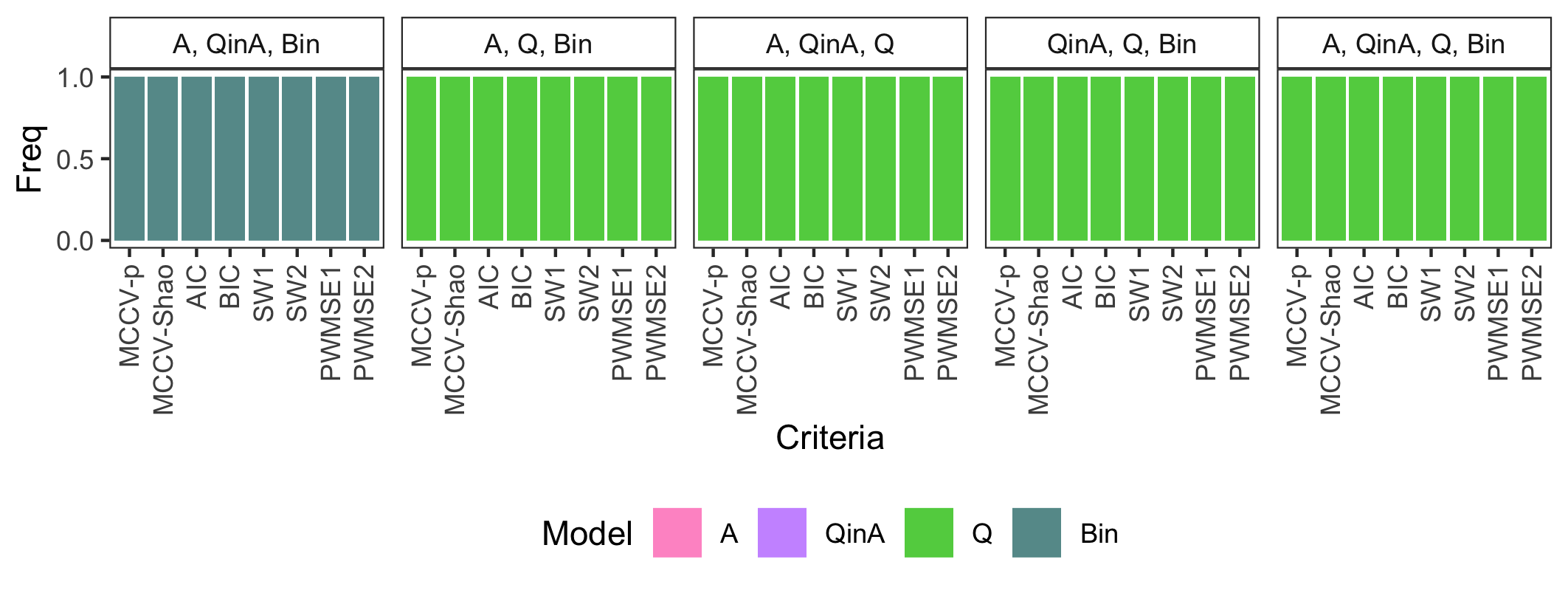}\\
\end{tabular}
}
\caption{Simulated Model Selection Outcomes}\label{fig:sim_model_select}
{\footnotesize {\it Notes:} In each panel, the title indicates the set of models being compared, the horizontal axis labels different model selection criteria, the height of a colored bar indicates the proportion of times that a particular model is selected among the set of models. PWMSE1 and PWMSE2 refer to PWMSE using $L^2$ norms of monthly and yearly differences with $h=10$ as the weights. The future period is set as $\tau=20$ The five panels on the left are for comparisons across nested models and the five on the right are for possibly non-nested models.}
\end{figure}

\begin{table}[htbp]
  \centering
  \fontsize{9}{9}\selectfont
  \caption{Existing Model Selection Criteria for the Temperature-Yield Relationship}
    \begin{tabular}{lcccccccc}\hline
    Model & $\widehat{R^2}$ & SNR   & MCCV$_\text{-p}$ & MCCV$_\text{-Shao}$ & AIC   & BIC   & SW$_1$   & SW$_2$ \\ \hline
    \textit{Unbalanced Panel} &       &       &       &       &       &       &       &  \\
    (a) no temperature variable 			& 9.18\% & 10.11\% & 71.26 & 73.69 & -33.31 & -33.25 & -29.94 & -25.94 \\
    (b) monthly averages 					& 20.19\% & 25.29\% & 64.43 & 67.52 & -34.84 & -34.78 & -31.15 & -26.76 \\
    (c) 1$^\circ$C daily bins 				& 26.88\% & 36.76\% & 59.79 & 62.18 & -35.88 & -35.78 & -30.49 & -24.11 \\
    (d) 3$^\circ$C step function 			& 29.34\% & 41.52\% & 57.68 & 59.84 & \emph{-36.29} & \emph{-36.21} & -32.16 & -27.26 \\
    (e) SHF degree days 					& 27.60\% & 38.13\% & 57.95 & 59.90 & -36.11 & -36.04 & -32.57 & -28.38 \\
    (f) piecewise: 0-29, 29+ 				& 28.99\% & 40.82\% & 57.78 & 59.54 & -36.13 & -36.06 & -32.65 & \emph{-28.52} \\
    (g) piecewise: 0-24, 24-26, 26+ 	& 28.24\% & 39.35\% & \emph{57.44} & \emph{59.23} & -36.21 & -36.15 & \emph{-32.68} & -28.49 \\
          &       &       &       &       &       &       &       &  \\
    \textit{Balanced Panel} &       &       &       &       &       &       &       &  \\
    (a) no temperature variable 			& 10.25\% & 11.42\% & 47.84 & 49.98 & -13.82 & -13.79 & -12.65 & -11.33 \\
    (b) monthly averages 					& 23.19\% & 30.19\% & 42.45 & 45.85 & -14.51 & -14.47 & -13.15 & -11.61 \\
    (c) 1$^\circ$C daily bins 				& 36.18\% & 56.69\% & 35.30 & 37.97 & -15.32 & -15.26 & -12.99 & -10.34 \\
    (d) 3$^\circ$C step function 			& 39.27\% & 64.66\% & 33.22 & 35.49 & \emph{-15.54} & \emph{-15.50} & -13.92 & -12.09 \\
    (e) SHF degree days 					& 38.25\% & 61.94\% & 35.83 & 37.87 & -15.33 & -15.30 & -14.07 & -12.64 \\
    (f) piecewise: 0-30, 30+				& 38.37\% & 62.25\% & \emph{33.10} & \emph{34.86} & -15.47 & -15.44 & \emph{-14.24} & \emph{-12.85} \\
    (g) piecewise: 0-29, 29-33, 33+ 	& 36.26\% & 56.88\% & 33.15 & 35.00 & -15.48 & -15.44 & -14.22 & -12.78 \\ \hline
    \multicolumn{9}{l}{\parbox[t]{16.5cm}{{\it Notes:} $\widehat{R^2}$ and SNR are calculated based on the following regression: $\widehat{\log(y_{it})} = \widehat{X_{it,\alpha}'}\beta_{\alpha} + \theta_1 \widehat{P_{it}} + \theta_2 \widehat{P_{it}^2} + \widehat{\epsilon_{it}}$, where the hatted variables are obtained by projecting the original variables out of county and year fixed effects and state quadratic trends. The SNR is formed by dividing the model sum of squares by the residual sum of squares. The columns for MCCV-p and MCCV-Shao present MSE for cross-validation with 1,000 simulations. We italicize the smallest model selection criterion in each column for the unbalanced and balanced sample, respectively. For legibility, we scale up the MSE for MCCV by 1,000 times and scale down the MSE for GICs by 10,000 times.}}    
    \end{tabular}\label{tab:ms_temp_yield}
\end{table}

\begin{landscape}
\begin{table}[htbp]
  \centering
  \fontsize{9}{9}\selectfont
  \caption{Proximity-weighted MSE for the Temperature-Yield Relationship}\label{tab:pwmse_yields}
    \begin{tabular}{lrrrrrrrrrrrrrrrrr} \hline
    \multicolumn{1}{r}{Norm:}		 & N &  & \multicolumn{3}{c}{M1} &       & \multicolumn{3}{c}{M2} &       & \multicolumn{3}{c}{Y1} &       & \multicolumn{3}{c}{Y2} \\ \cline{4-6} \cline{8-10} \cline{12-14} \cline{16-18}
    \multicolumn{1}{r}{$h$:}      &       &     & 1     & 10    & 100   &       & 1     & 10    & 100   &       & 1     & 10    & 100   &       & 1     & 10    & 100 \\ \hline
    {\it Unbalanced Panel}      \\
    (a) no temperature variable 			& 105.39 &       & 7.13  & 77.08 & 102.09 &       & 8.88  & 68.88 & 100.60 &       & 32.13 & 90.14 & 103.71 &       & 24.03 & 78.21 & 101.92 \\
    (b) monthly averages 					& 96.28 &       & 6.61  & 70.52 & 93.28 &       & 8.23  & 62.93 & 91.89 &       & 28.70 & 82.17 & 94.73 &       & 21.39 & 71.03 & 93.04 \\
    (c) 1$^\circ$C daily bins 			    & 85.19 &       & 5.77  & 62.34 & 82.53 &       & 7.08  & 55.43 & 81.27 &       & 24.72 & 72.55 & 83.80 &       & 18.43 & 62.53 & 82.28 \\
    (d) 3$^\circ$C step function 			& \emph{82.52} &       & 5.54  & \emph{60.33} & \emph{79.93} &       & 6.76  & \emph{53.56} & \emph{78.70} &       & 23.91 & \emph{70.24} & \emph{81.17} &       & 17.68 & \emph{60.45} & \emph{79.68} \\
    (e) SHF degree days 					& 82.96 &       & \emph{5.50} & 60.57 & 80.35 &       & \emph{6.68} & 53.62 & 79.08 &       & \emph{23.67} & 70.51 & 81.59 &       & \emph{17.43} & 60.54 & 80.07 \\
    (f) piecewise: 0-29, 29+ & 83.37 &       & 5.56  & 60.89 & 80.75 &       & 6.80  & 53.91 & 79.47 &       & 23.90 & 70.88 & 81.99 &       & 17.71 & 60.85 & 80.46 \\
    (g) piecewise: 0-24, 24-26, 26+ & 83.77 &       & 5.63  & 61.24 & 81.15 &       & 6.90  & 54.32 & 79.88 &       & 24.17 & 71.27 & 82.40 &       & 17.92 & 61.28 & 80.88 \\
\\ 
	{\it Balanced Panel}        \\
    (a) no temperature variable 			& 128.59 &       & 9.92  & 95.42 & 124.75 &       & 13.16 & 87.94 & 123.39 &       & 42.96 & 111.40 & 126.72 &       & 34.14 & 99.22 & 124.92 \\
    (b) monthly averages 					& 125.17 &       & 9.89  & 93.21 & 121.48 &       & 13.17 & 86.08 & 120.17 &       & 41.48 & 108.35 & 123.34 &       & 32.92 & 96.41 & 121.57 \\
    (c) 1$^\circ$C daily bins 				& 94.71 &       & 7.41  & 70.47 & 91.91 &       & 9.74  & 64.94 & 90.90 &       & 30.34 & 81.80 & 93.30 &       & 23.98 & 72.66 & 91.95 \\
    (d) 3$^\circ$C step function 			& \emph{89.51} &       & \emph{6.97} & \emph{66.57} & \emph{86.86} &       & \emph{9.16} & \emph{61.29} & \emph{85.89} &       & \emph{28.61} & \emph{77.28} & \emph{88.17} &       & \emph{22.59} & \emph{68.60} & \emph{86.89} \\
    (e) SHF degree days 					& 91.29 &       & 7.06  & 67.81 & 88.57 &       & 9.27  & 62.47 & 87.60 &       & 29.44 & 78.88 & 89.93 &       & 23.22 & 70.09 & 88.63 \\
    (f) piecewise: 0-30, 30+ & 91.33 &       & 7.11  & 67.91 & 88.63 &       & 9.34  & 62.55 & 87.64 &       & 29.35 & 78.88 & 89.97 &       & 23.18 & 70.03 & 88.66 \\
    (g) piecewise: 0-29, 29-33, 33+ & 91.31 &       & 7.11  & 67.90 & 88.60 &       & 9.34  & 62.52 & 87.62 &       & 29.25 & 78.84 & 89.95 &       & 23.11 & 69.98 & 88.63 \\ \hline
    \\
    \multicolumn{18}{l}{\parbox[t]{22.5cm}{{\it Notes:} The PWMSEs are obtained from cross-validation with 1,000 simulations. We italicize the smallest PWMSE in each column for the unbalanced and balanced sample, respectively. M1 and M2 are $L^\infty$ and $L^2$ norms of monthly differences; Y1 and Y2 are $L^\infty$ and $L^2$ norms of annual differences. For legibility, we scale up the PWMSE by 1,000 times.}}        
    \end{tabular}
\end{table}
\end{landscape}

\clearpage

\begin{figure}[htbp]
\begin{tabular}{ccc}
A. Monthly averages  & B. 1$^\circ$C bins & C. 3$^\circ$C step function \\~\\
\includegraphics[width=.32\linewidth]{tAvgm_hadcm3b1_crop}&
\includegraphics[width=.32\linewidth]{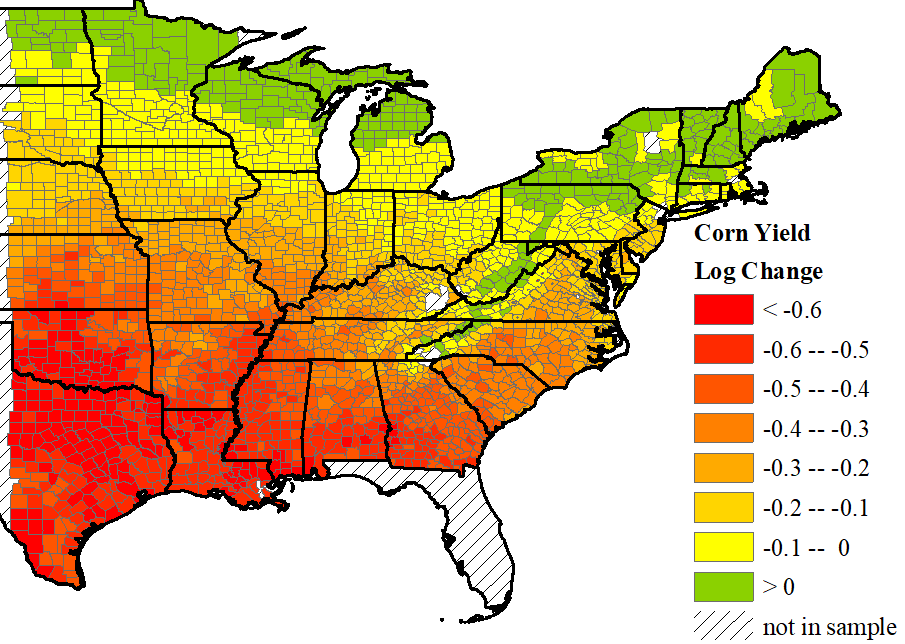}&
\includegraphics[width=.32\linewidth]{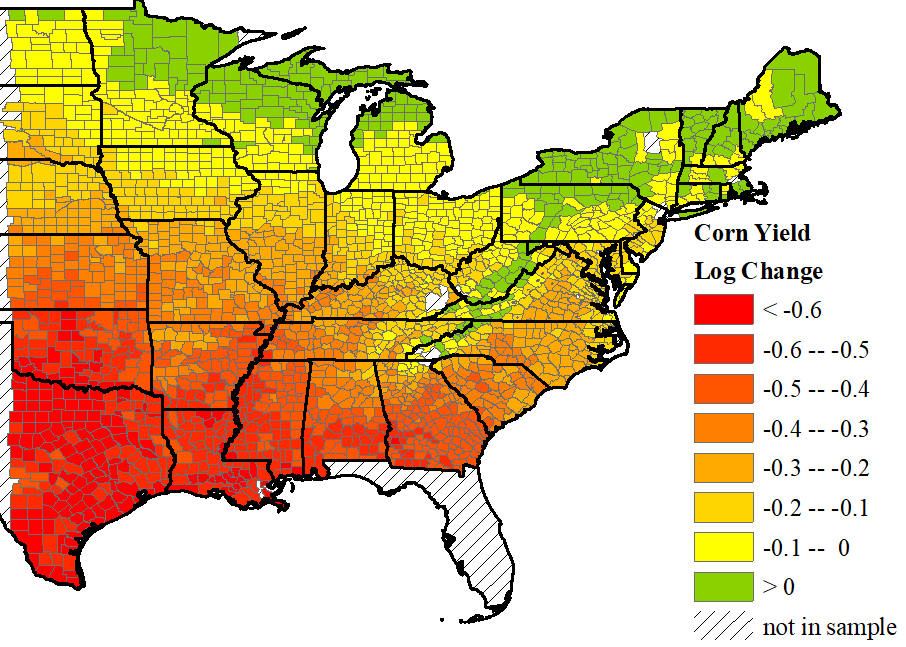}\\~\\
D. SHF degree days & E. One-knot spline & F. Two-knot spline \\~\\
\includegraphics[width=.32\linewidth]{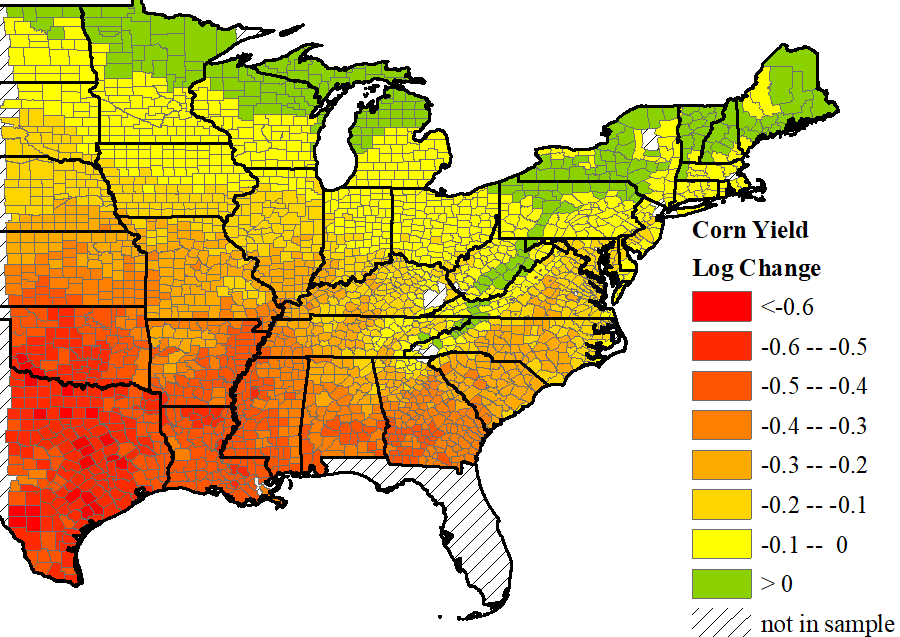}&
\includegraphics[width=.32\linewidth]{spline1_hadcm3b1_crop}&
\includegraphics[width=.32\linewidth]{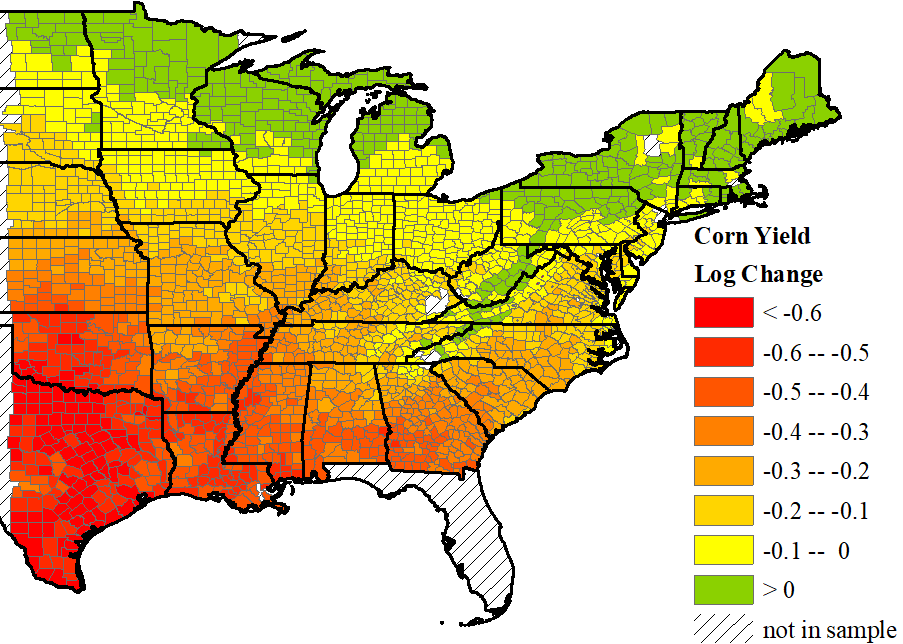} \\
\end{tabular}
\caption{Yield Impacts Projected under Future Climate}\label{fig:ag_projections}
{\footnotesize {\it Notes:} The county-level log changes in corn yields are obtained by applying different yield-response functions to the climate of 2050 projected under HadCM3-B1. The yield-response functions considered here correspond to models (b)-(g) in the unbalanced case. }
\end{figure}

\begin{figure}[htbp]
\begin{center}
\begin{tabular}{cc}
A. Unbalanced Panel & B. Balanced Panel \\
\includegraphics[width=.48\linewidth]{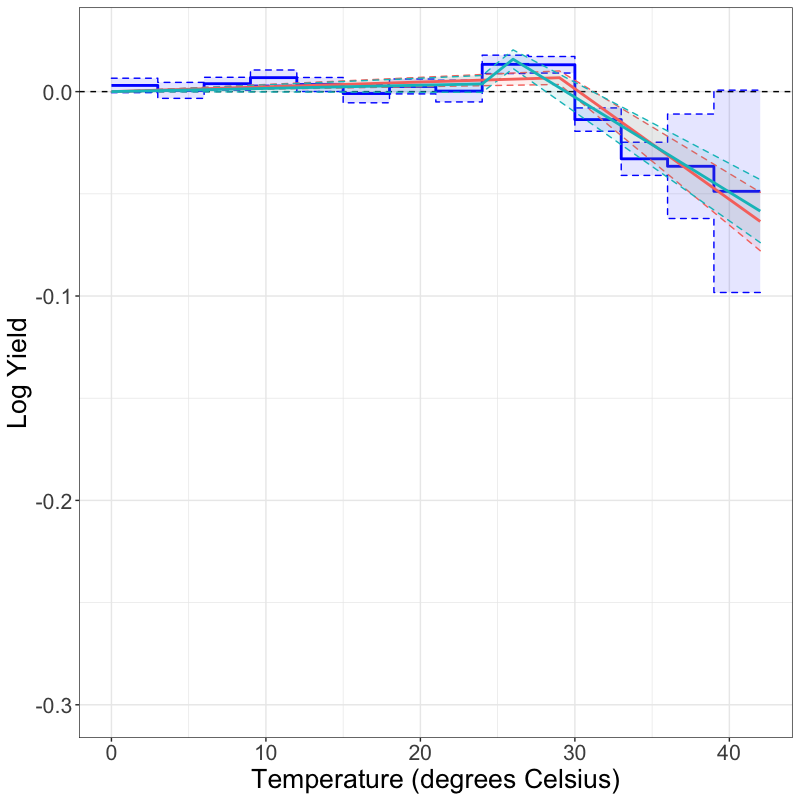} &
\includegraphics[width=.48\linewidth]{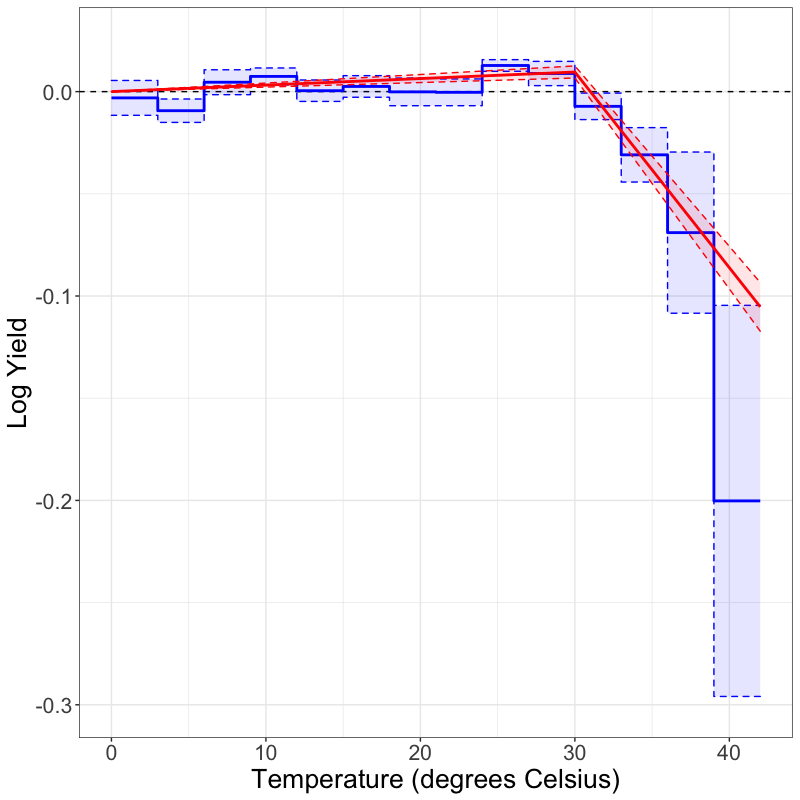}\\
\end{tabular}
\end{center}
\caption{Yield Response to Growing-season Temperature: Selected Models}\label{fig:yld_fn}
{\footnotesize {\it Notes:} The solid lines represent point estimates of the yield response functions, and the shallow bands are 95\% point-wise confidence intervals constructed by applying the delta method on state-clustered standard errors.  We caution that these confidence intervals do not account for the model selection step.}
\end{figure}

\newpage

\setcounter{page}{1} 

\begin{center}
    \huge{Online Appendix}
    
\end{center}
\startcontents[sections]
\printcontents[sections]{l}{1}{\setcounter{tocdepth}{2}}
\setcounter{figure}{0} \renewcommand{\thefigure}{A\arabic{figure}}
\setcounter{table}{0} \renewcommand{\thetable}{A\arabic{table}}
\setcounter{proposition}{0} \renewcommand{\theproposition}{A\arabic{proposition}}

\newpage

\section{Formal Definitions of Commonly Used Models in the Literature}\label{sec:model_def}
We provide formal definitions of some of the models commonly adopted in the empirical literature, 
based on the following representation of response functions:
\begin{align*}\mu_{\alpha}(\mathcal{W}_{it})&=X_{it,\alpha}'\beta_{\alpha}\\
X_{it,\alpha}&=\psi_{\alpha}\left(\sum_{h=1}^Hf_{\alpha}(W_{ith})^\prime\phi_{\alpha,h}\right).\end{align*}

\begin{enumerate}[(1)]
\item Mean temperatures:\\
 $f_{\alpha}(W_{ith})=W_{ith}$, 
 $\phi_{\alpha,h}=\frac{1}{H}$, 
 $\psi_\alpha=\sum_{h=1}^HW_{ith}/H$. 
\item Maximum temperatures:\\
  $f_{\alpha}(W_{ith})=W_{ith}$, 
 $\phi_{\alpha,h}=1\{W_{ith}=\max_h \{W_{ith}\}^H\}$, 
 $\psi_\alpha=\max_h \{W_{ith}\}^H$. 
\item Degree days:\\
$f_{\alpha}(W_{ith}) = \left( \max\{ W_{ith}-u,0\}, \max\{ \min \{ W_{ith}-l, u-l\}, 0\}   \right)'$, \\
$\phi_{\alpha,h} = (1,1)'$, $\psi_{\alpha}$ is the identity function and $u$ and $l$ are upper and lower thresholds ($u>l$).
\item Bins:\\
$f_{\alpha}(W_{ith})=\left(1\{W_{ith}\in[l_1,u_1)\},1\{W_{ith}\in[l_2,u_2)\},\dots, 1\{W_{ith}\in[l_{d_{f_{\alpha}}},l_{d_{f_{\alpha}}}]\}\right)^\prime,$\\
$\phi_{\alpha,h}=(1,1,\dots,1)^\prime$, $\psi_{\alpha}$ is the identity function and $\{[l_j,u_j)\}_{j=1}^{d_{f_\alpha}}$ is a set of intervals.  
\item Polynomials:\\
 $f_{\alpha}(W_{ith})=(W_{ith}, W_{ith}^2, ... , W_{ith}^k)'$ for $k$th polynomials, \\
 $\phi_{\alpha,h}=(\frac{1}{H},\frac{1}{H},...,\frac{1}{H})'$, $\psi_{\alpha}$ is the identity function.
\item Splines:\\
$f_\alpha (W_{ith}) = \left( s_0(W_{ith}) \cdot 1\{W_{ith} \in [t_0,t_1)\}, s_1(W_{ith}) \cdot \{W_{ith} \in [t_1,t_2) \}, ..., s_n (W_{ith}) \cdot \{W_{ith} \in [t_n,+\infty) \} \right)',$\\
$\phi_{\alpha,h}=(1,1,\dots,1)^\prime$, $\psi_{\alpha}$ is the identity function and $s_0, s_1, ... , s_n$ are a set of spline functions with associated knots of $t_0, t_1, ..., t_n$. 
\item Heat index:\\
Heat index typically involves $dim(W_{ith}) \geq 2$. For illustration, suppose we have $W_{ith} = (W_{1,ith}, W_{2,ith})$.
$f_\alpha (W_{ith}) = g(W_{1,ith},W_{2,ith})$ where $g$ is a transformation function that interacts $W_{1,ith}$ and $W_{2,ith}$ in a specific manner. 
$\phi_{\alpha,h}=\frac{1}{H}$, 
 $\psi_\alpha=\sum_{h=1}^HW_{ith}/H$. 
\end{enumerate}
\section{Consistent Model Selection of MCCV and GIC}\label{app:existing_criteria}
\subsection{Monte Carlo Cross-Validation}\label{app:mccv}


Here we provide the conditions under which Monte Carlo cross-validation is model-selection consistent. Before we proceed, we need to express the within-demeaned model in matrix form,  $\widetilde{\mathbb{V}}=(\widetilde{V}_{1},\dots,\widetilde{V}_{n})^\prime$. Further, write $U_{i}$, $i=1, \ldots, n$, to denote the error terms in the true DGP, which are assumed to be conditionally mean zero. Then we can express the within-demeaned version of model $\mathbf{M}_{\alpha}$ in matrix form as

\begin{align}
\widetilde{\mathbb{Y}}&=\widetilde{{\mathbb{X}}}_{\alpha}\beta_{\alpha}+\tbbU_{\alpha}.\nonumber
\end{align}
Similarly to \citet{Shao:1993}, we study the mean squared prediction error (MSPE) of $\mathbf{M}_{\alpha}$, which is estimated using $\{Y_{i}, X_{i}\}_{i=1}^{n}$, in predicting out-of-sample observations of $Y_{i}$, which we will refer to as $Z_{i}$.  Assume that the conditional variance of the error terms in the true DGP (which are conditionally mean zero) is equal to $E[U_{i}U_{i}'|\mathcal{W}_{i}]=\sigma^{2}I_{T}$, and also assume that $\{Y_{i},X_{i}\}$ are i.i.d. across $i$.  The expectation operators $E[ \cdot | \mathcal{W}_{i}]$ in the preceding sentence and $E[\cdot | \{\mathcal{W}_{i}\}_{i=1}^{n}]$ in the following displayed equation refer to the conditional distribution derived from the true joint distribution of $Y_{i}$ and $\mathcal{W}_{i}$. The MSPE of the fitted model $\mathbf{M}_{\alpha}$ is given by
\begin{align}
\Gamma_{\alpha,nT}&=\frac{1}{nT}E\left[\left.\sum_{i=1}^{n}\sum_{t=1}^{T}(\widetilde{Z}_{it}-\widetilde{X}_{it,\alpha}'\hat{\beta}_{\alpha})^{2}\right|\{\mathcal{W}_{i}\}_{i=1}^{n}\right] \nonumber \\
&=\frac{T-1}{T}\sigma^{2}+\underbrace{\frac{1}{nT}\sigma^{2}k_\alpha}_{\text{model dimension}}+\underbrace{\Delta_{_{\alpha},nT}}_{\text{``misspecification'' error}}, \nonumber
\end{align}
where $\Delta_{\alpha,nT}=\frac{1}{nT}\beta_{\star,\emph{o}}'\tbbX_{\star}'\left(I_{nT}-P_{\alpha}\right)\tbbX_{\star}\beta_{\star,\emph{o}}\geq 0$ and $P_{\alpha}$ is the projection matrix onto the column space of the design matrix $\tbbX_{\alpha}$.  The derivation of the above equality is included in Section~\ref{app:proofs} of the Appendix for the reader's convenience.  Several remarks are in order.  The homoskedasticity and serial uncorrelatedness of the idiosyncratic shocks are crucial to obtain a component of the MSPE that depends on the model dimension.  As in \citet{Shao:1993}, it is convenient to consider two categories of models, 
\begin{enumerate}[-]
\item Category I: $\Delta_{\alpha,nT}>0$,
\item Category II: $\Delta_{\alpha,nT}=0$, when $X_{it,\star}=R_{\star,\alpha}X_{it,\alpha}$. 
\end{enumerate}

The following standard conditions correspond to conditions in Shao (1993) which we have adapted to the fixed effects model with stochastic regressors.
\begin{condition}\label{cond:mccv} (MCCV Consistency)
\begin{enumerate}[1.]
\item (DGP and Models) For $i=1,2,\dots,n$, $t=1,2,\dots,T$, $Y_{it}=X_{it,\star}\beta_{\star,\emph{o}}+a_{i}+u_{it}$, where $u_{it}|\mathcal{W}_{i1},\dots,\mathcal{W}_{iT},a_{i}\overset{i.i.d.}{\sim} (0,\sigma^{2})$ across $i$ and $t$.  For some $\alpha=1,\dots,A$, $\mathbf{M}_{\alpha}=\mathbf{M}_{\star}$.
\item (Restriction on Category I Models) $\plim \inf_{n\rightarrow\infty}\Delta_{\alpha,nT}>0$ for $\mathbf{M}_{\alpha}$ in Category I.
\item (Regularity Conditions)
\begin{enumerate}[i.]\item  $\tbbX_{\alpha}'\tbbX_{\alpha}=O_{p}(n)$ and $\left(\tbbX_{\alpha}'\tbbX_{\alpha}\right)^{-1}=O_{p}(n^{-1})$ for $\alpha=1,2,\dots,A$,\\
 \item $\plim_{n\rightarrow\infty}\max_{i\leq n,t\leq T}w_{it,\alpha}=0$ $\forall \alpha=1,2,\dots,A$, where $w_{it,\alpha}$ is the it$^{th}$ diagonal element of $P_{\alpha}$,\\
\item $\max_{s\in\mathcal{\mathcal{R}}}\left\|\frac{1}{n_{v}}\sum_{i\in s^c}\sum_{t=1}^{T}\widetilde{X}_{it,\alpha}\widetilde{X}_{it,\alpha}'-\frac{1}{n_{c}}\sum_{i\in s}\sum_{t=1}^{T}\widetilde{X}_{it,\alpha}\widetilde{X}_{it,\alpha}'\right\|=o_{p}(1)$ for $\alpha=1,2,\dots,A$.
\end{enumerate}
\end{enumerate}
\end{condition}
Condition~\ref{cond:mccv}.1 imposes the i.i.d.\ assumption on the idiosyncratic shocks conditional on the time series of the high-frequency regressor $\{\mathcal{W}_{it}\}_{t=1}^T$ and $a_i$ across $i$ and $t$.\footnote{The derivation of $\Gamma_{\alpha,nT}$ in Appendix \ref{app:gamma_derivation} illustrates the crucial role of this condition in obtaining the term that depends on model dimension, $\frac{1}{nT}\sigma^2 k_{\alpha}$.} Note that the i.i.d. assumption is not imposed on the joint distribution of of $(Y_{it},X_{it})$. In fact, the regularity condition in \ref{cond:mccv}.3(i) is a high-level condition that ensures the applicability of a law of large numbers for $\tbbX_{\alpha}'\tbbX_{\alpha}/n$, and that it converges to an invertible matrix in probability for any $\alpha=1,\dots,A$. If we assume the cross-sectional i.i.d. assumption on $X_{it}$, then the existence of $E[\tilde{X}_{it}\tilde{X}_{it}']$ as well as its inverse would suffice. These conditions would also be sufficient for Condition \ref{cond:mccv}.3(ii)-(iii) \citep[\emph{see} Section 4.5,][for additional discussion]{Shao:1993}. As for Condition~\ref{cond:mccv}.2, it is a restriction that ensures that models in Category I are well-separated from models in Category II.  Note that $\Delta_{\alpha,nT}$ is the sum of squares of the elements of the vector of misspecification errors of $\mathbf{M}_{\alpha}$, $(I-P_{\alpha})\mathbb{X}_{\star}\beta_{\star,0}$. Since Category I models do not nest $\mathbf{M}_{\star}$, $\Delta_{\alpha,nT}$ will be greater than zero assuming $(\tilde{\mathbb{X}}_{\alpha}\tilde{\mathbb{X}}_{\alpha}')^{-1}$ exists.\footnote{By contrast, for models in Category II, $X_{it,\star}=R_{\star,\alpha}X_{it,\alpha}$ and therefore $\tilde{\mathbb{X}}_{\star}=\tilde{\mathbb{X}}_{\alpha}R_{\star,\alpha}$. As a result, for Category II models, $P_{\alpha}\tilde{\mathbb{X}}_{\star}\beta_{\star,0}=\tilde{\mathbb{X}}_{\alpha}(\tilde{\mathbb{X}}'\tilde{\mathbb{X}})^{-1}\tilde{\mathbb{X}}_{\alpha}'\tilde{\mathbb{X}}_{\alpha}R_{\star,\alpha}\beta_{\star,0}=\tilde{\mathbb{X}}_{\alpha}R_{\star,\alpha}'\beta_{\alpha}=\tilde{\mathbb{X}}_{\star}\beta_{\star,0}$. As a result, $\Gamma_{nT,\alpha}=0$ for models in Category II.} 

\begin{proposition}\label{prop:MCCV_consistency}
 Assume Condition \ref{cond:mccv}, and $n_{v}/n\rightarrow 1$ and $n_{c}=n-n_{v}\rightarrow\infty$, $b^{-1}n_{c}^{-2}n^2\rightarrow 0$.
\begin{enumerate}[(i)]
\item If $\mathbf{M}_{\alpha}$ is in Category I, then for some $R_{n}\geq 0$,
\begin{align}
\hat{\Gamma}_{\alpha,nT}^{MCCV}=\frac{1}{n_{v}Tb}\sum_{s\in\mathcal{R}}\tbbU_{s}'\tbbU_{s}+\Delta_{\alpha,nT}+o_{p}(1)+R_{n},
\end{align}
where $\tbbU_{s}=\widetilde{\mathbb{Y}}_{s}-\tbbX_{s}\beta$.
\item If $\mathbf{M}_{\alpha}$ is in Category II, then 
\begin{align}
\hat{\Gamma}_{\alpha,nT}^{MCCV}=\frac{1}{n_{v}Tb}\sum_{s\in\mathcal{R}}\tbbU_{s}'\tbbU_{s}+\frac{k_{\alpha}\sigma^{2}}{n_{c}T}+o_{p}(n_{c}^{-1}).
\end{align}
\item It follows that
\begin{align}
\lim_{n\rightarrow\infty}P(\widehat{\mathbf{M}}_{cv}=\mathbf{M}_{\star})=1.
\end{align}
\end{enumerate}
\end{proposition}
The proof is given in Appendix \ref{app:proofs}.  The above proposition establishes that if $\mathbf{M}_{\star}$ is under consideration, then MCCV with $n_{v}/n\rightarrow 1$ and $n_{c}\rightarrow\infty$, hereinafter MCCV-Shao, will select this model with probability tending to one in large samples.  Suppose $\mathbf{M}_{\star}$ is not considered, however some models that contain it (Category II) are in the set of candidate models.  Then the above proposition implies that the most parsimonious among those models in Category II will be selected with probability tending to one as $n \rightarrow \infty$ by the MCCV-Shao procedure.  However, the above does not ensure that if the models considered are all in Category I, i.e. none of the models considered contain $\mathbf{M}_{\star}$, that the most parsimonious model with the smallest $\lim_{n\rightarrow\infty}\Delta_{{\alpha},nT}$ will be selected with probability tending to one.  We explore this issue in simulations in Section \ref{sec:simulations_PWMSE}.  

In the absence of homoskedasticity and serial uncorrelatedness, it is well-known that it is very difficult to formally justify MCCV.  We will, however, examine its performance under weaker conditions in the numerical experiments.

\subsection{Generalized Information Criteria}\label{app:gic}
Here we introduce the generalized information criterion (GIC) for this problem.  In the linear fixed effects model, we do not need to specify a parametric family for the errors to define the estimator. However, it is computationally convenient to use the result that the linear FE estimator is identical to the conditional maximum likelihood estimator under the additional assumption of Gaussian errors, and the conditioning is on $\bar{Y}_{i}=\sum_{t=1}^{T}Y_{it}/T$, which is a sufficient statistic for the individual fixed effect \citep{arellano2003}. To be clear, we do not require the Gaussianity assumption or conditional maximum likelihood estimator for our theoretical results, but we take advantage of this equivalence with the linear FE estimator for convenience. We will use $f(\cdot)$ to denote the relevant density function. Let $Y_{i}=(Y_{i1},\dots,Y_{iT})^\prime$ and $X_{i,\alpha}=(X_{i1,\alpha},\dots,X_{iT,\alpha})$, the $i^{th}$ contribution to the conditional log-likelihood for $\mathbf{M}_{\alpha}$ is given by $\log \left( f(Y_{i}|X_{i,\alpha},a_{i,\alpha},\bar{Y}_{i};\beta_{\alpha},\sigma_{\alpha}^{2})\right)=\log \left(f(\tilde{Y}_{i}|\tilde{X}_{i,\alpha};\beta_{\alpha},\sigma_{\alpha}^{2})\right)\propto-(T-1)\log(\sigma_{\alpha}^{2})-\sum_{t=1}^{T}\left(\tilde{Y}_{it}-\tilde{X}_{it,\alpha}'\beta_{\alpha}\right)^2/\sigma_{\alpha}^2$.  The log-likelihood function is hence given by
\begin{align}
\ell_{nT}^{\alpha}(\beta_{\alpha},\sigma_{\alpha}^{2})&=-n(T-1)\log(\sigma_{\alpha}^{2})-\frac{\sum_{i=1}^{n}\sum_{t=1}^{T}\left(\tilde{Y}_{it}-\tilde{X}_{it,\alpha}'\beta_{\alpha}\right)^2}{\sigma_{\alpha}^2}.
\end{align}
In the following, we will work with the conditional profile likelihood function 
\begin{equation*}
\ell_{nT}^{\alpha}(\beta_{\alpha},\hat{\sigma}_{\alpha}^2(\beta_{\alpha})) \propto-n(T-1)\log\left(\hat{\sigma}_{\alpha}^2(\beta_{\alpha})\right),
\end{equation*}
where
\begin{equation}
\hat{\sigma}_{\alpha}^{2}(\beta_{\alpha})=\frac{1}{n(T-1)}\sum_{i=1}^{n}\sum_{t=1}^{T}(\tilde{Y}_{it}-\tilde{X}_{it,\alpha}'{\beta}_{\alpha})^{2}
\end{equation}
is the maximum likelihood estimator for $\sigma_{\alpha}^{2}$ given a fixed value of $\beta_{\alpha}$. Hereinafter, we let $\hat{\ell}_{nT}^\alpha\equiv\ell_{nT}^{\alpha}(\hat{\beta}_{\alpha},\hat{\sigma}_{\alpha}^2(\hat{\beta}_{\alpha}))= -n(T-1)\log(\hat{\sigma}_{\alpha}^{2}(\hat{\beta}_{\alpha}))$, where $$\hat{\beta}_{\alpha}=\arg\max_{\beta\in\mathcal{B}}\ell_{nT}^{\alpha}(\beta_{\alpha},\hat{\sigma}^{2}(\beta_{\alpha})).$$

The generalized information criterion (GIC) is given by the following
\begin{align}
GIC_{\alpha,\lambda_{nT}}&=\hat{\ell}_{nT}^{\alpha}-\lambda_{nT}k_{\alpha}.
\end{align}
The term $\lambda_{nT}$ penalizes model dimension.  The choice of $\lambda_{nT}=2$ corresponds to the AIC, whereas the choice $\lambda_{nT}=\log(nT)$ corresponds to the BIC, since $nT$ is the total number of observations.  One of the attractive features of information criteria is that we can formally justify their behavior under heteroskedasticity, spatial and/or time series dependence by viewing them as a misspecfication of the above log-likelihood.  Since we will deal with misspecification in this section, we introduce another definition, which is pseudo-consistency of a model selection procedure following \citet{Sin:1996}.  
\begin{definition}(Pseudo-Consistency of Model Selection) Let $\mathbb{M}=\{\mathbf{M}_{\alpha}\}_{\alpha=1}^{A}$ and $\mathbf{M}_{\star}$ is not nested in $\mathbf{M}_{\alpha}$ for any $\alpha=1,\dots,A$.  Then a model selection criterion $\textbf{C}$ is said to be pseudo-consistent if $\lim_{n\rightarrow\infty}P(\widehat{M}_{\textbf{C}}=\mathbf{M}_{\textbf{P}})=1$, where $\widehat{M}_{\textbf{C}}$ is the model selected by criterion $\textbf{C}$ and $\mathbf{M}_{\textbf{P}}$ is the most parsimonious model with the smallest Kullback-Leibler divergence \citep[][Chapter 2]{Claeskens:2008} from the true data-generating distribution among all models in $\mathbb{M}$.
\end{definition}

Building on previous results on the behavior of the quasi-log-likelihood ratio statistic \citep{Vuong:1989,Sin:1996}, we can establish conditions for GIC's consistency and pseudo-consistency in the context of climate change impact studies.  Without loss of generality, consider the choice between two models $\mathbf{M}_{\alpha}$ and $\mathbf{M}_{\gamma}$ with $k_{\alpha}<k_{\gamma}$.  Let $LR_{nT}^{\alpha,\gamma}=\hat{\ell}_{nT}^{\alpha}-\hat{\ell}_{nT}^{\gamma}$. Furthermore, write $\widehat{\mathbf{M}}_{\lambda_{nT}}$ to denote the model that minimizes the GIC given $\lambda_{nT}$, 
\begin{align}
&P(\widehat{\mathbf{M}}_{\lambda_{nT}}=\mathbf{M}_{\alpha})= P\left(GIC_{\alpha,\lambda_{nT}}>GIC_{\gamma,\lambda_{nT}}\right)
=P\left(LR_{nT}^{\alpha,\gamma}>\lambda_{nT}(k_{\alpha}-k_{\gamma})\right). \nonumber
\end{align}
\citet{Vuong:1989} establishes that the rate of convergence of the quasi-likelihood ratio statistic under the null hypothesis differs depending on whether the conditional densities under $\mathbf{M}_{\alpha}$ and $\mathbf{M}_{\gamma}$ agree at the pseudo-true parameter values or not.  In our setting, this is determined by whether the predicted values of the outcome at the pseudo-true parameters of the two models differ or coincide.

Since we allow for the possibility for misspecification in this section, we introduce the pseudo-true parameter value for a model $\mathbf{M}_{\alpha}$,
$$\beta_{\alpha}^*=\left(\bar{E}[\widetilde{X}_{i,\alpha}\widetilde{X}_{i,\alpha}']\right)^{-1}\bar{E}[\widetilde{X}_{i,\alpha}\widetilde{Y}_{i}],$$
where $\widetilde{X}_{i,\alpha}=(\widetilde{X}_{i1,\alpha},\widetilde{X}_{i2,\alpha},\dots,\widetilde{X}_{iT,\alpha})$ and $\widetilde{Y}_{i}=(\widetilde{Y}_{i1},\dots,\widetilde{Y}_{iT})'$. We also introduce $\widetilde{u}_{it,\alpha}=\tilde{Y}_{it,\alpha}-\tilde{X}_{it,\alpha}'\beta_{\alpha}^*$, $\widehat{{u}}_{it,\alpha}=\tilde{Y}_{it,\alpha}-\tilde{X}_{it,\alpha}'\hat{\beta}_{\alpha}$, $\widetilde{u}_{i,\alpha}=(\widetilde{u}_{i1,\alpha},\dots,\widetilde{u}_{iT,\alpha})'$ and $\widehat{u}_{i,\alpha}=(\widehat{u}_{i1,\alpha},\dots,\widehat{u}_{iT,\alpha})'$. For a random variable $W_i$, $\bar{E}[W_i]=\lim_{n\rightarrow\infty}\frac{1}{n}\sum_{i=1}^nE[W_i]$. We use $\sigma_{\alpha}^2(\beta_{\alpha}^*)=\bar{E}[\widetilde{u}_{i,\alpha}'\widetilde{u}_{i,\alpha}]$ to denote the population MSE for a model $\mathbf{M}_{\alpha}$.

\begin{align*}\widehat{D}_{\alpha}&=\frac{1}{n}\sum_{i=1}^n\widetilde{X}_{i,\alpha}\widetilde{X}_{i,\alpha}',&D_{\alpha}&=\bar{E}[\widetilde{X}_{i,\alpha}\widetilde{X}_{i,\alpha}'],\\
\widehat{C}_{\alpha}&=\frac{1}{n}\sum_{i=1}^n\widetilde{X}_{i,\alpha}\widetilde{Y}_{i},&C_{\alpha}&=\bar{E}\left[\widetilde{X}_{i,\alpha}\widetilde{Y}_{i}\right],\\
\hat{b}_{\alpha}&=\frac{1}{n}\sum_{i=1}^n\widetilde{X}_{i,\alpha}\widetilde{u}_{i,\alpha}',&b_{\alpha}&=\bar{E}[\widetilde{X}_{i,\alpha}\widetilde{u}_{i,\alpha}].\end{align*}
For two models $\mathbf{M}_{\alpha}$ and $\mathbf{M}_{\gamma}$, let \begin{align*}
B_{\alpha,\gamma}&=\bar{E}\left[\widetilde{X}_{i,\alpha}\widetilde{u}_{i,\alpha}\widetilde{u}_{i,\gamma}'\widetilde{X}_{i,\gamma}'\right], \nonumber\\
V_{\alpha,\gamma}&=\bar{E}[(\widetilde{u}_{i,\alpha}'\widetilde{u}_{i,\alpha}-\bar{E}[\widetilde{u}_{i,\alpha}'\widetilde{u}_{i,\alpha}])(\widetilde{u}_{i,\gamma}'\widetilde{u}_{i,\gamma}-\bar{E}[\widetilde{u}_{i,\gamma}'\widetilde{u}_{i,\gamma}])].\end{align*}
Note that by definition the matrix $B_{\gamma,\alpha}=B_{\alpha,\gamma}'$ and the scalar $V_{\alpha,\gamma}=V_{\gamma,\alpha}$. With a slight abuse of notation, we will use $B_{\alpha}$ and $V_{\alpha}$ to denote $B_{\alpha,\alpha}$ and $V_{\alpha,\alpha}$, respectively.

\begin{condition} The following conditions hold as $n\rightarrow\infty$ holding $T$ fixed:
\begin{enumerate}[1.]
\item $\widehat{D}_{\alpha}\overset{p}{\rightarrow} D_{\alpha}$ and $\widehat{D}_{\gamma}\overset{p}{\rightarrow} D_{\gamma}$, where $D_{\alpha}$ and $D_{\gamma}$ are finite, symmetric positive definite matrices.
\item $\widehat{C}_{\alpha}\overset{p}{\rightarrow}C_{\alpha}<\infty$ and $\widehat{C}_{\gamma}\overset{p}{\rightarrow}C_{\gamma}<\infty$.
\item $$\sqrt{n}\left(\begin{array}{c}\hat{b}_{\alpha}\\\hat{b}_{\gamma}\end{array}\right)\overset{d}{\rightarrow}N\left(0,\left(\begin{array}{cc}B_{\alpha}&B_{\alpha,\gamma}\\B_{\gamma,\alpha}&B_{\gamma}\end{array}\right)\right).$$ 
 \item $\hat{\sigma}_{\alpha}^2(\hat{\beta}_{\alpha})\overset{p}{\rightarrow}\sigma_{\alpha}^2(\beta_{\alpha}^*)$ and $\hat{\sigma}_{\gamma}^2(\hat{\beta}_{\gamma})\overset{p}{\rightarrow}\sigma_{\gamma}^2(\beta_{\gamma}^*)$, where $0<\sigma_{\alpha}^2(\beta_{\alpha}^*)<\infty$ and  $0<\sigma_{\gamma}^2(\beta_{\gamma}^*)<\infty$.
 \item $$\sqrt{n}\left(\begin{array}{c}\hat{\sigma}_{\alpha}^2(\beta_{\alpha}^*)-\sigma_{\alpha}^2(\beta_{\alpha}^*)\\\hat{\sigma}_{\gamma}^2(\beta_{\gamma}^*)-\sigma_{\gamma}^2(\beta_{\gamma}^*)\end{array}\right)\overset{d}{\rightarrow}N\left(0,\left(\begin{array}{cc}V_{\alpha}&V_{\alpha,\gamma}\\
 V_{\gamma,\alpha}&V_{\gamma}\end{array}\right)\right).$$
\end{enumerate}
\label{cond:gic}
\end{condition}
Conditions \ref{cond:gic}.1 and \ref{cond:gic}.2 ensure the consistency of $\hat{\beta}_{\alpha}$ and $\hat{\beta}_{\gamma}$ for their respective well-defined pseudo-true parameter values, $\beta_{\alpha}^*$ and $\beta_{\gamma}^*$. Conditions \ref{cond:gic}.1 and \ref{cond:gic}.3 ensure the joint asymptotic normality of  $\hat{\beta}_{\alpha}$ and $\hat{\beta}_{\gamma}$. Note that this condition does not require the variance-covariance matrix to be nonsingular. This is by design, since this condition is used to show the asymptotic distribution of the QLR statistic when the two models are nested relying on Lemma 3.2 in \citet{Vuong:1989}.  Condition \ref{cond:gic}.4 ensures the consistency of the sample mean-squared error to the population mean-squared error. Condition \ref{cond:gic}.5 ensures the joint asymptotic normality of $\hat{\sigma}_{\alpha}(\beta_{\alpha}^*)$ and $\hat{\sigma}_{\gamma}(\beta_{\gamma}^*)$. We choose to present the conditions in terms of consistency and asymptotic normality in order to illustrate that the results can allow for dependence across $i$ and $t$ subject to the applicability of appropriate laws of large numbers and central limit theorems. If additional sampling requirements are imposed, it is straightforward to obtain more primitive conditions that imply Condition \ref{cond:gic}. For instance, if we impose the cross-sectional i.i.d. assumption, then the existence of $D_{\alpha}$ and $D_{\gamma}$, assuming they are both positive definite, would be sufficient for Condition \ref{cond:gic}.1. The existence of $C_{\alpha}$ and $C_{\gamma}$ would imply Condition \ref{cond:gic}.2. As for Condition \ref{cond:gic}.3 (\ref{cond:gic}.5), the existence of $B_{\alpha}$ and $B_{\gamma}$ ($V_{\alpha}$ and $V_{\gamma}$) would be sufficient.

The following proposition gives sufficient conditions for $GIC_{\lambda_{nT}}$ to deliver (pseudo-)consistent model selection in our problem when considering three possible cases with two models.   In the first two cases, both models are equal in terms of Kullback-Leibler divergence from the true data-generating distribution.  In both cases, a (pseudo-) consistent GIC should select the more parsimonious model.  The third case is when one model is strictly better in terms of Kullback-Leibler divergence, in which case this model should be chosen by a (pseudo-) consistent GIC.  Recall that $x_{\omega_t,\alpha}$ is a realization of $X_{it,\alpha}$ given a particular realization of $\mathcal{W}_{it}$. Let $\tilde{\mathbf{x}}_{\{\omega_{t}\}_{t=1}^{T},\alpha}$ denote the within-demeaned version of $x_{\omega_t,\alpha}$ given $T$ realizations of $\mathcal{W}_{it}$, i.e. $\{\omega_{t}\}_{t=1}^{T}$.  
\begin{proposition}\label{prop:GICs} Assume Condition \ref{cond:gic} holds.  The following statements hold as $n\rightarrow\infty$, holding $T$ fixed.
\begin{enumerate}[1.] 
\item Suppose $\sigma_{\alpha}^2(\beta_{\alpha}^*)=\sigma_{\gamma}^2(\beta_{\gamma}^*)$ and $\tilde{\mathbf{x}}_{\cdot,\alpha}'\beta_{\alpha}^*=\tilde{\mathbf{x}}_{\cdot,\gamma}'\beta_{\gamma}^*$ hold. Then
\begin{align} 
&P(\widehat{\mathbf{M}}_{\lambda_{nT}}=\mathbf{M}_{\alpha})=P\left(GIC_{\alpha,\lambda_{nT}}>GIC_{\gamma,\lambda_{nT}}\right)
=P\left(LR_{n}^{\alpha,\gamma}>\lambda_{nT}(k_{\alpha}-k_{\gamma})\right)\rightarrow 1, \nonumber
\end{align}
if ${\lambda_{nT}\rightarrow\infty}$.
\item Suppose $\sigma_{\alpha}^2(\beta_{\alpha}^*)=\sigma_{\gamma}^2(\beta_{\gamma}^*)$ and $\tilde{\mathbf{x}}_{\cdot,\alpha}'\beta_{\alpha}^*\neq \tilde{\mathbf{x}}_{\cdot,\gamma}'\beta_{\gamma}^*$ hold. Then
\begin{align} 
&P(\widehat{\mathbf{M}}_{\lambda_{nT}}=\mathbf{M}_{\alpha})=P(GIC_{\alpha,\lambda_{nT}}>GIC_{\gamma,\lambda_{nT}})=P\left(\frac{1}{\sqrt{nT}}LR_{nT}^{\alpha,\gamma}>\frac{\lambda_{nT}}{\sqrt{n}}(k_{\alpha}-k_{\gamma})\right)\rightarrow 1, \nonumber 
\end{align}
if $\lambda_{nT}/\sqrt{n}\rightarrow\infty$.
\item \begin{enumerate}[a.]\item Suppose that $\sigma_{\alpha}^2(\beta_{\alpha}^*)<\sigma_{\gamma}^2(\beta_{\gamma}^*)$ holds. Then
{\small{\begin{align} &P(\widehat{\mathbf{M}}_{\lambda_{nT}}=\mathbf{M}_{\alpha})=P(GIC_{\alpha,\lambda_{nT}}>GIC_{\gamma,\lambda_{nT}})=P\left(\frac{1}{nT}LR_{nT}^{\alpha,\gamma}>\frac{\lambda_{nT}}{n}(k_{\alpha}-k_{\gamma})\right)\rightarrow 1, \nonumber\end{align}}}
if $\lambda_{nT}/n\rightarrow 0$.
\item Suppose that $\sigma_{\alpha}^2(\beta_{\alpha}^*)>\sigma_{\gamma}^2(\beta_{\gamma}^*)$ holds. Then
{\small{\begin{align} &P(\widehat{\mathbf{M}}_{\lambda_{nT}}=\mathbf{M}_{\gamma})=P(GIC_{\alpha,\lambda_{nT}}>GIC_{\gamma,\lambda_{nT}})=P\left(\frac{1}{n}LR_{nT}^{\alpha,\gamma}<\frac{\lambda_{nT}}{n}(k_{\alpha}-k_{\gamma})\right)\rightarrow 1, \nonumber\end{align}}}
if $\lambda_{nT}/n\rightarrow 0$.
\end{enumerate}
\end{enumerate}
\end{proposition}
The proof of the above proposition is provided in Appendix \ref{proof:prop2}. It builds on \citet{Vuong:1989}. The result is a special case of what has been shown in \citet{Sin:1996} and \citet{Hong:2012}.  The above theorem shows that for a GIC to be (pseudo-) consistent in all cases, then $\lambda_{nT}$ has to fulfill three conditions as $n\rightarrow\infty$: (1) $\lambda_{nT}\rightarrow\infty$, (2) $\lambda_{nT}/\sqrt{n}\rightarrow\infty$, (3) $\lambda_{nT}/n\rightarrow 0$. The first two conditions ensure that when two models have the same population MSE, the smaller model is chosen. The third condition ensures that when one model dominates the other in terms of population MSE that the term that depends on the model dimension converges to zero, such that the best model in terms of smallest population MSE is chosen, whether it is $\mathbf{M}_{\alpha}$ or $\mathbf{M}_{\gamma}$.


\section{Derivations and Proofs for Appendix \ref{app:existing_criteria}}\label{app:proofs}
\subsection{MSPE Derivation}\label{app:gamma_derivation}
Let $\tilde{\mathbb{Z}}$ denote a vector of $n$ ``future'' values of the within-demeaned outcome whereas $\hat{\beta}_{\alpha}$ was estimated using the sample $\tbbY$ and $\tbbX_{\alpha}$.
\begin{align}\widehat{\Gamma}_{\alpha,nT}&=\frac{1}{nT}\left\|\widetilde{\mathbb{Z}}-\tbbX_{\alpha}\hat{\beta}_{\alpha}\right\|^{2}\nonumber\\
&=\frac{1}{nT}\left\|\tbbU_{z}+\tbbX_{\star}\beta_{\star,\emph{o}}-\underbrace{\tbbX_{\alpha}\left(\tbbX_{\alpha}'\tbbX_{\alpha}\right)^{-1}\tbbX_{\alpha}'}_{\equiv P_{\alpha}}\widetilde{\mathbb{Y}}\right\|^{2}\nonumber\\
&=\frac{1}{nT}\left\|\tbbU_{z}+(I_{nT}-P_{\alpha})\tbbX_{\star}\beta_{\star,\emph{o}}-P_{\alpha}\tbbU\right\|^{2}\end{align}
where $\tbbU_{z}$ denotes the within-individual demeaned error term of the observations $\widetilde{\mathbb{Z}}$.\\ 
Let $\Gamma_{\alpha,nT}$ denote the expectation of $\widehat{\Gamma}_{\alpha,nT}$ conditional on $\{\mathcal{W}_{i}\}_{i=1}^{n}$
\begin{align}\Gamma_{\alpha,nT}&=\frac{1}{nT}E[\tbbU_{z}'\tbbU_{z}|\{\mathcal{W}_{i}\}_{i=1}^{n}]+E[\tbbU'P_{\alpha}\tbbU|\{\mathcal{W}_{i}\}_{i=1}^{n}]+\frac{1}{nT}\beta_{\star,\emph{o}}'\tbbX_{\star}'(I_{nT}-P_{\alpha})\tbbX_{\star}\beta_{\star,\emph{o}}\end{align}
The first term on the right hand size of the equality equals $E\left[\sum_{i=1}^{n}\sum_{t=1}^{T}\tilde{u}_{it}^{2}\right]/nT=E[\tilde{u}_{it}^{2}]=\sigma^{2}(T-1)/T$.  The second term can be simplified as follows
\begin{align}
&\frac{1}{nT}E\left[\tbbU'\mathbb{X}_{\alpha}(\tbbX_{\alpha}'\tbbX_{\alpha})^{-1}\tbbX_{\alpha}'\tbbU|\{\mathcal{W}_{i}\}_{i=1}^{n}\right]=\frac{1}{nT}tr\left(E\left[\tbbU\widetilde{\mathbb{U}}'\mathbb{X}_{\alpha}(\tbbX_{\alpha}'\tbbX_{\alpha})^{-1}\tbbX_{\alpha}'|\{\mathcal{W}_{i}\}_{i=1}^{n}\right]\right)\nonumber\\
=&\frac{1}{nT}\sigma^{2}tr(\underbrace{(I_{n}\otimes(I_{T}-\mathcal{J}_{T}/T))}_{\equiv I_{n}\otimes Q_{T}}\tbbX_{\alpha}(\tbbX_{\alpha}'\tbbX_{\alpha})^{-1}\tbbX_{\alpha}')=\frac{1}{nT}\sigma^{2}k_{\alpha}.\end{align}
where the last equality follows by noting $(I_{n}\otimes Q_{T})\tbbX_{\alpha}=\tbbX_{\alpha}$ as well as properties of the trace.

\subsection{Proof of Proposition \ref{prop:MCCV_consistency}}
\begin{proof}
The proof is adapted from \citet{Shao:1993} to the setting of a fixed effects model with stochastic high-frequency regressors.  Following \citet{Shao:1993}, we first show the results for Balanced Incomplete Cross-Validation (BICV) with stochastic regressors, then we extend the results to MCCV.  Let $\mathbf{B}$ be a collection of $b$ subsets of $\{1,\dots,n\}$ that have size $n_{v}$ such that (i) for each $i$, $1\leq i\leq n$, the same number of subsets of $\mathbf{B}$ include it, (ii) for each pair $(i,j)$ for $i,j\in\{1,\dots,n\}$, the same number of subsets of $\mathbf{B}$ include it.  From (3.1) in \citet{Shao:1993} and the balance property of $\mathbf{B}$,
\begin{align}\hat{\Gamma}_{\alpha,nT}^{BICV}\geq \frac{1}{n_{v}Tb}\sum_{s\in\mathbf{B}}\|\tbbY_{s}-\tbbX_{s,\alpha}\hat{\beta}_{\alpha}\|^{2}=n^{-1}\|\tbbY-\tbbX_{\alpha}\hat{\beta}_{\alpha}\|^{2}=(nT)^{-1}\tbbU'\tbbU+\Delta_{\alpha,nT}+o_{p}(1)\end{align}
where the last equality follows from the proof of (3.5) in \citet{Shao:1993}.  (i) in this proposition follows by letting $R_{n}=\hat{\Gamma}_{\alpha,nT}^{BICV}-\|\tbbY-\tbbX_{\alpha}\hat{\beta}_{\alpha}\|^{2}/n$.\\
By Condition \ref{cond:mccv}.3.(iii) with $s\in\mathbf{B}$ in lieu of $s\in\mathcal{R}$, it follows for every $s\in\mathbf{B}$,  
\begin{align}&\frac{1}{n}\tbbX_{\alpha}'\tbbX_{\alpha}-\frac{1}{n_{v}}\tbbX_{\alpha,s}'\tbbX_{\alpha,s}=\frac{1}{n}\left[\tbbX_{\alpha,s^{c}}'\tbbX_{\alpha,s^{c}}+\tbbX_{\alpha,s}'\tbbX_{\alpha,s}-\frac{n}{n_{v}}\tbbX_{\alpha,s}'\tbbX_{\alpha,s}\right]\nonumber\\
=&\frac{1}{n}\left[\tbbX_{\alpha,s^{c}}'\tbbX_{\alpha,s^{c}}+\tbbX_{\alpha,s}'\tbbX_{\alpha,s}-\frac{n_{c}+n_{v}}{n_{v}}\tbbX_{\alpha,s}'\tbbX_{\alpha,s}\right]=\frac{1}{n}\left[\tbbX_{\alpha,s^{c}}'\tbbX_{\alpha,s^{c}}-\frac{n_{c}}{n_{v}}\tbbX_{\alpha,s}'\tbbX_{\alpha,s}\right]\nonumber\\
=&\frac{n_{c}}{n}\left[\frac{1}{n_{c}}\tbbX_{\alpha,s^{c}}'\tbbX_{\alpha,s^{c}}-\frac{1}{n_{v}}\tbbX_{\alpha,s}'\tbbX_{\alpha,s}\right]=o_{p}\left(\frac{n_{c}}{n}\right)\end{align}
With some further manipulations,
\begin{align}\left(\frac{1}{n_{v}}\tbbX_{\alpha,s}'\tbbX_{\alpha,s}\right)^{-1}\frac{1}{n}\tbbX_{\alpha}'\tbbX_{\alpha}-I&=o_{p}\left(\frac{n_{c}}{n}\right)\left(\frac{1}{n_{v}}\tbbX_{\alpha,s}'\tbbX_{\alpha,s}\right)^{-1}\nonumber\\
\left(\frac{1}{n_{v}}\tbbX_{\alpha,s}'\tbbX_{\alpha,s}\right)^{-1}-\left(\frac{1}{n}\tbbX_{\alpha}'\tbbX_{\alpha}\right)^{-1}&=o_{p}\left(\frac{n_{c}}{n}\right)\left(\frac{1}{n_{v}}\tbbX_{\alpha,s}'\tbbX_{\alpha,s}\right)^{-1}\left(\frac{1}{n}\tbbX_{\alpha}'\tbbX_{\alpha}\right)^{-1}\nonumber\\
\left(\tbbX_{\alpha,s}'\tbbX_{\alpha,s}\right)^{-1}-\frac{n}{n_{v}}\left(\tbbX_{\alpha}'\tbbX_{\alpha}\right)^{-1}&=o_{p}\left(\frac{n_{c}}{n}\right)\left(\tbbX_{\alpha,s}'\tbbX_{\alpha,s}\right)^{-1}\underbrace{\left(\frac{1}{n}\tbbX_{\alpha}'\tbbX_{\alpha}\right)^{-1}}_{=O_{p}(1)\text{ by (3.3)}}
\end{align}
Hence, together with Condition \ref{cond:mccv}.3.i, the above implies that
\begin{align}(\tbbX_{\alpha,s}'\tbbX_{\alpha,s})^{-1}-\frac{n}{n_{v}}(\tbbX_{\alpha}'\tbbX_{\alpha})^{-1}=o_{p}\left(\frac{n_{c}}{n}\right)(\tbbX_{\alpha,s}'\tbbX_{\alpha,s})^{-1}\end{align}
For $P_{\alpha,s}=\tbbX_{\alpha,s}(\tbbX_{\alpha,s}'\tbbX_{\alpha,s})^{-1}\tbbX_{\alpha,s}'$,
\begin{align}P_{\alpha,s}
=&\frac{n}{n_{v}}\tbbX_{\alpha,s}\left(\tbbX_{\alpha}'\tbbX_{\alpha}\right)^{-1}\tbbX_{\alpha,s}'+\frac{n}{n_{v}}\tbbX_{\alpha,s}\left(\tbbX_{\alpha}'\tbbX_{\alpha}\right)^{-1}\tbbX_{\alpha,s}'+o\left(\frac{n_{c}}{n}\right)\tbbX_{\alpha,s}(\tbbX_{\alpha,s}'\tbbX_{\alpha,s})^{-1}\tbbX_{\alpha,s}'\nonumber\\
=&\frac{n}{n_{v}}Q_{\alpha,s}+o_p\left(\frac{n_{c}}{n}\right)P_{\alpha,s}\end{align}
Given that $n_{v}/n=O(1)$, it follows that 
\begin{align}Q_{\alpha,s}&=P_{\alpha,s}\left(\frac{n_{v}}{n}+o_p\left(\frac{n_{c}}{n}\right)\right)\hspace{2cm}\label{(A.3)}\end{align}
From the balance property of $\mathbf{B}$,
\begin{align}&\frac{1}{n_{v}Tb}\sum_{s\in\mathbf{B}}\textbf{r}_{\alpha,s}'Q_{\alpha,s}\mathcal{r}_{\alpha,s}=\frac{1}{n_{v}Tb}\sum_{s\in\mathbf{B}}\sum_{i\in s}\sum_{t=1}^{T}w_{it,\alpha}r_{it,\alpha}^{2}=\frac{1}{n_{v}Tb}\left(\frac{n_{v}b}{n}-\frac{n_{v}b}{n}\frac{n_{v}-1}{n-1}\right)\sum_{i=1}^{n}\sum_{t=1}^{T}w_{it,\alpha}r_{it,\alpha}^{2}\nonumber\\
=&\frac{1}{T}\left(\frac{1}{n}-\frac{n_{v}-1}{n(n-1)}\right)\sum_{i=1}^{n}\sum_{t=1}^{T}w_{it,\alpha}r_{it,\alpha}^{2}\nonumber\end{align}
where $\textbf{r}_{\alpha,s}=\tbbY_{s}-\tbbX_{\alpha,s}\hat{\beta}_{\alpha}$ and $r_{it,\alpha}=\tilde{Y}_{it}-\tilde{X}_{it,\alpha}'\hat{\beta}_{\alpha}$.\\
By \eqref{(A.3)} and $n_{v}/n\rightarrow 1$ and $n_{c}\rightarrow\infty$, let $c_{n}=n_{v}(n+n_{c})n_{c}^{-2}$,
\begin{align}\frac{c_{n}}{n_{v}Tb}\|P_{\alpha,s}\textbf{r}_{\alpha,s}\|^{2}&=\left[\frac{n_{v}}{n}+o_{p}\left(\frac{n_{c}}{n}\right)\right]^{-1}\frac{c_{n}}{n_{v}Tb}\sum_{s\in\mathbf{B}}\textbf{r}_{\alpha,s}'Q_{\alpha,s}\textbf{r}_{\alpha,s}\nonumber\\
&=\left[\frac{n}{n_{v}}+o_{p}\left(\frac{n_{c}}{n}\right)\right]\frac{n_{v}(n+n_{c})n_{c}^{-2}}{n_{v}Tb}\sum_{s\in\mathbf{B}}\textbf{r}_{\alpha,s}'Q_{\alpha,s}\textbf{r}_{\alpha,s}\nonumber\\
&=\left[1+o_{p}\left(\frac{n_{c}}{n}\right)\right]\frac{n+n_{c}}{n_{c}(n-1)T}\sum_{i=1}^{n}\sum_{t=1}^{T}w_{it,\alpha}r_{it,\alpha}^{2}\hspace{2cm}\label{(A.4)}\end{align}
Now we can write $\hat{\Gamma}_{\alpha,n}^{BICV}=D_{\alpha}+B_{\alpha}$, where
\begin{align}&\hat{\Gamma}_{\alpha,n}^{BICV}=\frac{1}{n_{v}Tb}\sum_{s\in\mathbf{B}}\|(I_{n_{v}T}-Q_{\alpha,s})^{-1}(Y_{s}-\tbbX_{\alpha,s}\hat{\beta}_{\alpha})\|^{2}=\frac{1}{n_{v}Tb}\sum_{s\in\mathbf{B}}\textbf{r}_{\alpha,s}'(I_{n_{v}T}-Q_{\alpha,s})^{-2}\textbf{r}_{\alpha,s}\nonumber\\
=&\frac{1}{n_{v}Tb}\sum_{s\in\mathbf{B}}\|(I_{n_{v}T}-Q_{\alpha,s})^{-1}(Y_{s}-\tbbX_{\alpha,s}\hat{\beta}_{\alpha})\|^{2}\nonumber\\
=&\underbrace{\frac{1}{n_{v}Tb}\sum_{s\in\mathbf{B}}\textbf{r}_{\alpha,s}'(I_{n_{v}T}-Q_{\alpha,s})^{-1}U_{\alpha,s}(I_{n_{v}T}-Q_{\alpha,s})^{-1}\textbf{r}_{\alpha,s}}_{\equiv D_{\alpha}}\nonumber\\
&+\underbrace{\frac{1}{n_{v}Tb}\sum_{s\in\mathbf{B}}\textbf{r}_{\alpha,s}'(I_{n_{v}T}-Q_{\alpha,s})^{-1}(I_{n_{v}T}-U_{\alpha,s})(I_{n_{v}T}-Q_{\alpha,s})^{-1}\textbf{r}_{\alpha,s}}_{\equiv B_{\alpha}}\nonumber\end{align}
where \begin{align}Z_{\alpha,s}&=(I_{n_{v}T}-Q_{\alpha,s})(I+c_{n}P_{\alpha,s})(I_{n_{v}T}-Q_{\alpha,s})\nonumber\end{align}
From the balance property of $\mathbf{B}$ and \eqref{(A.4)}
\begin{align}D_{\alpha}&=\frac{1}{n_{v}Tb}\sum_{s\in\mathbf{B}}\|\textbf{r}_{\alpha,s}\|^2+\frac{c_{n}}{n_{v}Tb}\sum_{s\in\mathbf{B}}P_{\alpha,s}\|\textbf{r}_{\alpha,s}\|^2\nonumber\\
&=\frac{1}{nT}\|\tbbY-\tbbX_{\alpha}\hat{\beta}_{\alpha}\|^{2}+\left[1+o_{p}\left(\frac{n_{c}}{n}\right)\right]\frac{n+n_{c}}{n_{c}(n-1)T}\sum_{i=1}^{n}\sum_{t=1}^{T}w_{it,\alpha}r_{it,\alpha}^{2}\hspace{2cm}\label{(A.7)}\end{align}
Assume $\mathbf{M}_{\alpha}$ is in Category II.  Then by \eqref{(A.7)} and $\sum_{i=1}^{n}\sum_{t=1}^{T}w_{it,\alpha}r_{it,\alpha}^{2}=k_{\alpha}\sigma^{2}+o_{p}(1)$
\begin{align}D_{\alpha}&=\frac{1}{n}\tbbU'(I-P_{\alpha})\tbbU+\left[1+o_{p}\left(\frac{n_{c}}{n}\right)\right]\frac{n+n_{c}}{n_{c}(n-1)T}\left[k_{\alpha}\sigma^{2}+o_{p}(1)\right]\nonumber\\
&=\frac{1}{n}\tbbU'\tbbU-\frac{1}{n}\tbbU'P_{\alpha}\tbbU+\left[1+o_{p}\left(\frac{n_{c}}{n}\right)\right]\frac{n+n_{c}}{n_{c}(n-1)T}\left[k_{\alpha}\sigma^{2}+o_{p}(1)\right]\nonumber\\
&=\frac{1}{n}\tbbU'\tbbU+\frac{k_{\alpha}\sigma^{2}}{n_{c}T}+o_{p}\left(\frac{1}{n_{c}}\right)\nonumber\end{align}
It remains to show that $B_{\alpha}=o_{p}(n_{c}^{-1})$. From \eqref{(A.3)}
\begin{align}(I_{n_{v}T}-Q_{\alpha,s})P_{\alpha,s}(I_{n_{v}T}-Q_{\alpha,s})&=\left(1-\frac{n_{v}}{n}+o\left(\frac{n_{c}}{n}\right)\right)P_{\alpha,s}(I_{n_{v}T}-Q_{\alpha,s})=\left(1-\frac{n_{v}}{n}+o\left(\frac{n_{c}}{n}\right)\right)^{2}P_{\alpha,s}\nonumber\\
&=\left(\frac{n_{c}}{n}+o_{p}\left(\frac{n_{c}}{n}\right)\right)^{2}P_{\alpha,s}\end{align}
Thus, 
\begin{align}\left(\frac{n}{n_{c}}\right)^{2}(I_{n_{v}T}-Q_{\alpha,s})P_{\alpha,s}(I_{n_{v}T}-Q_{\alpha,s})&=(1+o(1))^2P_{\alpha,s}\geq \frac{1}{2}P_{\alpha,s}
\end{align}
for $s\in\mathbf{B}$ and $n$ sufficiently large.  Pre- and post-multiplying the above by $(I_{n_{v}T}-Q_{\alpha,s})^{-1}$ yields
\begin{align}(I_{n_{v}T}-Q_{\alpha,s})^{-1}P_{\alpha,s}(I_{n_{v}T}-Q_{\alpha,s})^{-1}\leq 2\left(\frac{n}{n_{c}}\right)^{2}P_{\alpha,s}\end{align}
Similarly by \eqref{(A.3)}
\begin{align}Z_{\alpha,s}&=\left\{I_{n_{v}T}-\left[\frac{n_{v}}{n}+o_{p}\left(\frac{n_{c}}{n}\right)\right]P_{\alpha,s}\right\}(I_{n_{v}T}+c_{n}P_{\alpha,s})\left\{I_{n_{v}T}-\left[\frac{n_{v}}{n}+o_{p}\left(\frac{n_{c}}{n}\right)\right]\right\}\nonumber\\
&=I_{n_{v}T}+\left[o_{p}\left(\frac{n_{c}}{n}\right)\right]^2(1+c_{n})P_{\alpha,s}\end{align}
since $c_{n}(1-n_{v}/n)^{2}=(2-n_{v}/n)n_{v}/n$.
Using \eqref{(A.7)}
\begin{align}&(I_{n_{v}T}-Q_{\alpha,s})^{-1}(I_{n_{v}T}-Z_{\alpha,s})(I_{n_{v}T}-Q_{\alpha,s})^{-1}\nonumber\\
=&\left[o_{p}\left(\frac{n_{c}}{n}\right)\right]^{2}(1+c_{n})(I_{n_{v}T}-Q_{\alpha,s})^{-1}P_{\alpha,s}(I_{n_{v}T}-Q_{\alpha,s})^{-1}\leq o_{p}(1)(1+c_{n})P_{\alpha,s}.\nonumber\end{align}
Thus,
\begin{align}B_{\alpha}\leq o_{p}(1)(1+c_{n})\left(\frac{1}{n_{v}Tb}\sum_{s\in\mathbf{B}}\|P_{\alpha,s}\textbf{r}_{\alpha,s}\|^{2}\right)=o_{p}\left(\frac{1}{n_{c}}\right)\end{align}
since from the above $(c_{n}/n_{v}Tb)\sum_{s\in\mathbf{B}}\|P_{\alpha,s}\textbf{r}_{\alpha,s}\|^{2}=O_{p}(n_{c}^{-1})$, which proves (ii) in the proposition for BICV.  (iii) follows in a straightforward manner from (i) and (ii).

The extension of the proof to MCCV is straightforward from Theorem 2 in \citet{Shao:1993} assuming the sufficient conditions given in Condition \ref{cond:mccv}. 
\end{proof}

\subsection{Proof of Proposition \ref{prop:GICs}}\label{proof:prop2}
First, note that Conditions \ref{cond:gic}.1-\ref{cond:gic}.3 imply that  
\begin{align}\sqrt{n}\left(\begin{array}{c}\hat{\beta}_{\alpha}-\beta_{\alpha}^*\\\hat{\beta}_{\gamma}-\beta_{\gamma}^*\end{array}\right)&=\sqrt{n}\left(\begin{array}{cc}\widehat{D}_{\alpha}^{-1}&\underset{{k_{\alpha}\times k_{\gamma}}}{\mathbf{0}}\\\underset{{k_{\gamma}\times k_{\alpha}}}{\mathbf{0}}&\widehat{D}_{\gamma}^{-1}\end{array}\right)\left(\begin{array}{c}\hat{b}_{\alpha}\\\hat{b}_{\gamma}\end{array}\right)\nonumber\\
=&\left(\begin{array}{cc}D_{\alpha}^{-1}&\underset{{k_{\alpha}\times k_{\gamma}}}{\mathbf{0}}\\\underset{{k_{\gamma}\times k_{\alpha}}}{\mathbf{0}}&D_{\gamma}^{-1}\hat{b}_{\gamma}\end{array}\right)\sqrt{n}\left(\begin{array}{c}\hat{b}_{\alpha}\\\hat{b}_{\gamma}\end{array}\right)+o_p(1)\overset{d}{\rightarrow}N\left(0,\left(\begin{array}{cc}\Sigma_{\alpha}&\Sigma_{\alpha,\gamma}\\\Sigma_{\gamma,\alpha}&\Sigma_{\gamma}\end{array}\right)\right)\end{align}
where $\Sigma_{\alpha}=D_{\alpha}^{-1}B_{\alpha}D_{\alpha}^{-1}$, $\Sigma_{\alpha,\gamma}=D_{\alpha}^{-1}B_{\alpha,\gamma}D_{\gamma}^{-1}$, $\Sigma_{\gamma}=D_{\gamma}^{-1}B_{\gamma}D_{\gamma}^{-1}$. The first equality follows by definition of the FE estimators and their pseudo-true parameter values. The second equality follows by Conditions \ref{cond:gic}.1 and \ref{cond:gic}.3. The convergence in distribution follows by Condition \ref{cond:gic}.3.

Next, we perform the following Taylor series expansion of $\ell_{nT}(\beta_{\alpha})\equiv \ell_{nT}(\beta_{\alpha},\sigma^2(\beta_{\alpha}))$ 
\begin{align}\ell_{nT}^{\alpha}(\beta_{\alpha}^*)&=\ell_{nT}^{\alpha}(\hat{\beta}_{\alpha})+\frac{1}{2}\left(\hat{\beta}_{\alpha}-\beta_{\alpha}^*\right)\left.\frac{\partial^2 \ell_{nT}^\alpha(\beta_{\alpha})}{\partial \beta_{\alpha}\partial\beta_{\alpha}'}\right|_{\beta_{\alpha}=\hat{\beta}_{\alpha}}\left(\hat{\beta}_{\alpha}-\beta_{\alpha}^*\right)+o_p(1)\nonumber\\
&=\ell_{nT}^{\alpha}(\hat{\beta}_{\alpha})-n\left(\hat{\beta}_{\alpha}-\beta_{\alpha}^*\right)'\frac{D_{\alpha}}{\sigma_{\alpha}^2(\beta_{\alpha}^*)}\left(\hat{\beta}_{\alpha}-\beta_{\alpha}^*\right)+o_p(1)\end{align}
where the last equality follows by Lemma \ref{lem:Hessian}. By similar arguments,
\begin{align}\ell_{nT}^{\gamma}(\beta_{\gamma}^*)
&=\ell_{nT}^{\gamma}(\hat{\beta}_{\gamma})-n\left(\hat{\beta}_{\gamma}-\beta_{\gamma}^*\right)'\frac{D_{\gamma}}{\sigma_{\gamma}^2(\beta_{\gamma}^*)}\left(\hat{\beta}_{\gamma}-\beta_{\gamma}^*\right)+o_p(1)\end{align}
As a result, 
\begin{align}LR_{nT}^{\alpha,\gamma}&=\ell_{nT}^{\alpha}(\beta_{\alpha}^*)-\ell_{nT}^\gamma(\beta_{\gamma}^*)+n\left(\hat{\beta}_{\alpha}-\beta_{\alpha}^*\right)'\frac{D_{\alpha}}{\sigma_{\alpha}^2(\beta_{\alpha}^*)}\left(\hat{\beta}_{\alpha}-\beta_{\alpha}^*\right)-n\left(\hat{\beta}_{\gamma}-\beta_{\gamma}^*\right)'\frac{D_{\gamma}}{\sigma_{\gamma}^2(\beta_{\gamma}^*)}\left(\hat{\beta}_{\gamma}-\beta_{\gamma}^*\right)+o_p(1).\label{eq:general_LR_expression}\end{align}

\noindent \textbf{1.} Note that in this case, $\ell_{nT}^{\alpha}(\beta_{\alpha}^*)=\ell_{nT}^{\gamma}(\beta_{\gamma}^*)$, since $\widetilde{x}_{\cdot,\alpha}'\beta_{\alpha}^*= \widetilde{x}_{\cdot,\gamma}'\beta_{\gamma}^*$. By Lemma 3.2 in \citet{Vuong:1989}, with a symmetric real $Q$ given by
$$Q=\left(\begin{array}{cc}\frac{D_{\alpha}}{\sigma_{\alpha}^2(\beta_{\alpha}^*)}&\underset{{k_{\alpha}\times k_{\gamma}}}{\mathbf{0}}\\\underset{{k_{\gamma}\times k_{\alpha}}}{\mathbf{0}}&-\frac{D_{\gamma}}{\sigma_{\gamma}^2(\beta_{\gamma}^*)}\end{array}\right),$$
$LR_{nT}^{\alpha,\gamma}$ converges in distribution to a weighted sum of chi-squares with parameters $k_{\alpha}+k_{\gamma}$ and $\lambda$, the vector of eigenvalues of $$Q\left(\begin{array}{cc}\Sigma_{\alpha}&\Sigma_{\alpha,\gamma}\\\Sigma_{\gamma,\alpha}&\Sigma_{\gamma}\end{array}\right).$$
As a result, $LR_{nT}^{\alpha,\gamma}=O_{p}(1)$, whereas $\lambda_{nT}\rightarrow\infty$. Since $k_{\alpha}<k_{\gamma}$ by definition, $\lambda_{nT}(k_{\alpha}-k_{\gamma})\rightarrow-\infty$. It follows that $P(LR_{nT}^{\alpha,\gamma}>\lambda_{nT}(k_{\alpha}-k_{\gamma}))\rightarrow 1$.

\noindent\textbf{2}. Recall that in this case, $\sigma_{\alpha}^2(\beta_{\alpha}^*)=\sigma_{\gamma}^2(\beta_{\gamma}^*)$, but
$\widetilde{x}_{\cdot,\alpha}'\beta_{\alpha}^*\neq \widetilde{x}_{\cdot,\gamma}'\beta_{\gamma}^*$. As a result, from \eqref{eq:general_LR_expression} we have 
\begin{align}\frac{1}{\sqrt{n}}LR_{nT}^{\alpha,\gamma}&=-\frac{(T-1)}{\sqrt{n}}(\ell_{nT}^{\alpha}(\beta_{\alpha}^*)-\ell_{nT}^{\gamma}(\beta_{\gamma}^*))+o_p(1)=\sqrt{n}(T-1)\left(\log(\hat{\sigma}_{\gamma}^2(\beta_{\gamma}^*)-\log(\hat{\sigma}_{\alpha}^2(\beta_{\alpha}^*))\right)+o_p(1)
\end{align}
By Conditions \ref{cond:gic}.4-\ref{cond:gic}.5 and the $\delta$-method, it follows that $\frac{1}{\sqrt{n}}LR_{nT}^{\alpha,\gamma}=O_p(1)$. The result follows from noting that since $k_{\alpha}<k_{\gamma}$, $\frac{\lambda_{nT}}{\sqrt{n}}(k_{\alpha}-k_{\gamma})\rightarrow-\infty$ as $\lambda_{nT}/\sqrt{n}\rightarrow\infty$.\\

\noindent\textbf{3.a.} Here, we note that by definition
\begin{align}\frac{1}{n}LR_{nT}^{\alpha,\gamma}&=-(T-1)\left(\log(\hat{\sigma}^2_{\alpha}(\hat{\beta}_{\alpha}))-\log(\hat{\sigma}^2_\gamma(\hat{\beta}_{\gamma}))\right)=(T-1)\left(\log(\sigma_{\gamma}^2(\beta_{\gamma}^*))-\log(\sigma_{\alpha}^2(\beta_{\alpha}^*))\right)+o_p(1),\end{align}
where the second equality follows by Condition \ref{cond:gic}.4 and the continuous mapping theorem. Since in this case $\sigma_{\alpha}^2(\beta_{\alpha}^*)<\sigma_{\gamma}^2(\beta_{\gamma}^*)$, $(T-1)\left(\log(\sigma_{\gamma}^2(\beta_{\gamma}^*))-\log(\sigma_{\alpha}^2(\beta_{\alpha}^*))\right)>0$. The result follows from noting that $\frac{\lambda_{nT}}{n}(k_{\alpha}-k_{\gamma})\rightarrow 0$ and $\frac{1}{n}LR_{nT}^{\alpha,\gamma}\overset{p}{\rightarrow}(T-1)\left(\log(\sigma_{\gamma}^2(\beta_{\gamma}^*))-\log(\sigma_{\alpha}^2(\beta_{\alpha}^*))\right)>0$.\\

\noindent \textbf{3.b.} The result follows by similar arguments as in 3.a noting that in this case the result follows from $\frac{1}{n}LR_{nT}^{\alpha,\gamma}\overset{p}{\rightarrow}(T-1)\left(\log(\sigma_\gamma^2(\beta_{\gamma}^*))-\log(\sigma_\alpha^2(\beta_{\alpha}^2))\right)<0$.
\qed

\subsection{Supplementary Lemma}

\begin{lemma}\label{lem:Hessian}
 Under Conditions \ref{cond:gic}.1, \ref{cond:gic}.2 and \ref{cond:gic}.4, as $n\rightarrow\infty$ holding $T$ fixed
\begin{enumerate}[(i)]\item $\left.\frac{1}{n}\frac{\partial^2\ell^{\alpha}_{nT}(\beta_{\alpha})}{\partial\beta_{\alpha}\partial\beta_{\alpha}'}\right|_{\beta_{\alpha}=\hat{\beta}_{\alpha}}\overset{p}{\rightarrow}-\frac{2D_{\alpha}}{\sigma_{\alpha}^2(\beta_{\alpha}^*)}$
\item $\left.\frac{1}{n}\frac{\partial^2\ell^{\gamma}_{nT}(\beta_{\gamma})}{\partial\beta_{\gamma}\partial\beta_{\gamma}'}\right|_{\beta_{\gamma}=\hat{\beta}_{\gamma}}\overset{p}{\rightarrow}-\frac{2D_{\gamma}}{\sigma_{\gamma}^2(\beta_{\gamma}^*)}$.
\end{enumerate}
\end{lemma}
\begin{proof}
(i) 
\begin{align}\frac{1}{n}\frac{\partial \ell_{nT}^\alpha(\beta_{\alpha})}{\partial \beta_{\alpha}}&=(T-1)\frac{2}{\sum_{i=1}^n\sum_{t=1}^T(\tilde{Y}_{it}-\tilde{X}_{it,\alpha}'\beta_{\alpha})^2/(n(T-1))}\frac{1}{n(T-1)}\sum_{i=1}^n\sum_{t=1}^T\widetilde{X}_{it,\alpha}(\widetilde{Y}_{it}-\widetilde{X}_{it,\alpha}'\beta_{\alpha})\nonumber\\
&=\frac{2}{\sum_{i=1}^n\sum_{t=1}^T(\tilde{Y}_{it}-\tilde{X}_{it,\alpha}'\beta_{\alpha})^2/(T-1)}\sum_{i=1}^n\sum_{t=1}^T\widetilde{X}_{it,\alpha}(\widetilde{Y}_{it}-\widetilde{X}_{it,\alpha}'\beta_{\alpha})\nonumber\\
\frac{1}{n}\frac{\partial^2 \ell_{nT}^\alpha(\beta_{\alpha})}{\partial \beta_{\alpha}\partial\beta_{\alpha}'}&=-\frac{2}{\sum_{i=1}^n\sum_{t=1}^T(\tilde{Y}_{it}-\widetilde{X}_{it,\alpha}'\beta_{\alpha})^2/(T-1)}\sum_{i=1}^n\sum_{t=1}^T\widetilde{X}_{it}\widetilde{X}_{it}'\nonumber\\
&+\frac{4}{T-1}\frac{\sum_{i=1}^n\sum_{t=1}^T\widetilde{X}_{it,\alpha}(\widetilde{Y}_{it}-\widetilde{X}_{it,\alpha}'\beta_{\alpha})\sum_{i=1}^n\sum_{s=1}^T(\tilde{Y}_{it}-\tilde{X}_{it}'\beta_{\alpha})\tilde{X}_{it,\alpha}'}{\left(\sum_{i=1}^n\sum_{t=1}^T(\tilde{Y}_{it}-\tilde{X}_{it}'\beta_{\alpha})^2\right)^2/(T-1)}\label{eq:H_ll}
\end{align}
Note that under Conditions \ref{cond:gic}.1-\ref{cond:gic}.2 and Slutzky's theorem, it follows that as $n\rightarrow\infty$,
\begin{align}\frac{1}{n}\sum_{i=1}^n\sum_{t=1}^T\widetilde{X}_{it,\alpha}(\widetilde{Y}_{it}-\widetilde{X}_{it,\alpha}'\hat{\beta}_{\alpha})&\overset{p}{\rightarrow}C_{\alpha}-D_{\alpha}D_{\alpha}^{-1}C_{\alpha}=0\end{align}
As a result, the second term in \eqref{eq:H_ll} is $o_p(1)$, since its numerator once scaled by $n^{-2}$ is $o_{p}(1)$ and its denominator, once scaled by $n^{-2}$, converges in probability to $(\sigma_{\alpha}^2(\beta_{\alpha}^*))^2>0$ under Condition \ref{cond:gic}.4.

As a result, it follows that under Conditions \ref{cond:gic}.1, \ref{cond:gic}.2 and \ref{cond:gic}.4, as $n\rightarrow$
\begin{align}
\left.\frac{1}{n}\frac{\partial^2\ell^{\alpha}_{nT}(\beta_{\alpha})}{\partial\beta_{\alpha}\partial\beta_{\alpha}'}\right|_{\beta_{\alpha}=\hat{\beta}_{\alpha}}&\overset{p}{\rightarrow}-\frac{2D_{\alpha}}{\sigma_{\alpha}^2(\beta_{\alpha}^*)}
\end{align}
(ii) follows by symmetric arguments.
\end{proof}

\section{Predicted Values at Pseudo-true Parameter Values}\label{app:pseudo-true}
Proposition \ref{prop:GICs} establishes that BIC would be (pseudo-)inconsistent in settings where two models do not yield the same predicted values of the outcome, yet have similar MSE. In standard nonlinear model selection problems, existing literature \citep{Sin:1996,Hong:2012} establishes that BIC is pseudo-inconsistent when choosing between non-nested, misspecified models. Our simulation results demonstrate that in the context of mixed-frequency panel data BIC is (pseudo)-inconsistent when two models are nested but neither nests the true model. In this section, we show how the predicted values at the pseudo-true parameters are generally different, when two misspecified models are nested in the context of our model selection problem, and thereby explain the BIC pseudo-inconsistency in this setting.
\subsection{Relationship between Pseudo-true Parameter Values of Misspecified Models}
Given that all models considered here use regressors that are functions of different summary statistics of the same time series, we formalize the (pseudo-)true parameter values of the models under consideration. We first introduce the within-demeaning notation for linear fixed effects estimation.  For $V_{it}$, $\widetilde{V}_{it}=V_{it}-\bar{V}_{i}$, where $\bar{V}_{i}=\sum_{t=1}^{T}V_{it}/T$.  For $\mathbf{M}_{\alpha}$, the within-transformation is given by\
\begin{align}Y_{it}&=X_{it,\alpha}'\beta_{\alpha}+a_{i,\alpha}+u_{it,\alpha} \; , \nonumber\\
\tilde{Y}_{it}&=\tilde{X}_{it,\alpha}'\beta_{\alpha}+\tilde{u}_{it,\alpha} \; . \end{align}
The probability limit of the FE estimator of the above model, which we refer to as the pseudo-true parameter vector of $\mathbf{M}_{\alpha}$, is denoted by $\beta_{\alpha}^{*}$ and is given in \eqref{omvarbias}. We further assume that the estimation problem is sufficiently regular, and we also assume strict exogeneity of the high-frequency regressor, $E[u_{it}|\mathcal{W}_{i1},\dots,\mathcal{W}_{iT},a_{i}]=0$. For a random variable $W_{i}$, let $\bar{E}[W_{i}]=\lim_{n\rightarrow\infty}\sum_{i=1}^nE[W_i]/n$. Then, assuming sufficient conditions for the application of a law of large numbers (see Condition \ref{cond:PWMSE} in Section \ref{subsec:CCP-MSE} and Step 2 of the proof of Proposition \ref{prop:CCP-MSE} in Appendix \ref{sec:proof_prop_3}),
\begin{align}
\label{omvarbias}
\beta_{\alpha}^{*}&=plim_{n\rightarrow\infty}\left(\sum_{i=1}^{n}\sum_{t=1}^{T}\tilde{X}_{it,\alpha}\tilde{X}_{it,\alpha}'\right)^{-1}\sum_{i=1}^{n}\sum_{t=1}^{T}\tilde{X}_{it,\alpha}\tilde{Y}_{it}\nonumber\\
&=\left(\bar{E}\left[\sum_{t=1}^T\tilde{X}_{it,\alpha}\tilde{X}_{it,\alpha}'\right]\right)^{-1}\bar{E}\left[\sum_{t=1}^T\tilde{X}_{it,\alpha}\tilde{X}_{it,\star}'\right]\beta_{\star,\emph{o}}.
\end{align}
Equation \eqref{omvarbias} is the counterpart of the omitted variable bias formula in this problem. In the context of mixed-frequency time series, this issue has also been recognized as neglected nonlinearity in \cite{miller2014mixed}.
To gain some intuition for \eqref{omvarbias}, consider the case where both $X_{it,\star}$ and $X_{it,\alpha}$ are scalar. Then
\begin{align}\beta_{\alpha}^{*}&=\frac{\bar{E}[\sum_{t=1}^T\tilde{X}_{it,\alpha}\tilde{X}_{it,\star}]}{\bar{E}[\sum_{t=1}^T\tilde{X}_{it,\alpha}^{2}]}\beta_{\star,\emph{o}}=\rho_{*,\alpha}\sqrt{\frac{\bar{E}[\sum_{t=1}^T\tilde{X}_{it,\star}^{2}]}{\bar{E}[\sum_{t=1}^T\tilde{X}_{it,\alpha}^{2}]}}\beta_{\star,\emph{o}}, 
\end{align}
where $\rho_{*,\alpha}=\bar{E}[\sum_{t=1}^T\tilde{X}_{it,\alpha}\tilde{X}_{it,\star}]/\sqrt{\bar{E}[\sum_{t=1}^T\tilde{X}_{it,\alpha}^{2}]\bar{E}[\tilde{X}_{it,\star}^{2}]}$ is the within-correlation coefficient between ${X}_{it,\alpha}$ and ${X}_{it,\star}$.  Under the assumption that $\beta_{\star,\emph{o}}$ is non-zero, the sign and the magnitude of $\beta_{\alpha}^{*}/\beta_{\star,\emph{o}}$ will depend on the within-correlation between ${X}_{it,\alpha}$ and $X_{it,\star}$ as well as the ratio of their variances.  If the within-correlation between the two variables is positive, then $\beta_{\alpha}^{*}$ and $\beta_{\star,o}$ will have the same sign, otherwise $\beta_{\alpha}^{*}$ will have the opposite sign of $\beta_{\star,\emph{o}}$.  Suppose that $X_{it,\alpha}$ and $X_{it,\star}$ have equal within-variance, then $\beta_{\alpha}^{*}$ will tend to be smaller in magnitude the weaker the within-correlation between $X_{it,\alpha}$ and $X_{it,\star}$.  This example of attenuation bias is similar to the classical measurement error problem.  If $X_{it,\alpha}$ has greater within-variance than $X_{it,\star}$, then the attenuation is greater.  

Returning to the general (non-scalar) case, if $\mathbf{M}_{\alpha}$ contains $\mathbf{M}_{\star}$, i.e. $X_{it,\star}=R_{\star,\alpha}X_{it,\alpha}$ for some $R_{\star,\alpha}$, then
\begin{align}
\label{eq:nestedpseudo}
\beta_{\alpha}^{*}&=R_{\star,\alpha}'\beta_{\star,\emph{o}} .
\end{align}
For instance, if $\mathbf{M}_{\alpha}$ is the quarterly mean model, and $\mathbf{M}_{\star}$ is the annual mean model, 
\begin{align}
\beta_{\alpha}^{*}&=\left(\begin{array}{c}\frac{|Q_{1}|}{H}\\\vdots\\\frac{|Q_{4}|}{H}\end{array}\right)\beta_{\star,\emph{o}} .
\end{align}
Since $X_{it,\alpha}$ and $X_{it,\star}$ are summary statistics of $\mathcal{W}_{it}$, if $\beta_{\star,\emph{o}}$ is non-zero, then we expect all elements of $\beta_{\alpha}^{*}$ to be non-zero, unless $R_{\star,\alpha}$ has zero rows.  This is different from the standard variable selection problem, where the pseudo-true parameter value for models that contain $\mathbf{M}_{\star}$ will have zero elements for variables that are not in the DGP.
\subsection{Predicted Values at the Pseudo-True Parameters}\label{supp:predicted_values}
Let $\tilde{Y}_{it,\alpha}^{*}(\mathcal{W}_{i})\equiv\tilde{X}_{it,\alpha}'\beta_{\alpha}^{*}$ denote the within-demeaned predicted value of the outcome for individual $i$ in period $t$ given $\mathcal{W}_{i}$ using the pseudo-true parameter vector of $\mathbf{M}_{\alpha}$.  Consider two models, $\mathbf{M}_{\alpha}$ and $\mathbf{M}_{\gamma}$, where both models contain $\mathbf{M}_{\star}$, i.e. $\tilde{X}_{it,\star}=R_{\star,\alpha}\tilde{X}_{it,\alpha}=R_{\star,\gamma}\tilde{X}_{it,\gamma}$.  By Eq. \eqref{eq:nestedpseudo}, $\beta_{\alpha}^{*}=R_{\star,\alpha}'\beta_{\star,\emph{o}}$ for $\mathbf{M}_{\alpha}$ when $\mathbf{M}_{\star}$ is nested in it.  As a result,
\begin{align}
\tilde{Y}_{it,\alpha}^{*}({\mathcal{W}}_{i})&=\tilde{X}_{it,\alpha}'\beta_{\alpha}^{*}=\tilde{X}_{it,\alpha}'R_{\star,\alpha}'\beta_{\star,\emph{o}}=\tilde{X}_{it,\star}'\beta_{\star,\emph{o}} , \nonumber\\
\tilde{Y}_{it,\gamma}^{*}({\mathcal{W}}_{i})&=\tilde{X}_{it,\gamma}'\beta_{\gamma}^{*}=\tilde{X}_{it,\gamma}'R_{\star,\gamma}'\beta_{\star,\emph{o}}=\tilde{X}_{it,\star}'\beta_{\star,\emph{o}}.\end{align}
  Hence, in this case, both models yield identical predictions given $\mathcal{W}_{i}$ using their respective pseudo-true parameter vectors.   This result holds regardless of the relationship between the two models as long as $\mathbf{M}_{\star}$ is nested in both of them.

Note that if $\mathbf{M}_{\alpha}$ is nested in $\mathbf{M}_{\gamma}$, but the DGP is not contained in either model, they may still have different predictions using their respective pseudo-true parameter vectors. To see this, consider
\begin{align}
\tilde{Y}_{it,\gamma}^{*}({\mathcal{W}}_{i})&=\tilde{X}_{it,\gamma}'\beta_{\gamma}^{*}=\tilde{X}_{it,\gamma}'\left(\bar{E}[\tilde{X}_{it,\gamma}\tilde{X}_{it,\gamma}']\right)^{-1}\bar{E}[\tilde{X}_{it,\gamma}X_{it,\star}']\beta_{\star,\emph{o}} \; ,\\
\tilde{Y}_{it,\alpha}^{*}({\mathcal{W}}_{i})&=\tilde{X}_{it,\alpha}'\beta_{\alpha}^{*}=\tilde{X}_{it,\alpha}'\left(\bar{E}[\tilde{X}_{it,\alpha}\tilde{X}_{it,\alpha}']\right)^{-1}\bar{E}[\tilde{X}_{it,\alpha}\tilde{X}_{it,\star}']\beta_{\star,\emph{o}}\nonumber\\
&=\tilde{X}_{it,\gamma}'R_{\alpha,\gamma}'\left(R_{\alpha,\gamma}\bar{E}[\tilde{X}_{it,\gamma}\tilde{X}_{it,\gamma}']R_{\alpha,\gamma}'\right)^{-1}R_{\alpha,\gamma}\bar{E}[\tilde{X}_{it,\gamma}\tilde{X}_{it,\star}']\beta_{\star,\emph{o}} \; .\end{align}
Note that $\tilde{Y}_{it,\gamma}^{*}(\mathcal{W}_{i})={Y}_{it,\alpha}^{*}(\mathcal{W}_{i})$ is true if
\begin{align}R_{\alpha,\gamma}'\left(R_{\alpha,\gamma}\bar{E}[\tilde{X}_{it,\gamma}\tilde{X}_{it,\gamma}']R_{\alpha,\gamma}'\right)^{-1}R_{\alpha,\gamma}&=\left(\bar{E}[\tilde{X}_{it,\gamma}\tilde{X}_{it,\gamma}']\right)^{-1}, \label{equality}
\end{align}
which would hold in general if $R_{\alpha,\gamma}$ were symmetric and invertible. However, by definition it is not a square matrix.  As a result, $\tilde{Y}_{it,\gamma}^{*}(\mathcal{W}_{i})\neq {Y}_{it,\alpha}^{*}(\mathcal{W}_{i})$ in general if neither of the models contain the true DGP. 

We now consider a simple example to illustrate this point.  Suppose that $\mathbf{M}_{\alpha}$ is the annual mean model and $\mathbf{M}_{\gamma}$ is the quarterly mean model. Then $\bar{E}[\tilde{X}_{it,\gamma}\tilde{X}_{it,\gamma}']$ is the within variance-covariance matrix of the quarterly means, and $\bar{E}[\tilde{X}_{it,\alpha}^2]$ is the within-variance of the annual mean, which is a weighted average of the quarterly means.  Clearly, the ``variability'' is not in general the same for the higher- and lower-frequency mean, unless we impose some restrictive assumptions.  For instance, if we require that the within-variance is the same for all quarterly means and that there is no within-covariance between the quarterly means, then $\bar{E}[\tilde{X}_{it,\gamma}\tilde{X}_{it,\gamma}']=\bar{E}[\tilde{X}_{it,\alpha}^{2}]I_{k_{\gamma}}$, where $\bar{E}[\tilde{X}_{it,\alpha}^{2}]>0$.  This would imply that the within-variance of summer and winter average temperatures are the same and that there is no inter-seasonal correlation in temperature.  These are unrealistic assumptions that we entertain to illustrate our point.  In this example, \eqref{equality} simplifies to
\begin{align}
R_{\alpha,\gamma}'\left(R_{\alpha,\gamma}R_{\alpha,\gamma}'\right)^{-1}R_{\alpha,\gamma}&=I_{k_{\gamma}}, \nonumber\\
\frac{1}{\sum_{j=1}^{4}|Q_{j}|^{2}/H^{2}}R_{\alpha,\gamma}'R_{\alpha,\gamma}&=I_{k_{\gamma}}.
\end{align}
The above equality is trivially fulfilled if $R_{\alpha,\gamma}$ is proportional to the identity matrix, which would imply that both models are identical.  But this is not true in this simple example.  If we further simplify the problem by assuming that $|Q_{j}|=H/4$ for $j=1,\dots,4$, then $\mathcal{R}_{\alpha,\gamma}=\frac{1}{4}\textbf{1}_{k}'$, where $\textbf{1}_{k}$ is a $k\times 1$ vector with all elements equal to one.  It follows that the above equality clearly does not hold, since its left-hand side would simplify to $\frac{1}{4}\textbf{1}_{k}\textbf{1}_{k}'$.  Hence, even in this simple example, it is difficult to show that it is possible to obtain identical predictions of the outcome variable given $\mathcal{W}_{i}$ when considering two models that do not nest $\mathbf{M}_{\star}$. The above inights help explain the pseudo-inconsistency of BIC in our simulations.

\section{On PWMSE: Weight Specifications and Additional Simulations}\label{app:pwmse}

\subsection{Extended Simulation Results}\label{sec:PWMSE_sims}
Constructing the feasible PWMSE requires forming weights to measure proximity, which involves choosing a norm and a tuning parameter. Here, we consider six different norms of the temperature differences between $T-r$ and $T+\tau$: $L^\infty$ norm of daily differences (D1),  $L^2$ norm of daily differences (D2),  $L^\infty$ norm of monthly differences (M1),  $L^2$ norm of monthly differences (M2), $L^\infty$ norm of annual differences (Y1), and $L^2$ norm of annual differences (Y2). 
To be precise on the definitions of these norms, we first introduce additional notation. 
\begin{enumerate}[-]
\item $W_{i,T-r,h}$: the historical realization in location $i$ in day $h$ in the $r$th year before the last observable year $T$. 
\item $W_{i,T-r,m}$: the historical realization in location $i$ in month $m$ in the $r$th year before the last observable year $T$, averaged over daily realizations within the month, i.e.,  $W_{i,T-r,m} = \sum_{h_m=1,...,H_m} W_{i,T-r,h_m}/H_m$ where $h_m$ denotes the days within month $m$.
\item $W_{i,T-r}$: represent the historical realization in location $i$ in the $r$th year before the last observable year $T$, averaged over daily realizations within the year, i.e., $W_{i,T-r} = \sum_{h=1,...,H} W_{i,T-r,h}/H$.
\item $\mathcal{W}_{i,T+\tau,h}$, $\mathcal{W}_{i,T+\tau,m}$, $\mathcal{W}_{i,T+\tau}$: $\tau$ years forward projected future counterparts of $W_{i,T-r,h}$, $W_{i,T-r,m}$, and $W_{i,T-r}$.
\end{enumerate}
Using the notations above, we formally define the six norms considered in our simulation study:
\begin{enumerate}[(1)]
\item $L^\infty$ norm of daily differences (D1): $\max_{h} |W_{i,T-r,h}-\mathcal{W}_{i,T+\tau,h}|$;
\item $L^2$ norm of daily differences (D2): $\sqrt{\sum_{h} (W_{i,T-r,h}-\mathcal{W}_{i,T+\tau,h})^2}$;
\item $L^\infty$ norm of monthly differences (M1): $\max_{m} |W_{i,T-r,m}-\mathcal{W}_{i,T+\tau,m}|$;
\item $L^2$ norm of monthly differences (M2): $\sqrt{\sum_{m} (W_{i,T-r,m}-\mathcal{W}_{i,T+\tau,m})^2}$;
\item $L^\infty$ norm of yearly differences (Y1): $|W_{i,T-r}-\mathcal{W}_{i,T+\tau}|$;
\item $L^2$ norm of yearly differences (Y2): $(W_{i,T-r}-\mathcal{W}_{i,T+\tau})^2$.
\end{enumerate}
Based on each of these norms, we form the weights with different tuning parameters $h=1,10,100$. In addition, we consider a case where no weight is applied (denoted as N). 

Using simulated pseudo-true parameter values reported in Table \ref{tab:pseudo_true_param}, we compute the (population) PWMSE in \eqref{eq:MSE_criterion_sumtaustar} with different specifications of the weights $\pi(\mathcal{W}_{i,T-r},\mathcal{W}_{i,T+\tau}^f)$ for $r=1,...,T-1$. Table \ref{tab:sim_feasible_targets_plus} presents the models that minimize the feasible targets under each DGP using different weight specifications for $\tau=20$, accompanied by additional results for $\tau=15, 25$. In most cases, the feasible target is minimized by the true model. In a few occasions, however, the feasible target is minimized by a less parsimonious model that nests the true model. This occurs when the norm is based on temperature differences at a finer temporal resolution (e.g., D1 and D2) and the tuning parameter is small (e.g., $h=1$). In these situations, the calculation essentially puts much higher weight on historical observations with daily realizations that highly resemble those of the future year.

\begin{figure}[htbp]
{\fontsize{9}{5}\selectfont
\noindent
\begin{tabular}{cc}
\multicolumn{2}{c}{\bf DGP=$A$}\\
\multicolumn{2}{l}{$h=1$}\\
\includegraphics[width=.5\linewidth]{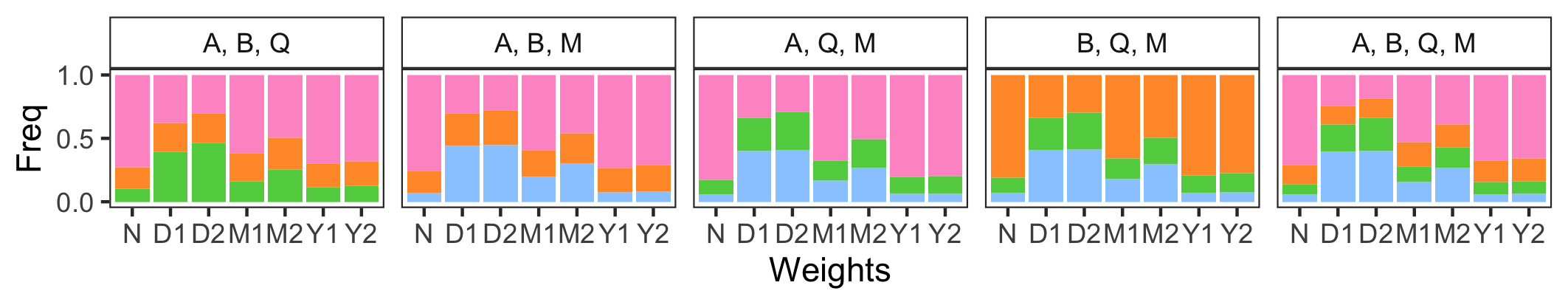}&
\includegraphics[width=.5\linewidth]{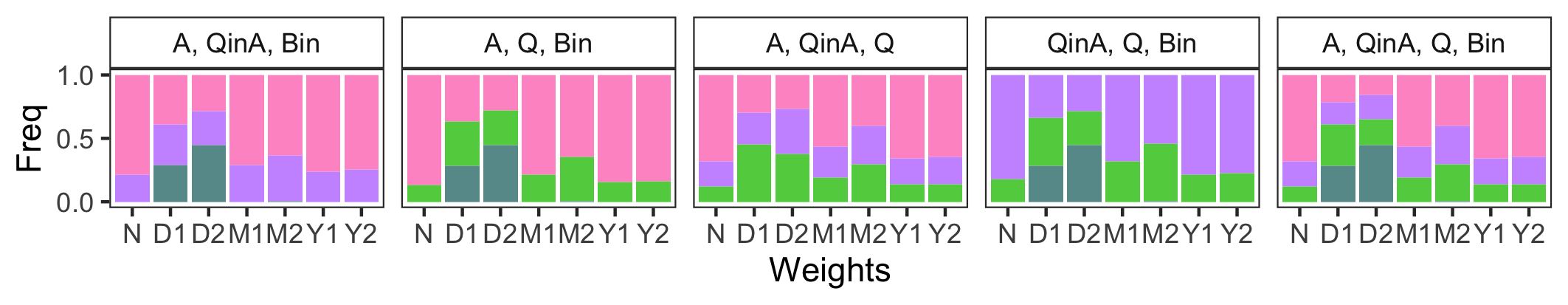}\\
\multicolumn{2}{l}{$h=10$}\\
\includegraphics[width=.5\linewidth]{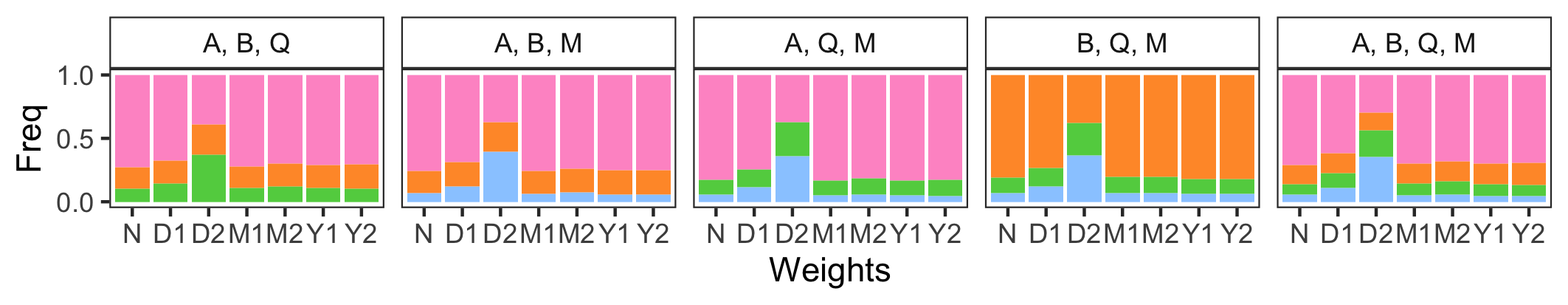}&
\includegraphics[width=.5\linewidth]{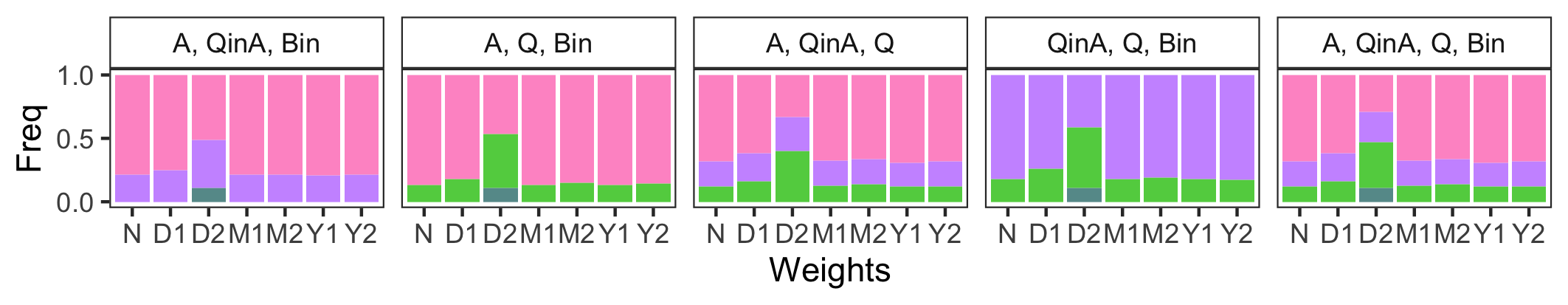}\\
\multicolumn{2}{l}{$h=100$}\\
\includegraphics[width=.5\linewidth]{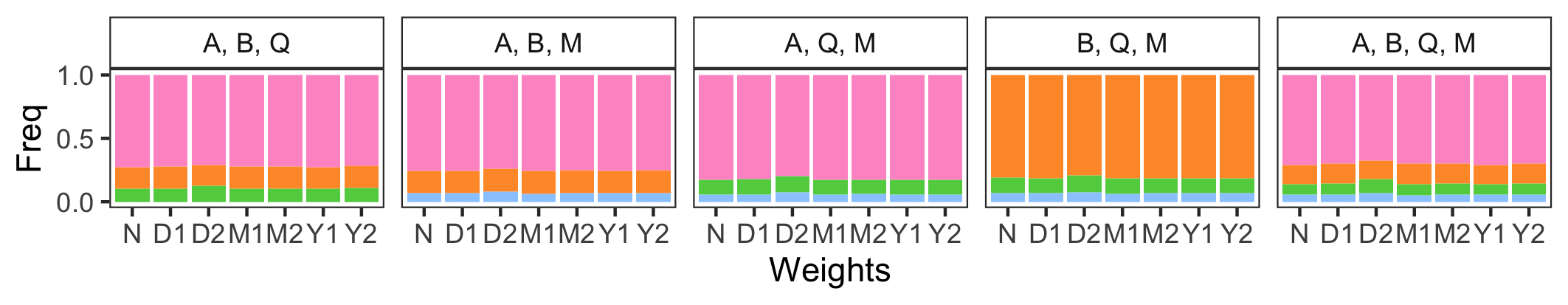}&
\includegraphics[width=.5\linewidth]{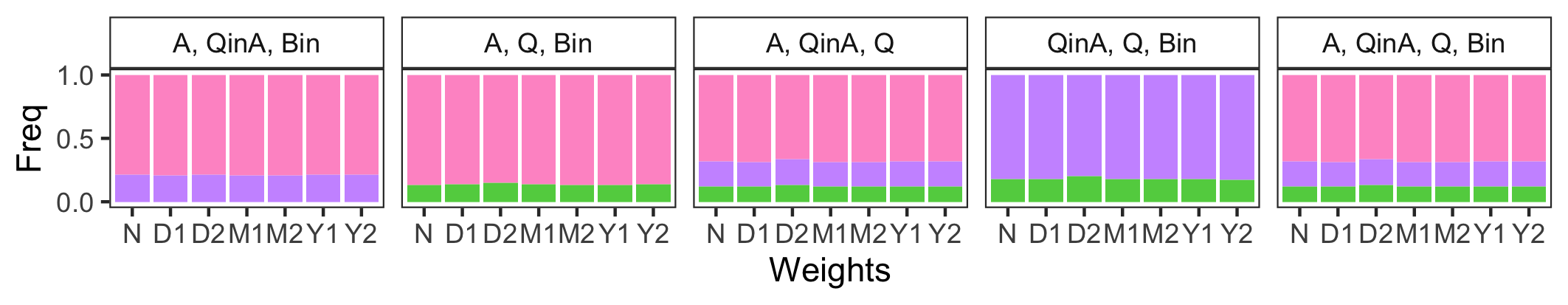}\\
\\
\multicolumn{2}{c}{\bf DGP=$QinA$}\\
\multicolumn{2}{l}{$h=1$}\\
\includegraphics[width=.5\linewidth]{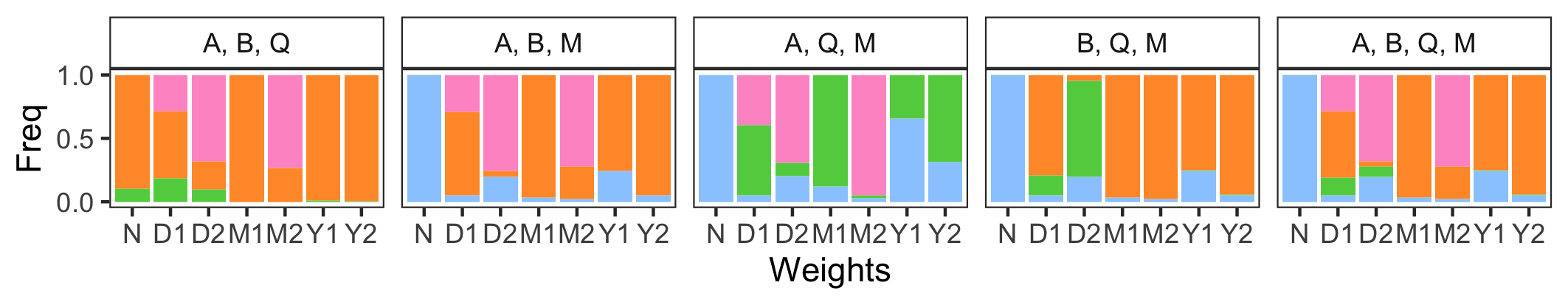}&
\includegraphics[width=.5\linewidth]{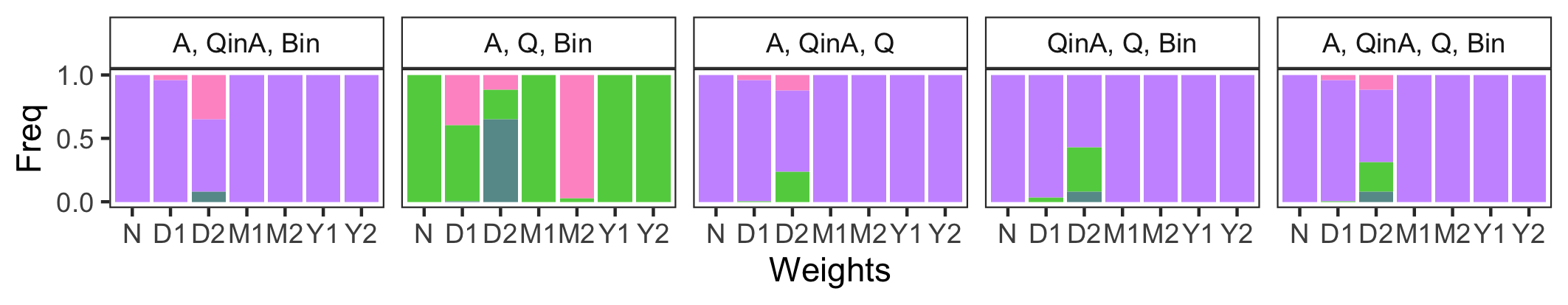}\\
\multicolumn{2}{l}{$h=10$}\\
\includegraphics[width=.5\linewidth]{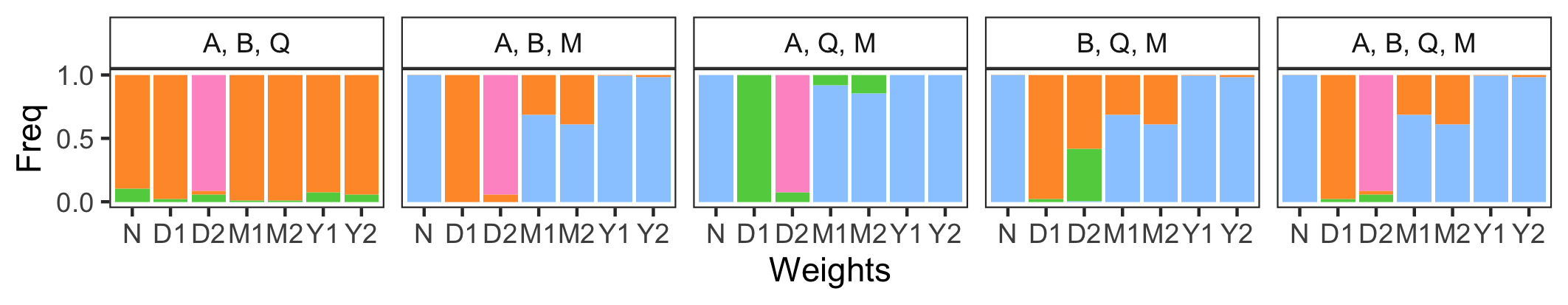}&
\includegraphics[width=.5\linewidth]{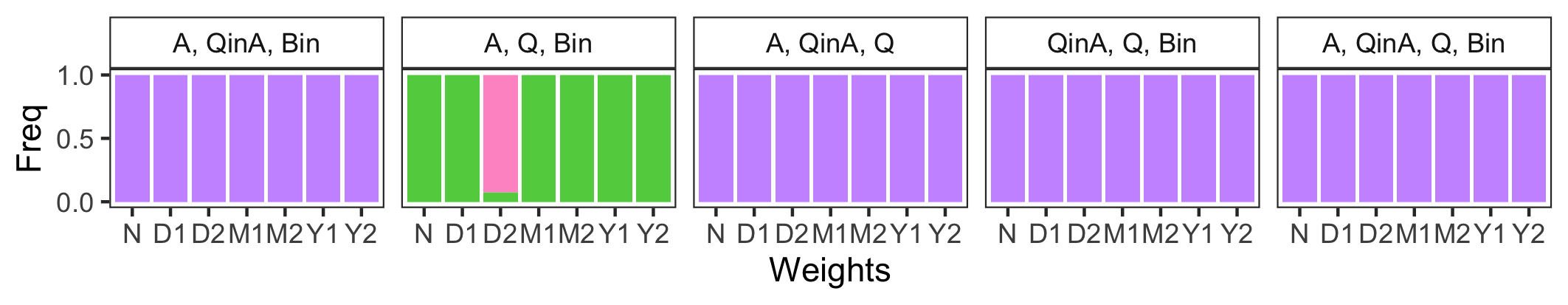}\\
\multicolumn{2}{l}{$h=100$}\\
\includegraphics[width=.5\linewidth]{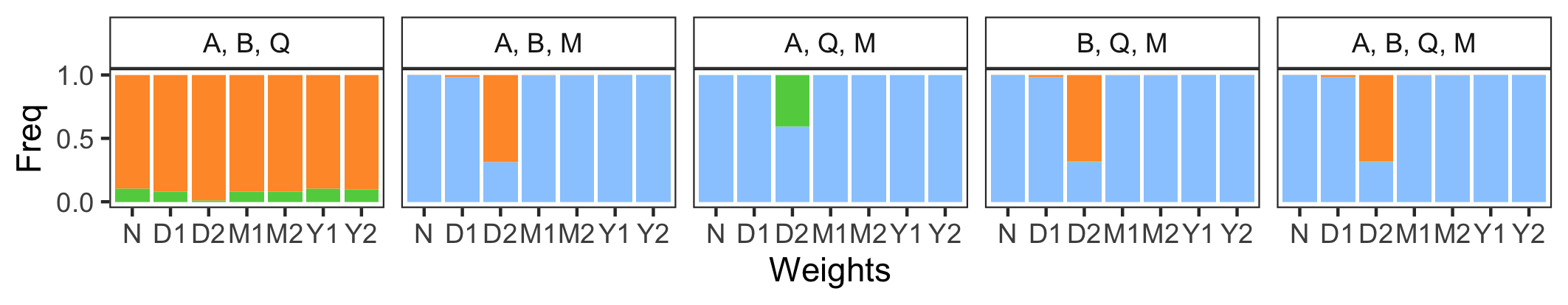}&
\includegraphics[width=.5\linewidth]{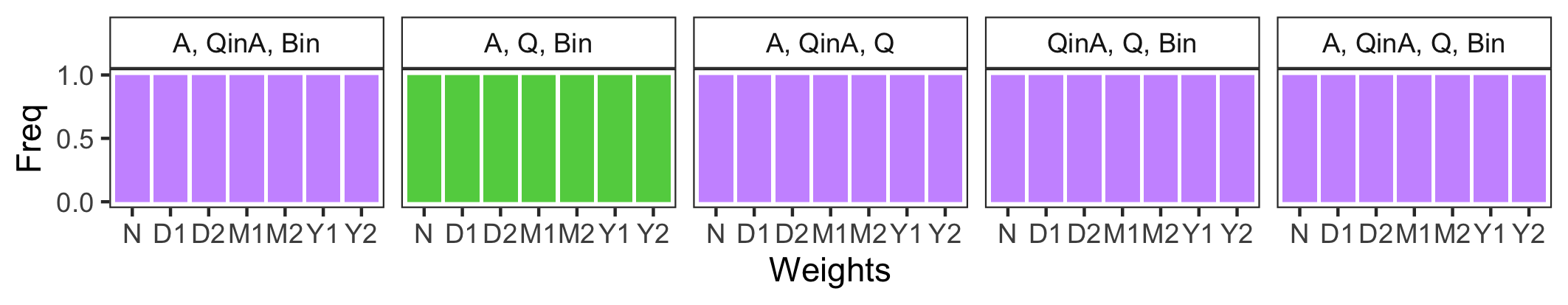}\\
\\
\multicolumn{2}{c}{\bf DGP=$Q$}\\
\multicolumn{2}{l}{$h=1$}\\
\includegraphics[width=.5\linewidth]{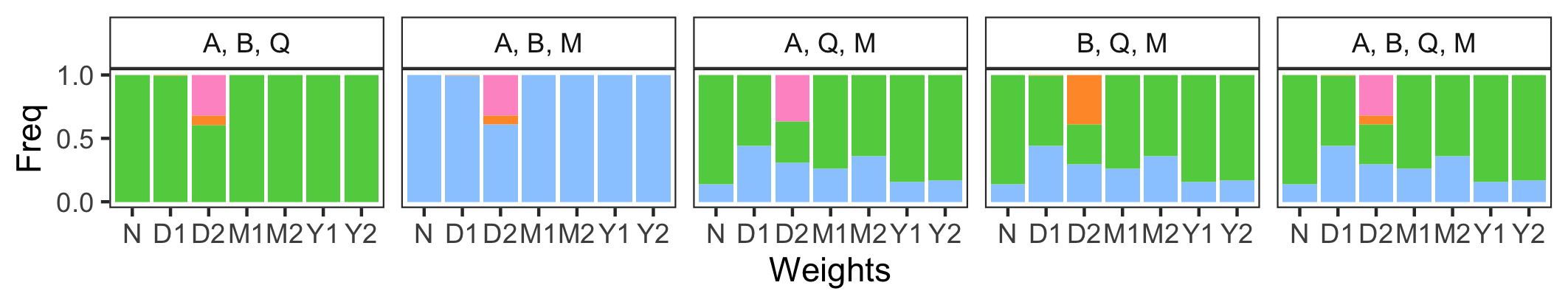}&
\includegraphics[width=.5\linewidth]{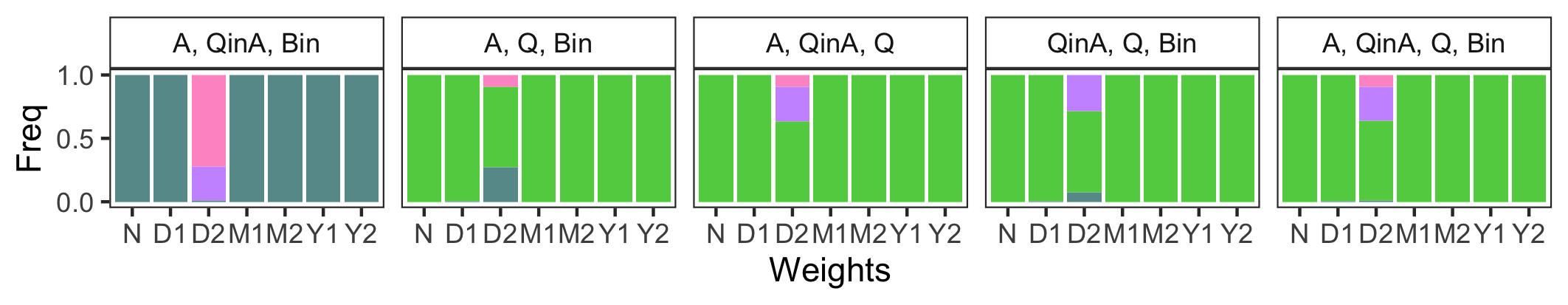}\\
\multicolumn{2}{l}{$h=10$}\\
\includegraphics[width=.5\linewidth]{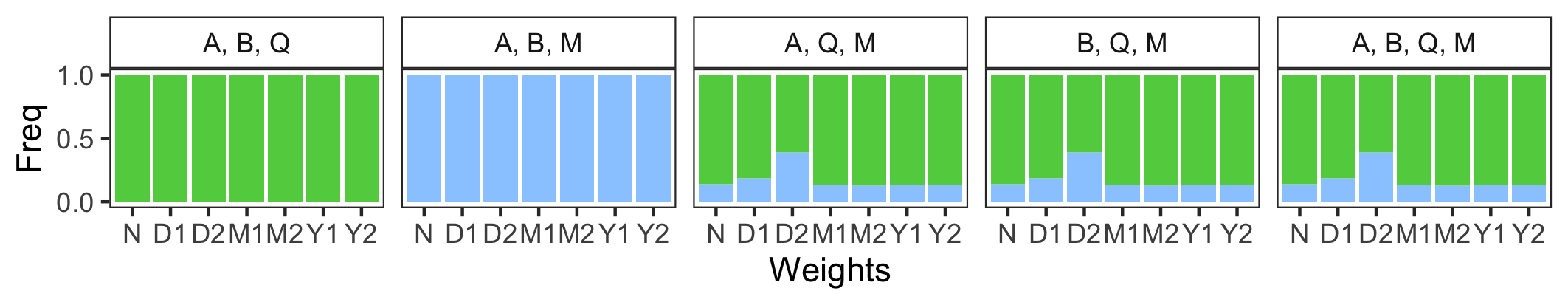}&
\includegraphics[width=.5\linewidth]{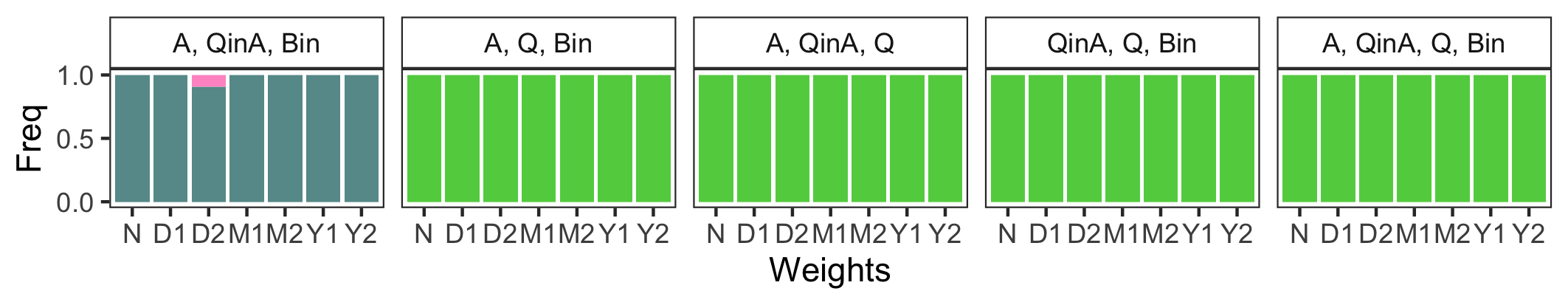}\\
\multicolumn{2}{l}{$h=100$}\\
\includegraphics[width=.5\linewidth]{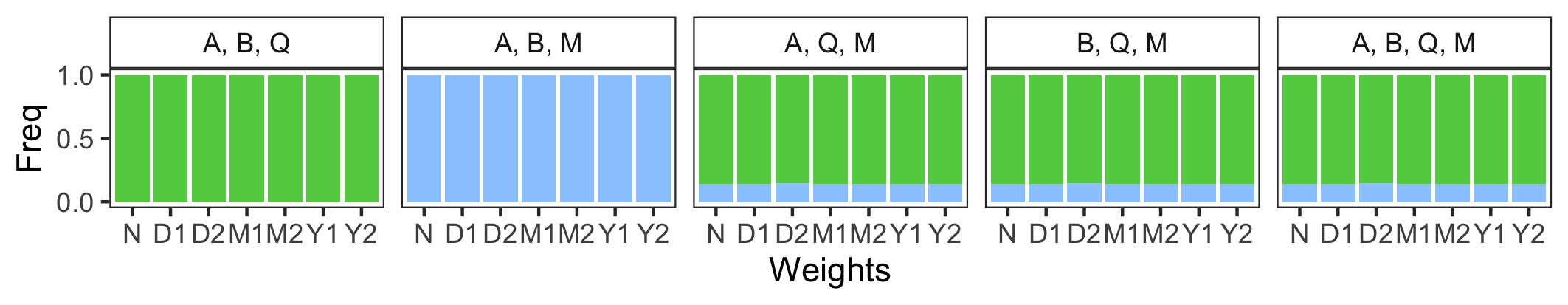}&
\includegraphics[width=.5\linewidth]{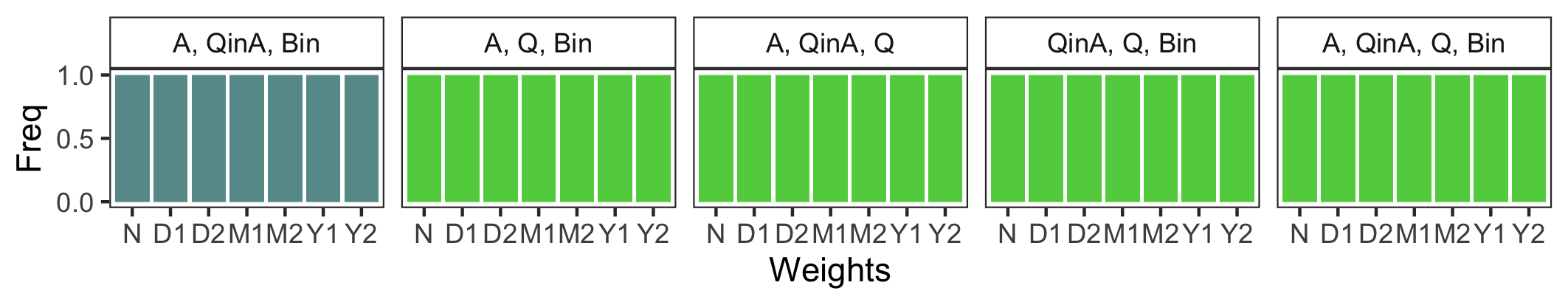}\\
\\
\includegraphics[width=.5\linewidth]{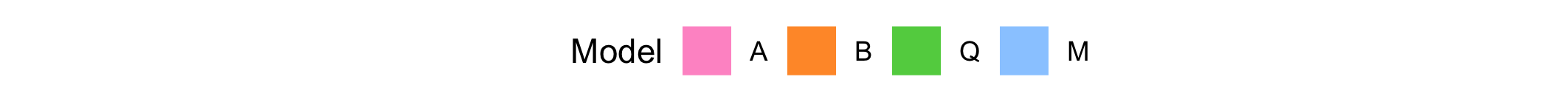}&
\includegraphics[width=.5\linewidth]{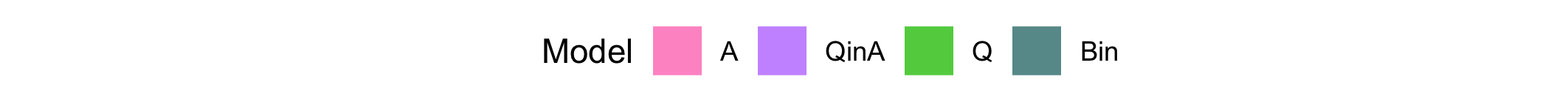}\\
\end{tabular}
}
\caption{Simulated Model Selection Outcomes for $\tau=20$.}\label{fig:sim_model_select_1997}
{\footnotesize {\it Notes:} In each panel, the title indicates the set of models being compared, the horizontal axis labels different norms being considered for forming weights, the height of a colored bar indicates the proportion of times that a particular model is selected among the set of models. The five panels on the left are for comparisons across nested models and the five on the right are for possibly non-nested models. The results are based on $\tau=20$.}
\end{figure}

We then turn to the full set of simulation results on model selection using our proposed PWMSE.
Figure \ref{fig:sim_model_select_1997} summarizes the simulation selection frequencies  for $\tau=20$. The figure is arranged in three blocks with each corresponding to a different DGP (i.e., $A$, $QinA$, and $Q$). Within each block, we generally consider two groups of model comparisons. The panels on the left are five sets of comparisons across nested models (i.e., $A$, $B$, $Q$, $M$), and the panels on the right are another five sets of comparisons across possibly non-nested models (i.e., $A$, $QinA$, $Q$, $Bin$). Within the nested/non-nested models, we compare across the four models as well as across all combinations of every three out of the four models. This systematic comparison design allows us to evaluate model selection performance when the true model is either included or excluded in the set of models being considered under each DGP. Within each panel, we show the simulation proportion of a model being selected with a colored bar using different weights defined using various norms (N, D1, D2, M1, M2, Y1, Y2) and tuning parameters ($h=1,10,100$).

Several patterns emerge from the results. First, when the true model is considered in a set of candidate models, in general, the true model is much more likely to be selected. This is particularly evident in cases when the DGP is $QinA$ or $Q$. The results are slightly weaker when DGP is $A$. Specifically, models that nest $A$ are selected for a non-negligible proportion of times even if $A$ is among the candidate models, especially when the norms are based on daily differences and the adopted tuning parameters $h$ are small.
Second, when the true model is not considered in a set of models, the candidate models that correspond to smaller values of the ideal target tend to be selected. This result can be seen in cases where (i) $B$ is selected from $\{B, Q, M\}$ and $QinA$ is selected from $\{QinA$, $Q$, $Bin\}$ when the DGP is $A$, (ii) $B$/$M$ is selected from $\{A, B, Q, M\}$ and $Q$ is selected from $\{A, Q, Bin\}$ when the DGP is $QinA$, and (iii) $M$ is selected from $\{A, B, M\}$ and $Bin$ is selected from $\{A, QinA, Bin\}$ when the DGP is $Q$.

Comparing across different tuning parameters, we note that the choice of norm is inconsequential for high values of the tuning parameter. This is unsurprising given that $h=100$ yields relatively equal weights no matter which norm is considered, rendering a situation that is very similar to the case of specifying no weight. In contrast, $h=1$ produces very small weights on those observations that do not resemble their future counterparts in terms of the norm definition. Since different norms characterize similarities between the past and future realizations from different aspects, with small $h$, the observations that have been weighted more in calculating PWMSE could also vary across the different norms adopted, potentially resulting in different model selection outcomes. This phenomenon is reflected in our simulation results when comparing model selection outcomes between D1/D2 versus Y1/Y2 under $h=1$.

This finding also indicates that the choice of norms can matter for the use of PWMSE especially if we assign much higher weights on those past realizations that are highly similar to their future counterparts. Among the six norms in our simulation study, the norms based on daily temperature differences tend to select larger models, which is most evident from the simulation results when DGP is A with small $h$. Since the true model in our simulation design does not exploit data variation at a very fine temporal scale, we conjecture that this may help explain why the norms Y1/Y2 result in higher simulation probabilities of selecting $A$ when it is the true model.

\begin{figure}[htbp]
{\fontsize{9}{5}\selectfont
\noindent
\begin{tabular}{cc}
\multicolumn{2}{c}{\bf DGP=$A$}\\
\multicolumn{2}{l}{$h=1$}\\
\includegraphics[width=.5\linewidth]{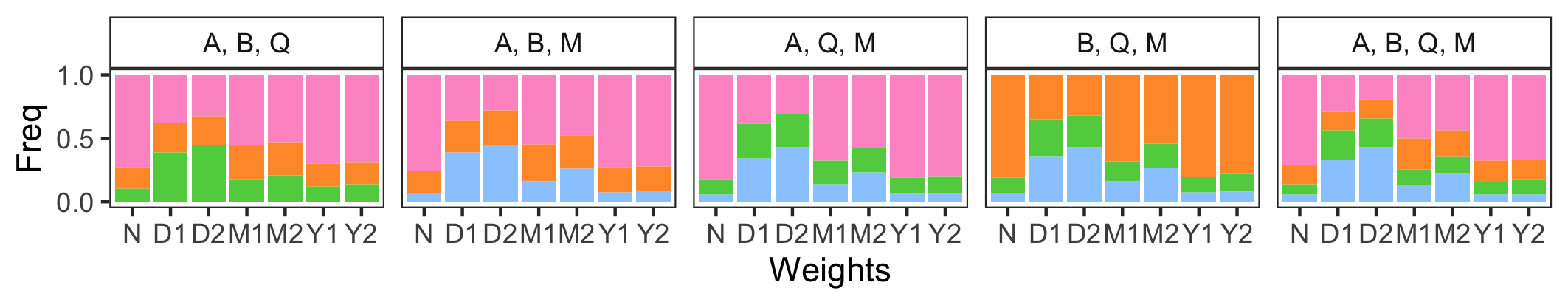}&
\includegraphics[width=.5\linewidth]{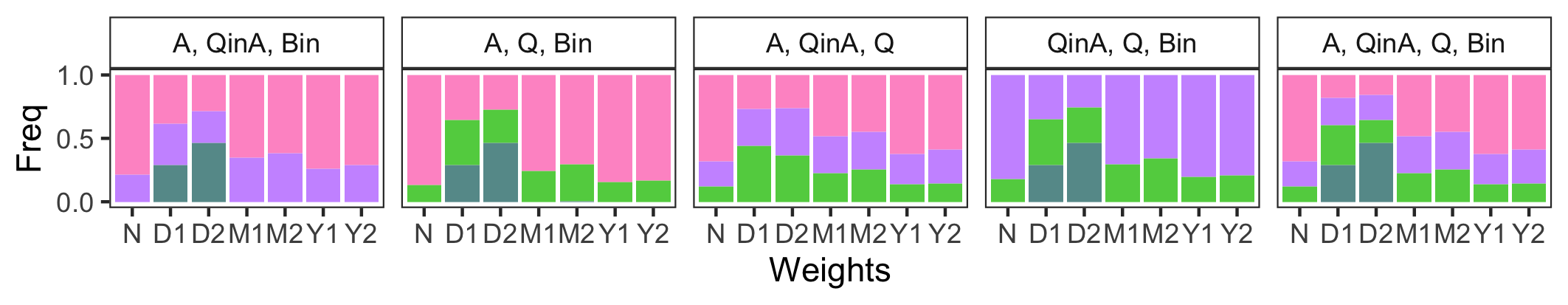}\\
\multicolumn{2}{l}{$h=10$}\\
\includegraphics[width=.5\linewidth]{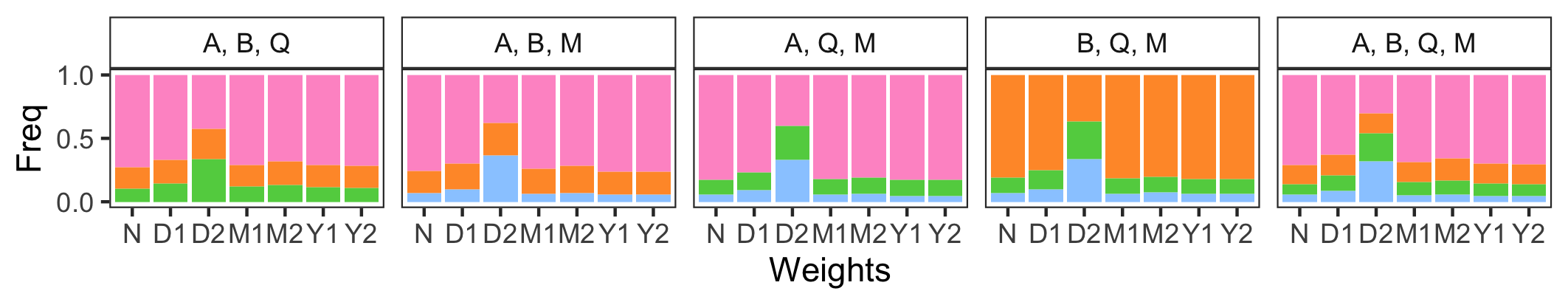}&
\includegraphics[width=.5\linewidth]{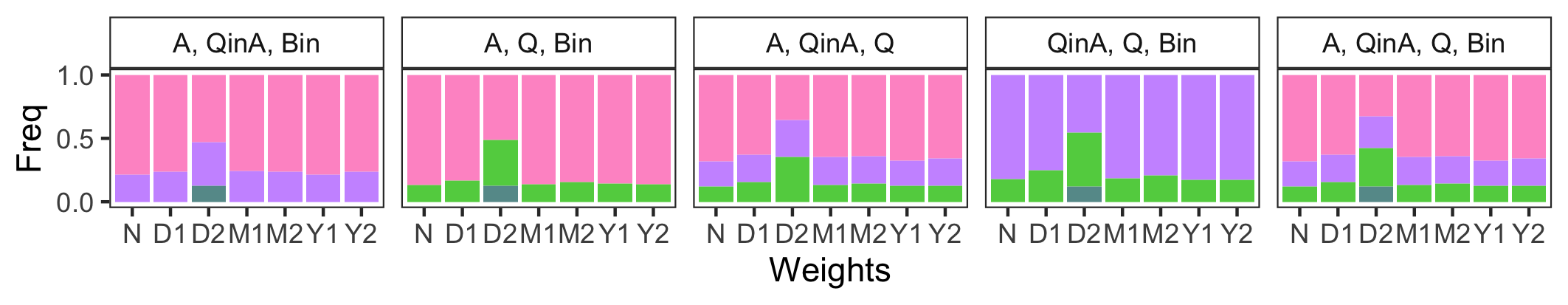}\\
\multicolumn{2}{l}{$h=100$}\\
\includegraphics[width=.5\linewidth]{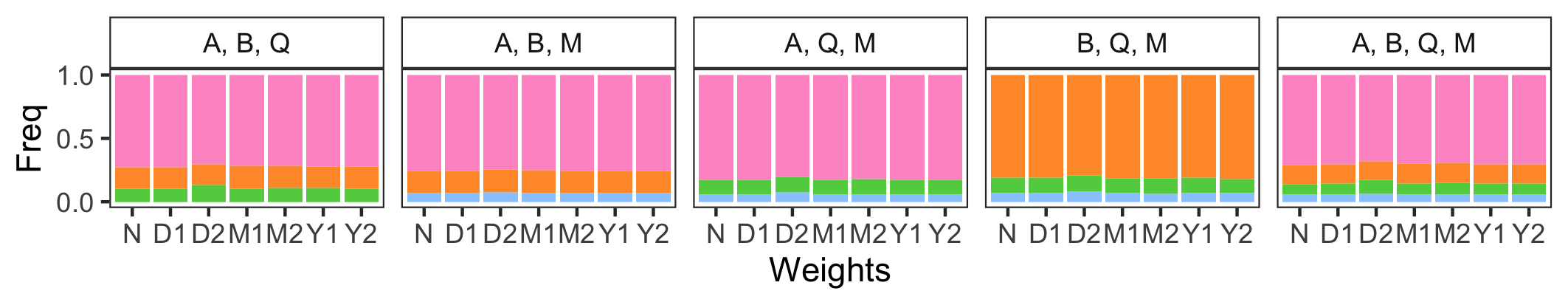}&
\includegraphics[width=.5\linewidth]{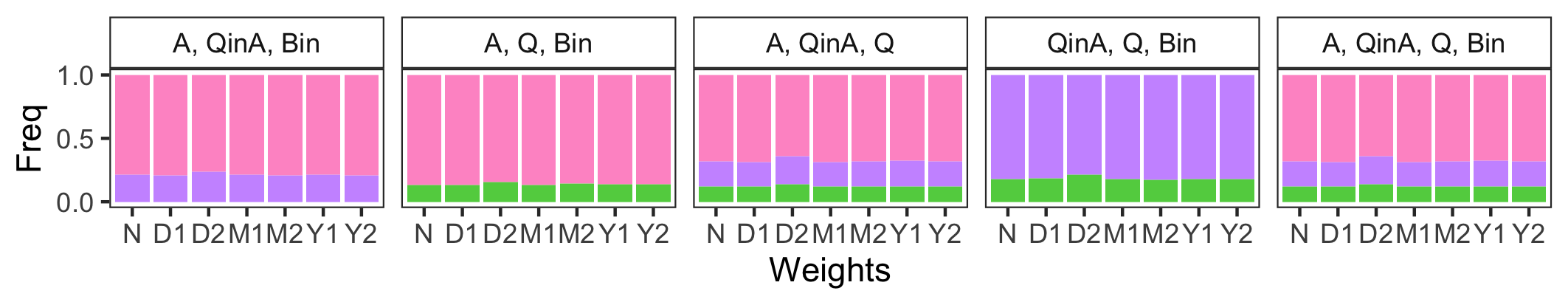}\\
\\
\multicolumn{2}{c}{\bf DGP=$QinA$}\\
\multicolumn{2}{l}{$h=1$}\\
\includegraphics[width=.5\linewidth]{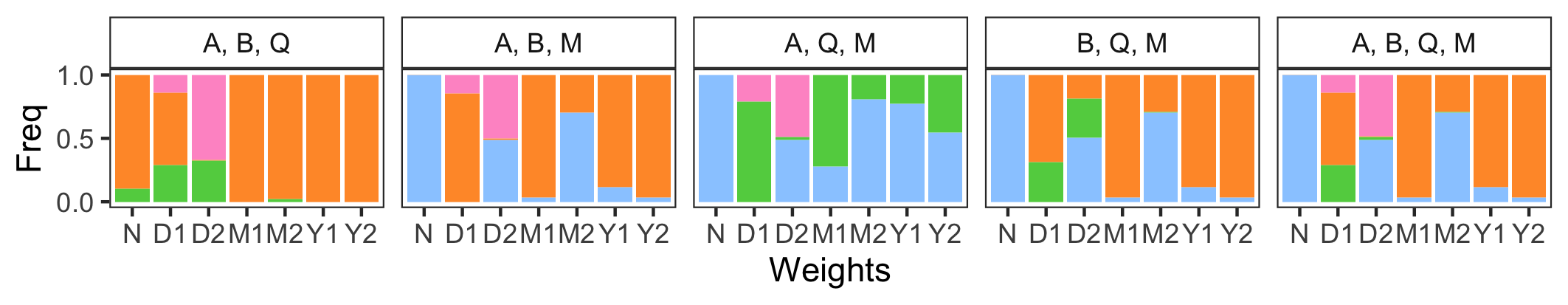}&
\includegraphics[width=.5\linewidth]{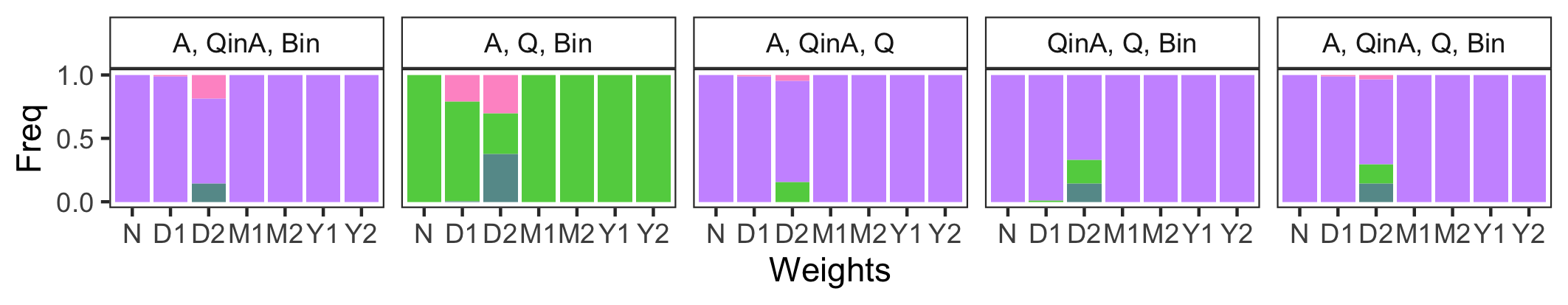}\\
\multicolumn{2}{l}{$h=10$}\\
\includegraphics[width=.5\linewidth]{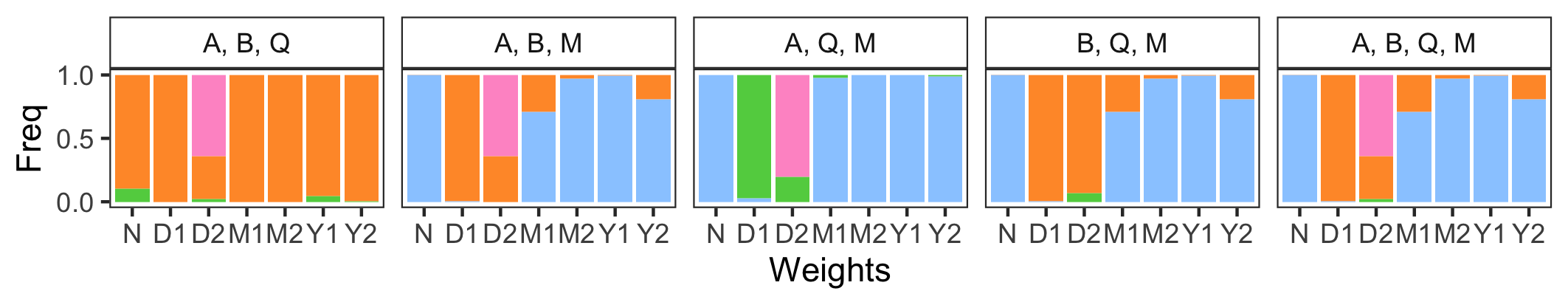}&
\includegraphics[width=.5\linewidth]{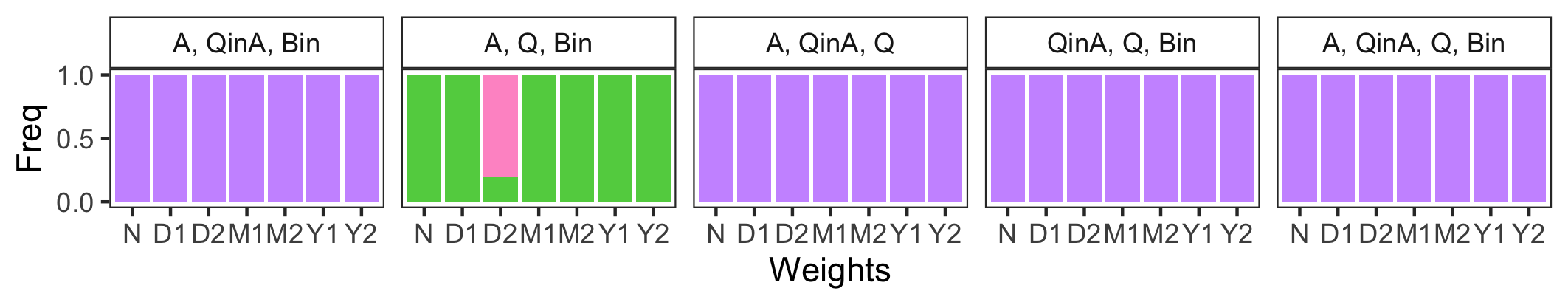}\\
\multicolumn{2}{l}{$h=100$}\\
\includegraphics[width=.5\linewidth]{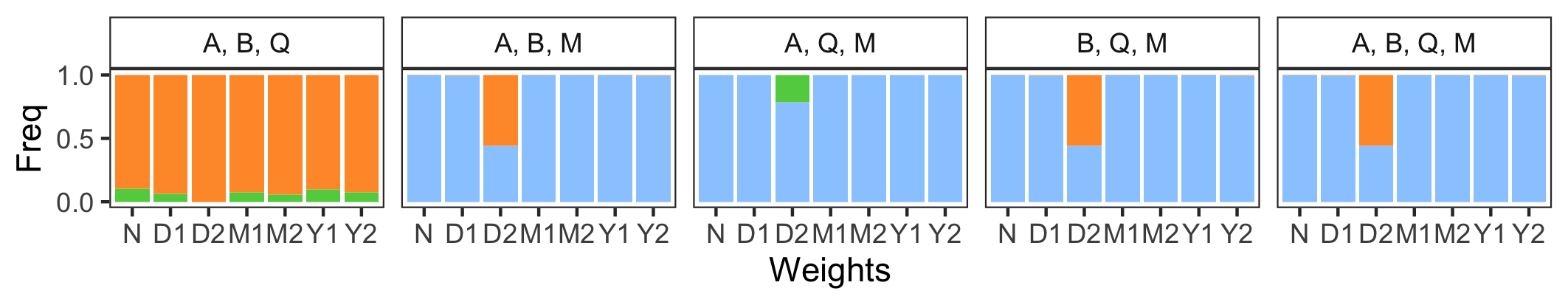}&
\includegraphics[width=.5\linewidth]{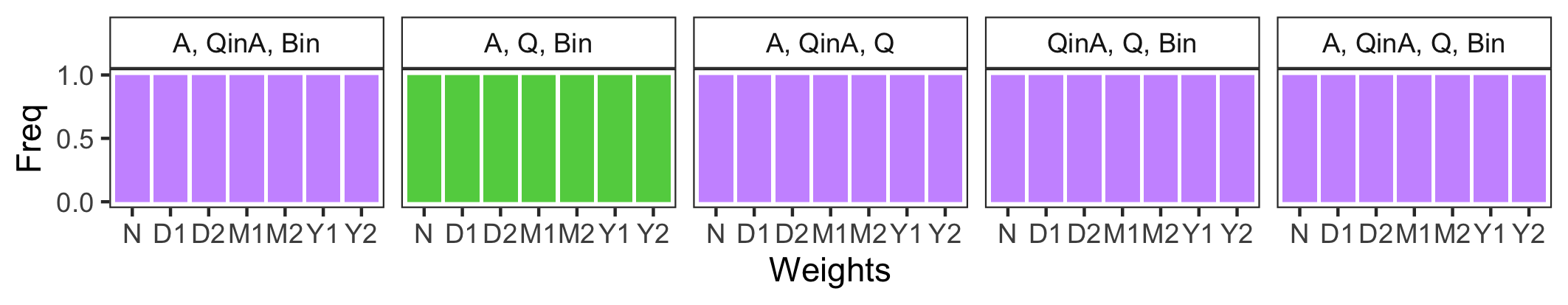}\\
\\
\multicolumn{2}{c}{\bf DGP=$Q$}\\
\multicolumn{2}{l}{$h=1$}\\
\includegraphics[width=.5\linewidth]{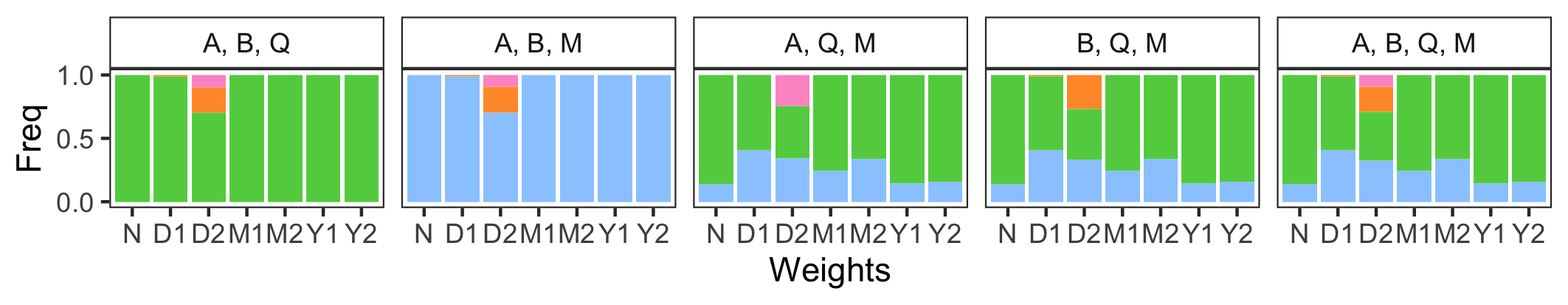}&
\includegraphics[width=.5\linewidth]{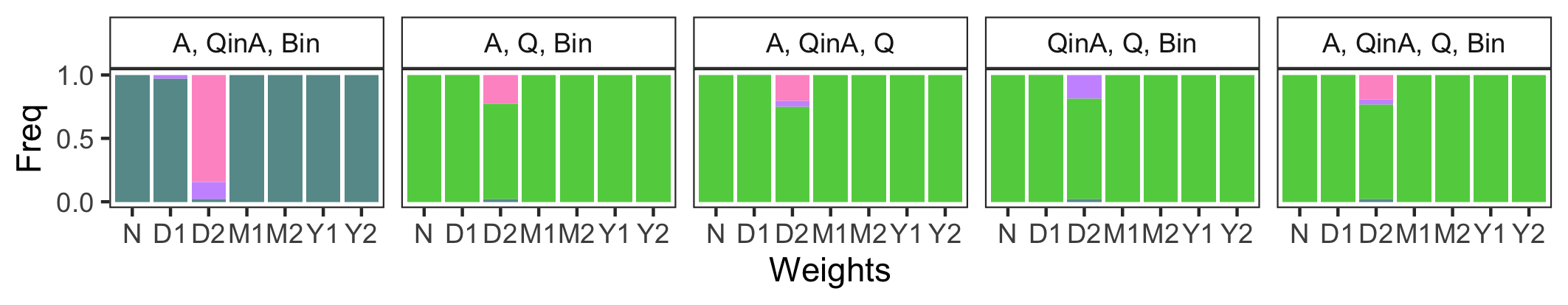}\\
\multicolumn{2}{l}{$h=10$}\\
\includegraphics[width=.5\linewidth]{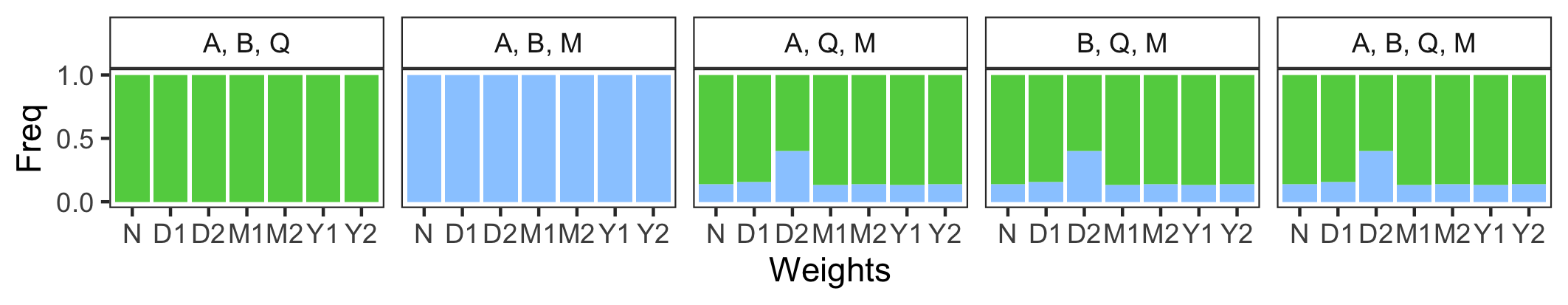}&
\includegraphics[width=.5\linewidth]{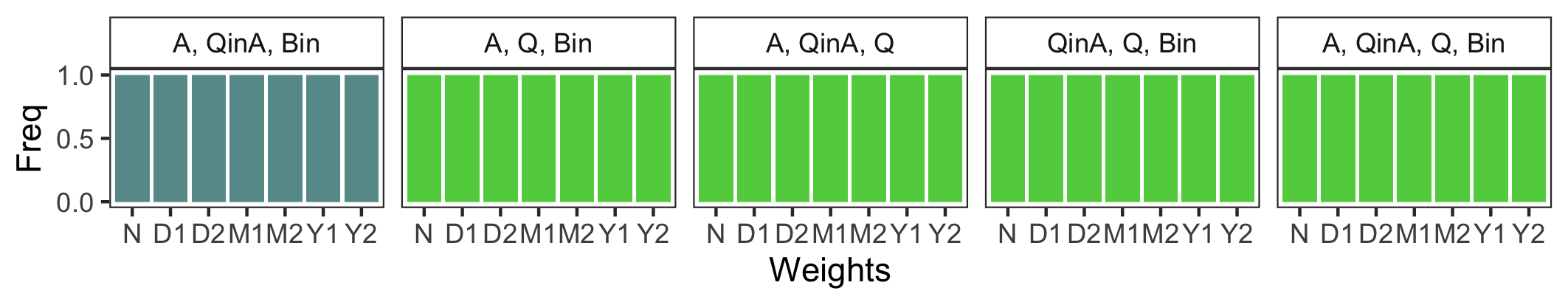}\\
\multicolumn{2}{l}{$h=100$}\\
\includegraphics[width=.5\linewidth]{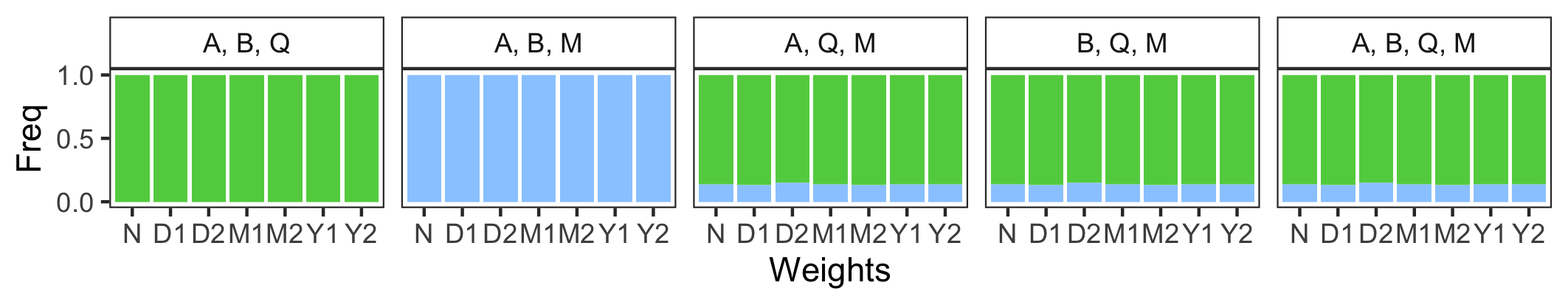}&
\includegraphics[width=.5\linewidth]{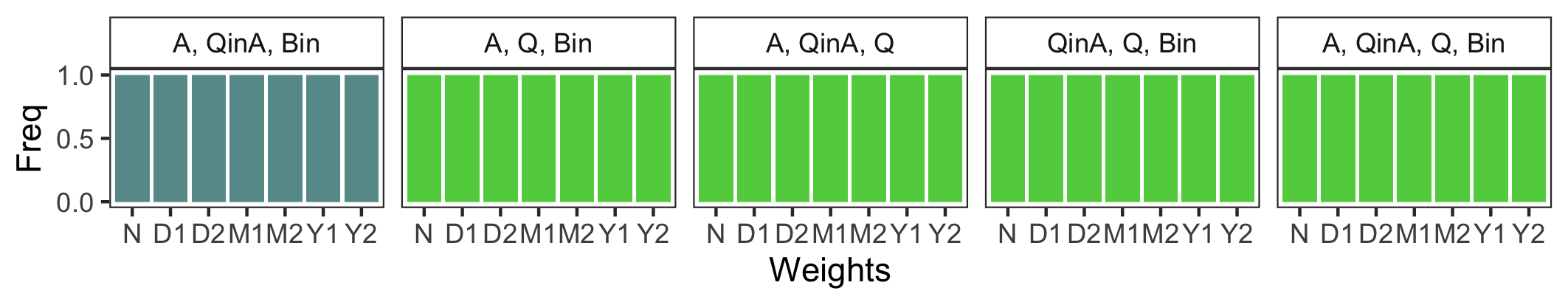}\\
\\
\includegraphics[width=.5\linewidth]{sample_nested_w_legend}&
\includegraphics[width=.5\linewidth]{sample_nonnested_w_legend}\\
\end{tabular}
}
\caption{Simulated Model Selection Outcomes for $\tau=15$}\label{fig:sim_model_select_1992}
{\footnotesize {\it Notes:} In each panel, the title indicates the set of models being compared, the horizontal axis labels different norms being considered for forming weights, the height of a colored bar indicates the proportion of times that a particular model is selected among the set of models. The five panels on the left are for comparisons across nested models and the five on the right are for possibly non-nested models. The results are based on $\tau=15$.}
\end{figure}

\begin{figure}[htbp]
{\fontsize{9}{5}\selectfont
\noindent
\begin{tabular}{cc}
\multicolumn{2}{c}{\bf DGP=$A$}\\
\multicolumn{2}{l}{$h=1$}\\
\includegraphics[width=.5\linewidth]{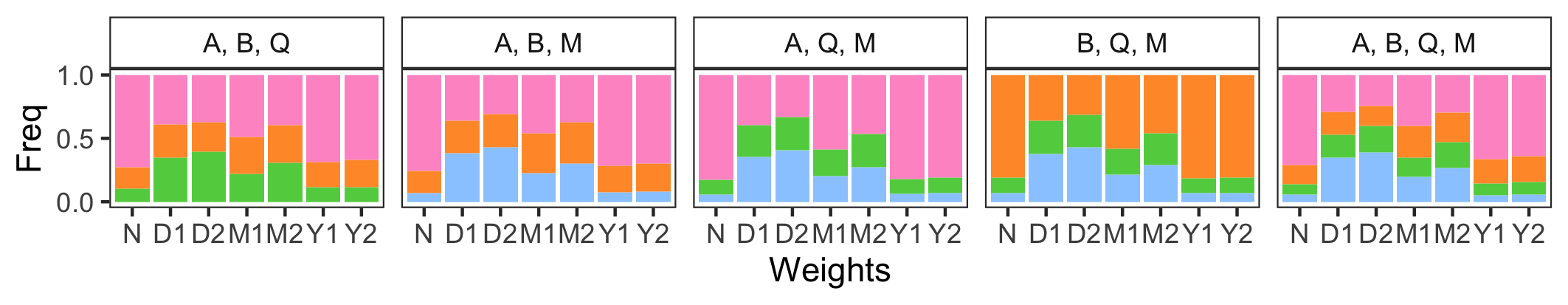}&
\includegraphics[width=.5\linewidth]{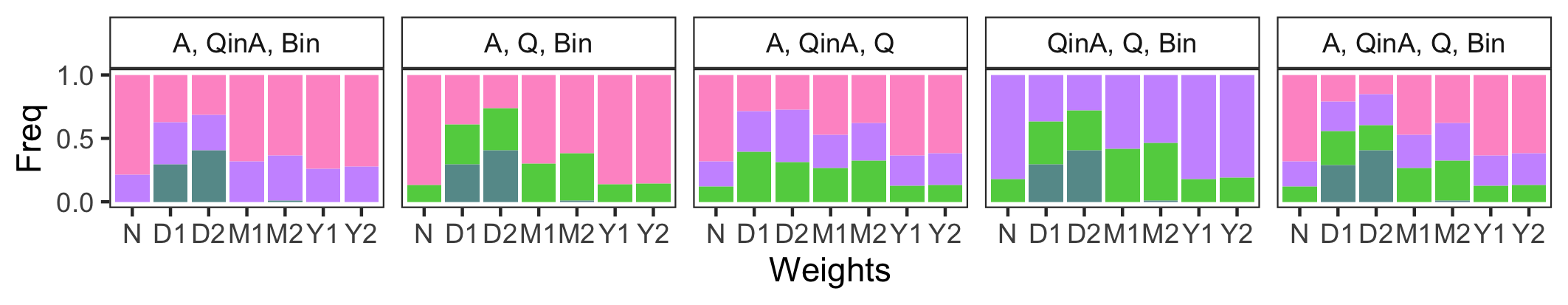}\\
\multicolumn{2}{l}{$h=10$}\\
\includegraphics[width=.5\linewidth]{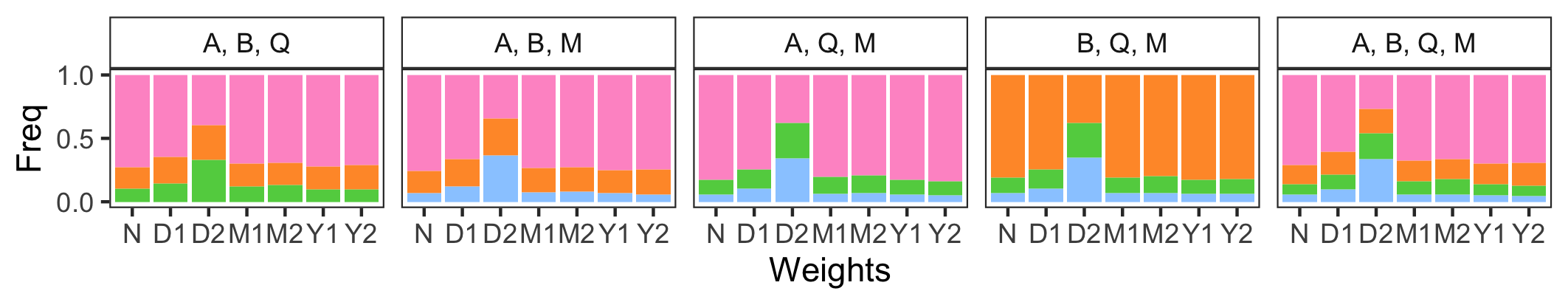}&
\includegraphics[width=.5\linewidth]{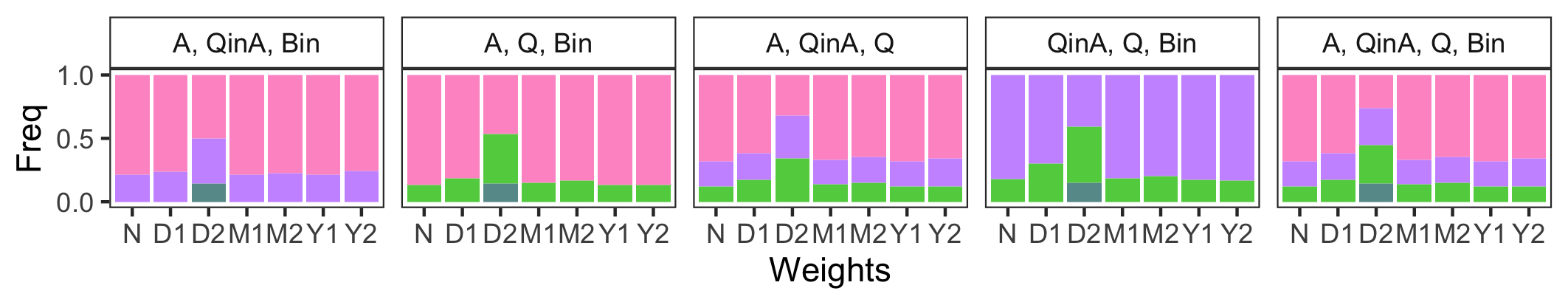}\\
\multicolumn{2}{l}{$h=100$}\\
\includegraphics[width=.5\linewidth]{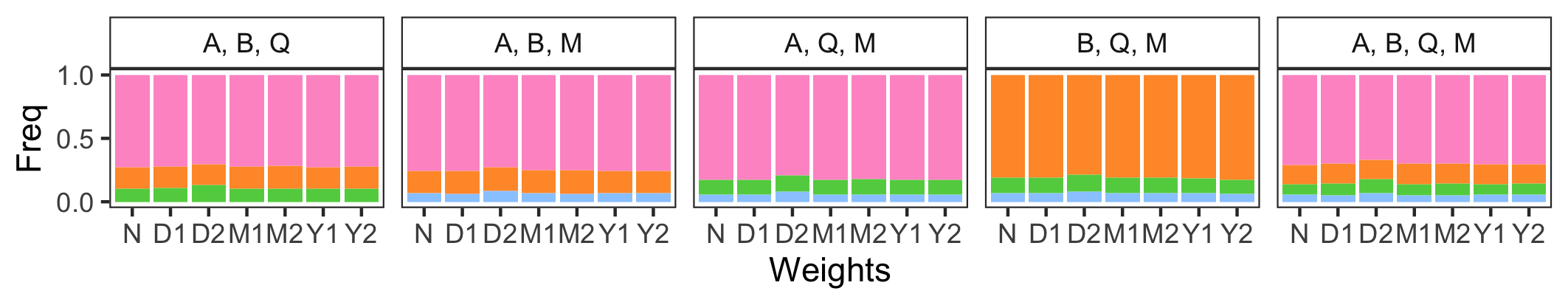}&
\includegraphics[width=.5\linewidth]{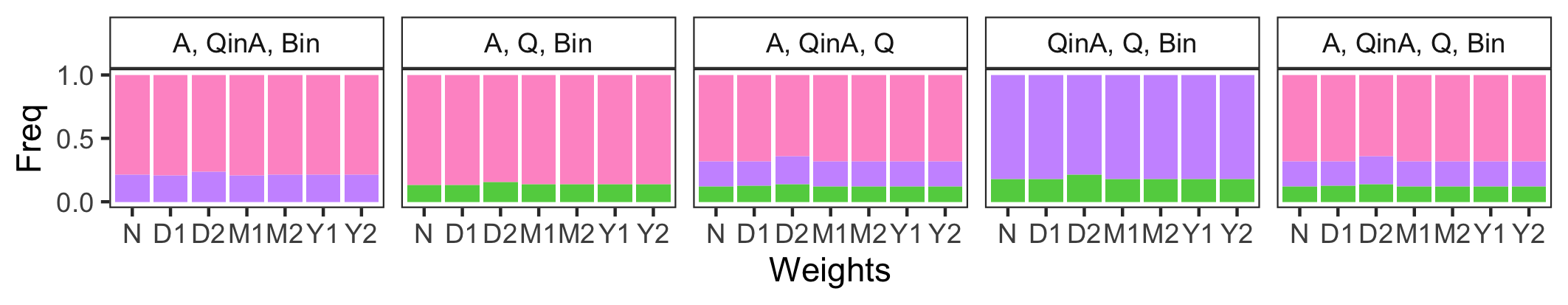}\\
\\
\multicolumn{2}{c}{\bf DGP=$QinA$}\\
\multicolumn{2}{l}{$h=1$}\\
\includegraphics[width=.5\linewidth]{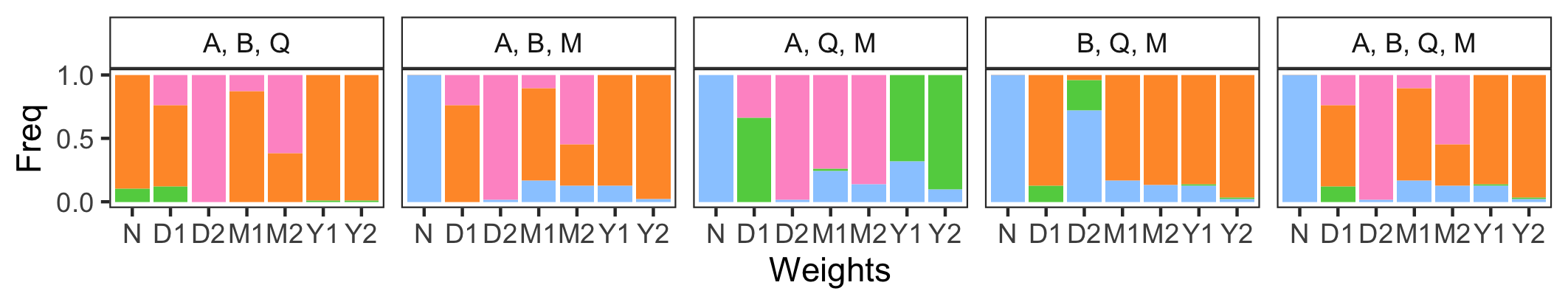}&
\includegraphics[width=.5\linewidth]{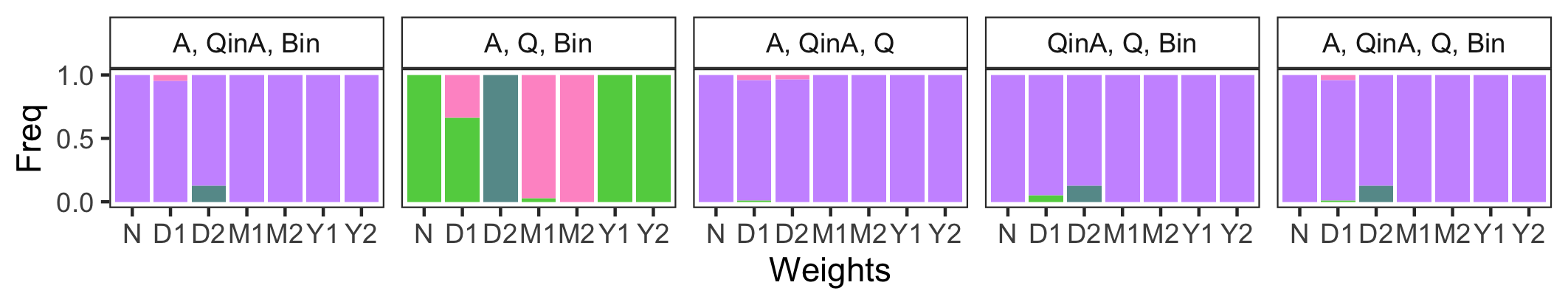}\\
\multicolumn{2}{l}{$h=10$}\\
\includegraphics[width=.5\linewidth]{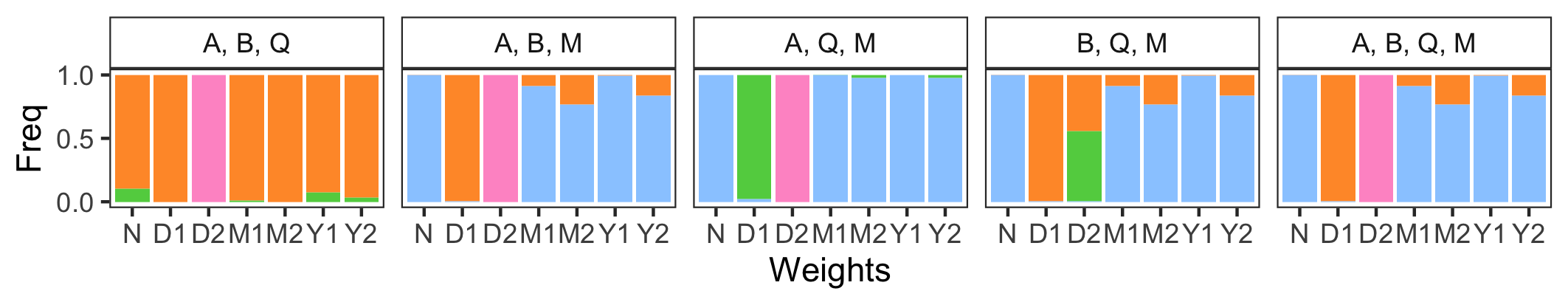}&
\includegraphics[width=.5\linewidth]{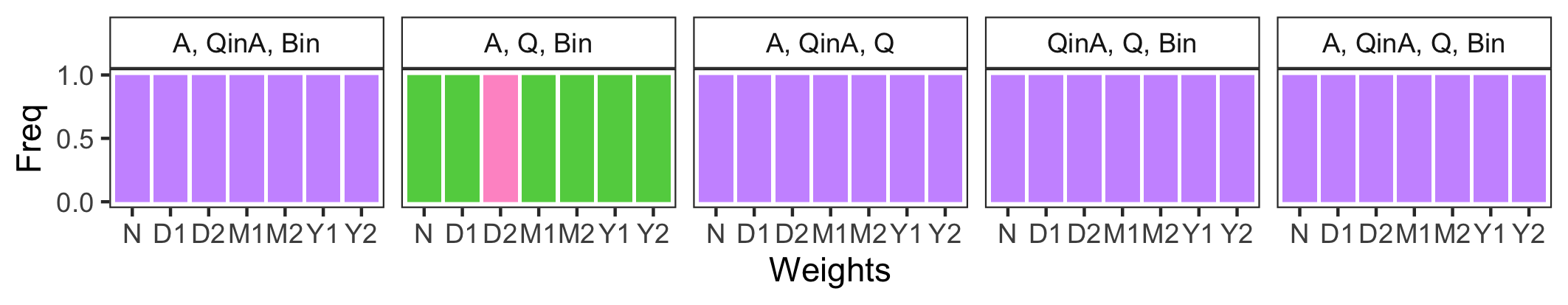}\\
\multicolumn{2}{l}{$h=100$}\\
\includegraphics[width=.5\linewidth]{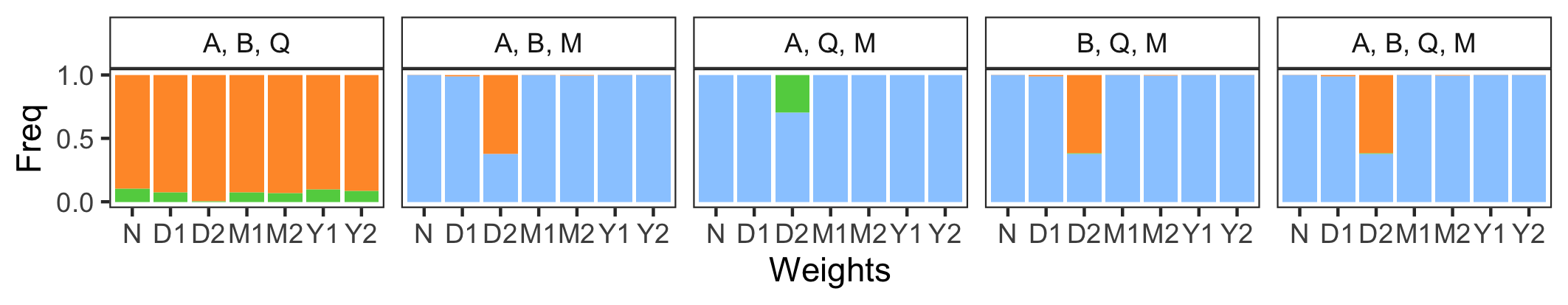}&
\includegraphics[width=.5\linewidth]{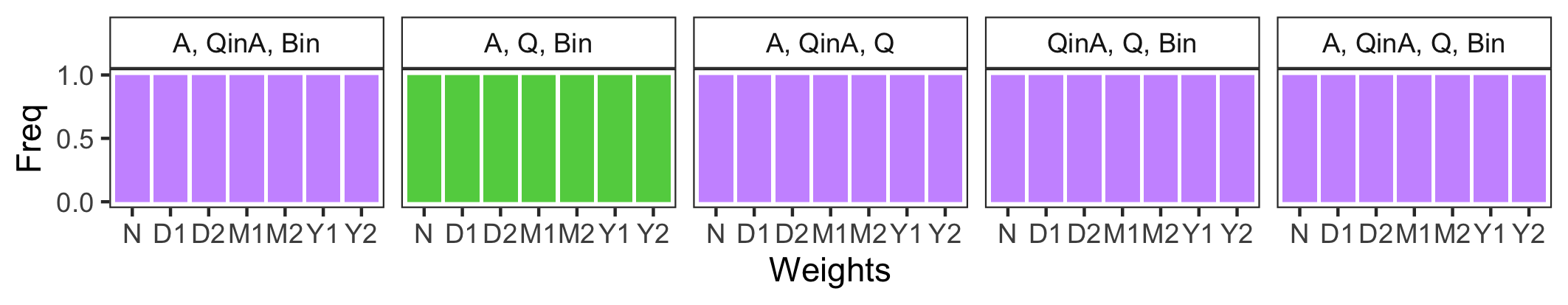}\\
\\
\multicolumn{2}{c}{\bf DGP=$Q$}\\
\multicolumn{2}{l}{$h=1$}\\
\includegraphics[width=.5\linewidth]{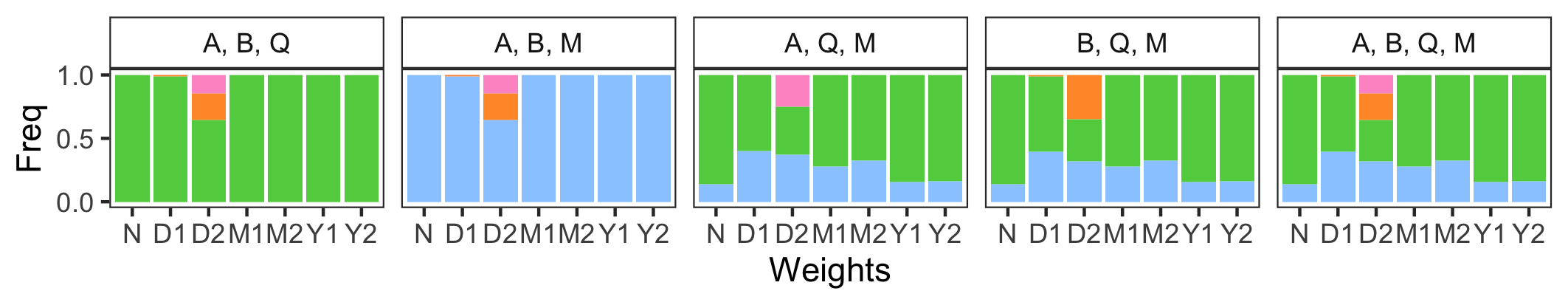}&
\includegraphics[width=.5\linewidth]{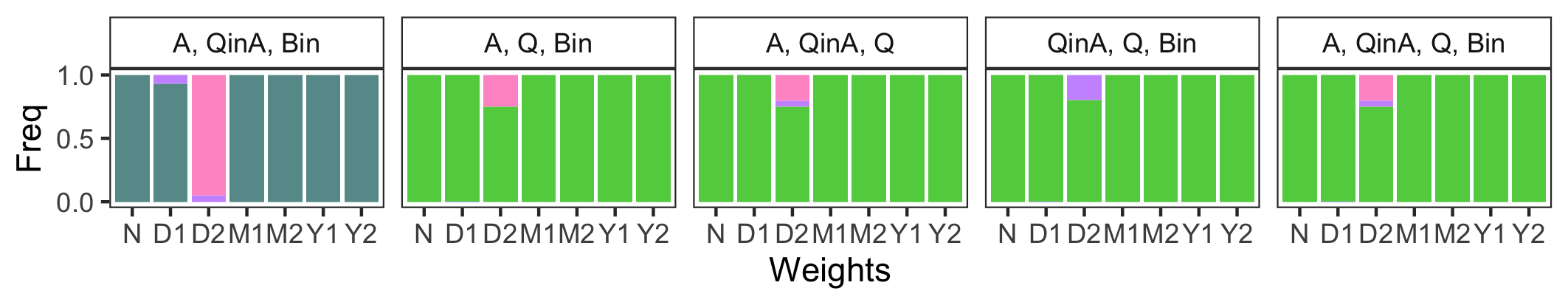}\\
\multicolumn{2}{l}{$h=10$}\\
\includegraphics[width=.5\linewidth]{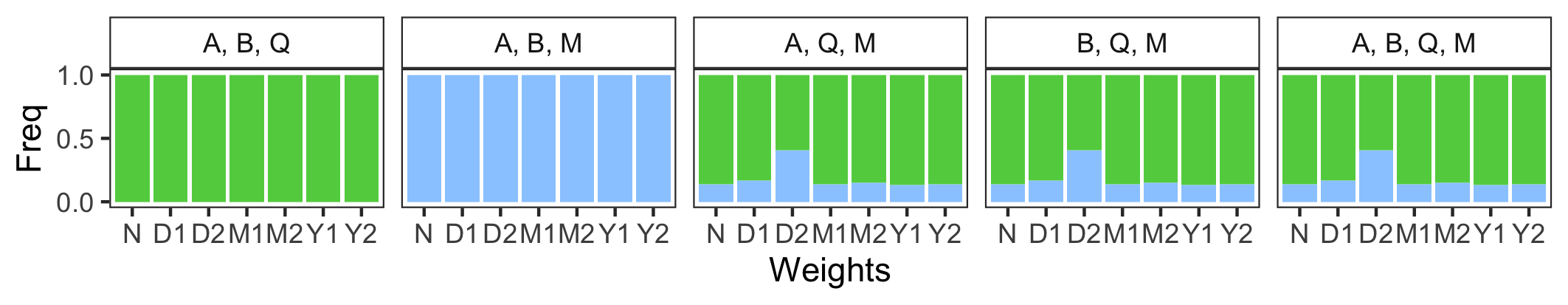}&
\includegraphics[width=.5\linewidth]{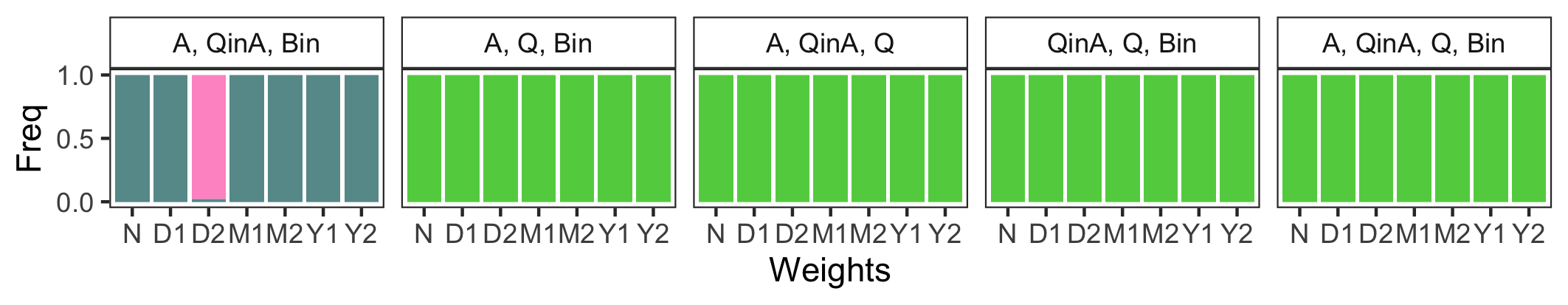}\\
\multicolumn{2}{l}{$h=100$}\\
\includegraphics[width=.5\linewidth]{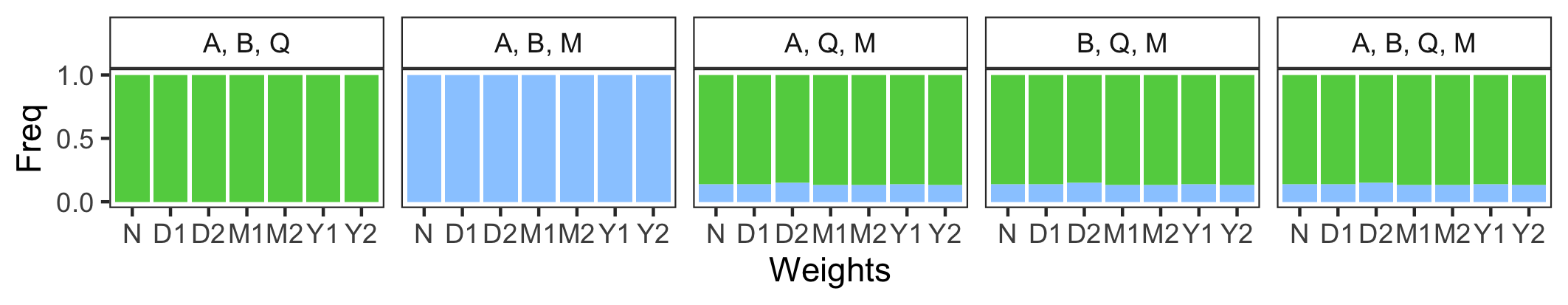}&
\includegraphics[width=.5\linewidth]{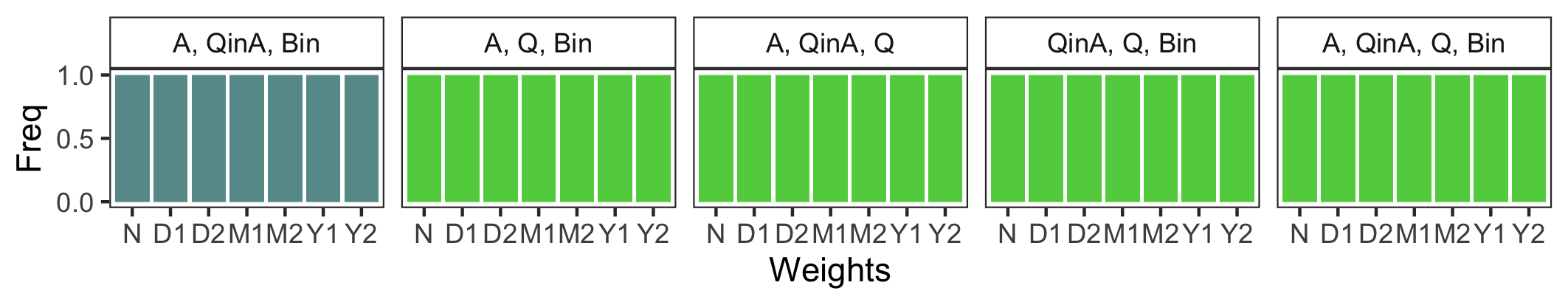}\\
\\
\includegraphics[width=.5\linewidth]{sample_nested_w_legend}&
\includegraphics[width=.5\linewidth]{sample_nonnested_w_legend}\\
\end{tabular}
}
\caption{Simulated Model Selection Outcomes for $\tau=25$}\label{fig:sim_model_select_2002}
{\footnotesize {\it Notes:} In each panel, the title indicates the set of models being compared, the horizontal axis labels different norms being considered for forming weights, the height of a colored bar indicates the proportion of times that a particular model is  selected among the set of models. The five panels on the left are for comparisons across nested models and the five on the right are for possibly non-nested models. The results are based on $\tau=25$.}
\end{figure}

We also present the simulation results for different future climates $\tau=15$ and $\tau=25$ in Figures \ref{fig:sim_model_select_1992} and \ref{fig:sim_model_select_2002}, respectively.  For these different future climates, the model selection results do not vary much when the true model or a model that nests it is under consideration. However, the selection results vary across $\tau$ substantially when all models are misspecified. This is particularly true when $h$ is small, which leads to bigger differences in the weighting function depending on the value of the norm. This is most evident when comparing results of DGP$=QinA$ with $h=1$ across different $\tau$. In particular, some results based on the norms of daily and monthly differences are substantially different for the different values of $\tau$ we consider. Based on these insights from the simulation results, we provide guidance for practitioners on how to specify the weighting function when using our proposed PWMSE criterion in practice.

\subsection{A Practitioner's Guide on the Specification of the Weights in the PWMSE}\label{sec:guide_weight}

In our proposed PWMSE, the weight consists of a norm and a tuning parameter. The norm measures the distance between past and future realizations. How we construct this norm reflects the specific dimension we would like to characterize proximity by. In our extended simulations, we consider three sets of norms based on temperature differences at the daily, monthly, and annual level, respectively. The measured proximity can be substantially different with norms constructed based on different temporal scales. For example, norms based on annual mean temperature ignore the within-year temperature variability, whereas such information would be incorporated in daily norms. Put simply, two years that have the same annual mean temperature may still be very different in terms of the daily norm.

The chosen aggregation level used for the norm pins down the temporal scale at which the differences between past and future realizations will be measured. Therefore, we recommend choosing the level such that it matches the temporal resolution of the empirical context. For instance, the DGPs in our simulations only leverage data variation at the annual (or seasonal) level. As reflected from our simulation results, the model selection results tend to be more consistent when we opt for norms based on differences at the annual (or monthly) level, holding the tuning parameters constant. However, if the empirical setting examines an outcome variable at a finer temporal scale, using norms based on daily differences may be more appropriate.

\begin{figure}[htbp]
\begin{center}
\includegraphics[width=.45\linewidth]{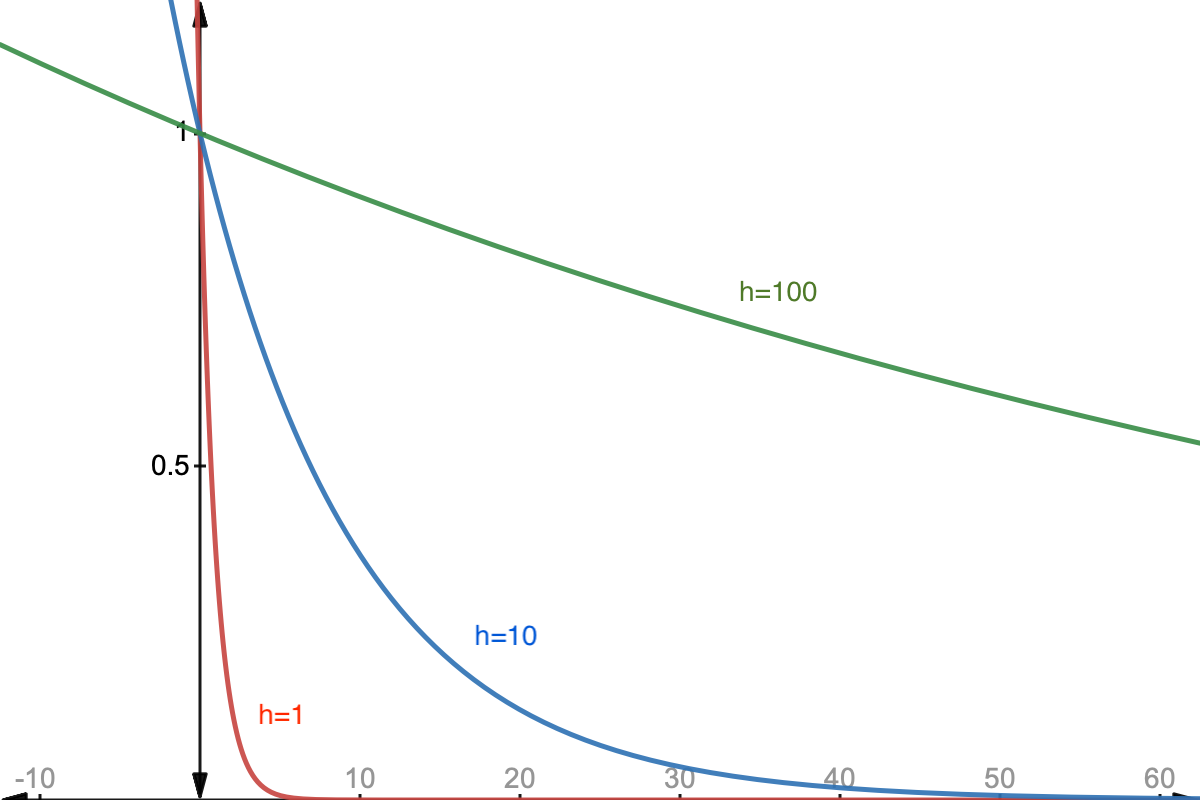}
\end{center}
\caption{Weights under Different Tuning Parameters}\label{fig:visual_h}
{\footnotesize {\it Notes:} The figure plots the function $y=e^{-x/h}$ on the space of $(x,y)$ with $h=1,10,100$, respectively.}
\end{figure}

For any norm adopted, the tuning parameter determines the relative weights on observations with different degrees of similarity to the projected climate. We plot a set of functions of $y=e^{-x/h}$ in Figure \ref{fig:visual_h} to illustrate the difference between different levels of $h$. Referring to the three levels of $h$ we considered in our simulation, $h=1$ would assign much higher weights only to those with little differences in their past and future realizations. In contrast, $h=100$ would produce relatively equal weights across observations. It is important to note that the relative differences between different levels of $h$ have to be examined under a particular empirical context, since the magnitude of the norms is not unit free and varies across settings. For the empirical context of temperature impacts, like the one in this paper, we recommend choosing tuning parameters no lower than one and no higher than 100. Nevertheless, we remind the practitioners that there is essentially a trade-off to be made. The practitioner has to balance between relying on more representative historical data and focusing only on those past realizations that are highly similar to their future counterparts.

\clearpage

\begin{figure}[htbp]
    \begin{center}
     \includegraphics[width=.6\linewidth]{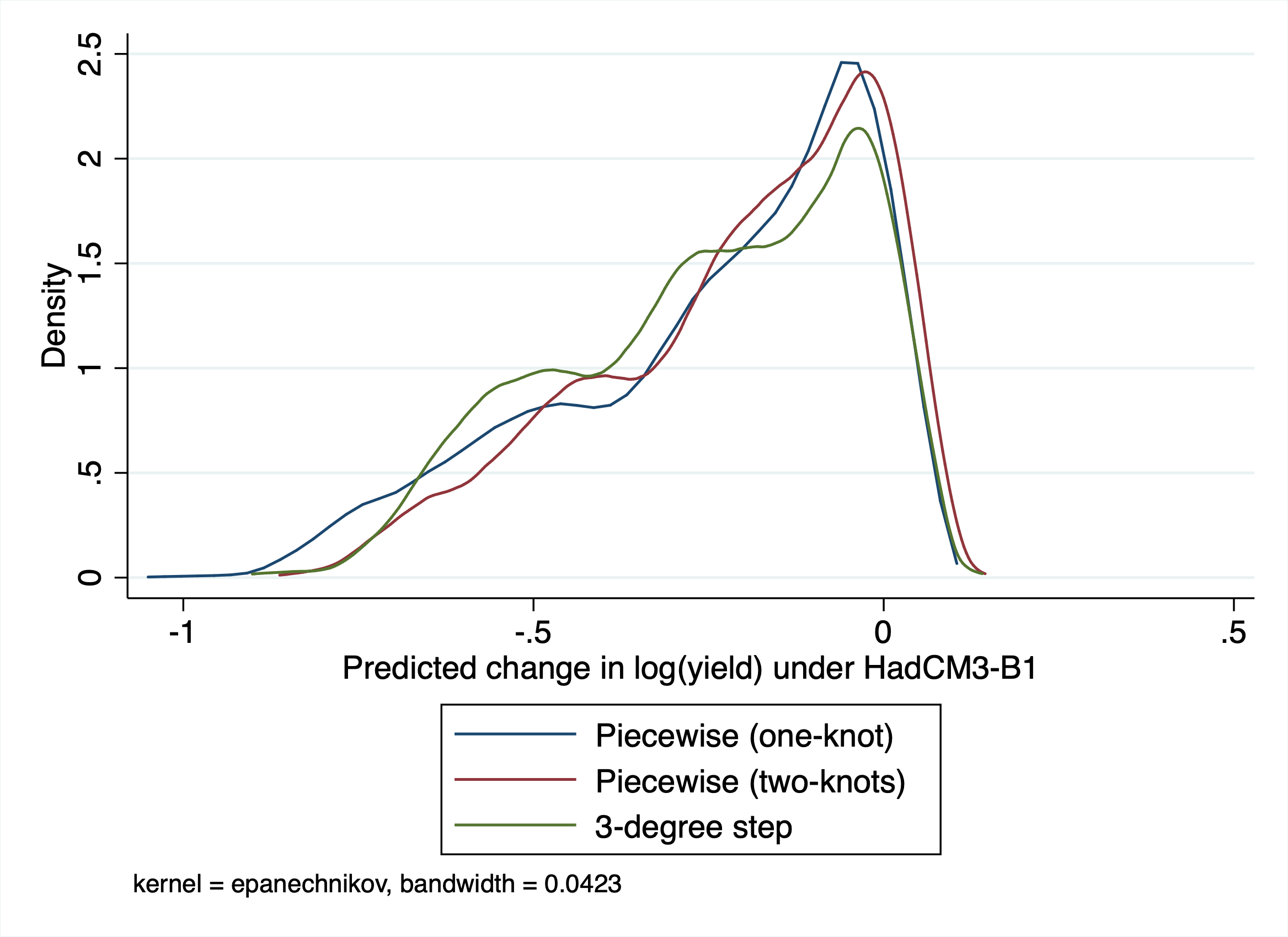}
    \end{center}
    \caption{Distribution of Predicted Yield Changes under HadCM3-B1}
    \label{fig:kdensity_hadcm3}
    {\footnotesize {\it Notes:} The figure plots the kernel density estimates of the county-level predicted yield changes under the HadCM3-B1 climate projection, based on the estimated one-knot, two-knots piecewise functions, and the 3$^\circ$C step function.}
\end{figure}

\clearpage
\begin{table}[htbp]
  \centering \fontsize{10}{10}\selectfont  
  \caption{Ideal Targets Evaluated at Simulation Mean of Coefficients for $\tau=15,25$}\label{tab:ideal_target_plus}
    \begin{tabular}{p{2cm}llllllll} \hline
 DGP:         & \multicolumn{2}{c}{A} &       & \multicolumn{2}{c}{QinA} &       & \multicolumn{2}{c}{Q} \\ \cline{2-3}\cline{5-6}\cline{8-9}
          & $\tau=$15    & $\tau=$25    &       & $\tau=$15    & $\tau=$25    &       & $\tau=$15    & $\tau=$25 \\ \hline
$\mathbf{M}_{\alpha}$:&          &       &       &       &       &       &       &         \\
    A     & 1.87$\times 10^{-7}$ & {2.15$\times 10^{-7}$} &       & 0.98  & 1.10  &       & 14.44  & 1.84  \\
    B     & {1.85$\times 10^{-7}$} & 3.17$\times 10^{-7}$ &       & 0.92  & 1.10  &       & 8.42  & 0.74  \\
    Q     & 2.77$\times 10^{-7}$ & 4.93$\times 10^{-7}$ &       & 0.94  & 1.11  &       & {2.77$\times 10^{-7}$} & {4.93$\times 10^{-7}$} \\
    M     & 4.27$\times 10^{-7}$ & 11.67$\times 10^{-7}$ &       & 0.92  & 1.54  &       & 4.27$\times 10^{-7}$ & 11.67$\times 10^{-7}$ \\
    QinA  & 2.10$\times 10^{-7}$ & 2.26$\times 10^{-7}$ &       & {2.10$\times 10^{-7}$} & {2.26$\times 10^{-7}$} &       & 14.27  & 1.83  \\
    Bin   & 0.07  & 0.09  &       & 4.27  & 4.89  &       & 1.81  & 1.13  \\ \hline
          \multicolumn{9}{l}{\parbox[t]{15cm}{\footnotesize {\it Notes:} The table presents the calculated ideal target $\frac{1}{n}\sum_{i=1}^n(\bar{\mu}_{\alpha}(\mathcal{W}^f_{i,T+\tau})-\bar{\mu}_{\alpha}(\mathcal{W}_{iT})-(\mu_{\star}(\mathcal{W}^f_{i,T+\tau})-\mu_{\star}(\mathcal{W}_{iT})))^2$, where $\bar{\mu}_{\alpha}(\mathcal{W}_{it}) = X_{it}' \bar{\beta}_\alpha^k$ for the simulated $\bar{\beta}_\alpha^k$ in Table \ref{tab:pseudo_true_param}.}}        
    \end{tabular}
\end{table}

\begin{table}[htbp]
  \fontsize{10}{10}\selectfont
  \centering
  \caption{Models Minimizing Feasible Targets}\label{tab:sim_feasible_targets_plus}
    \begin{tabular}{p{2cm}cccp{1cm}cccp{1cm}ccc} \hline
    DGP:   & \multicolumn{3}{c}{$A$} &       & \multicolumn{3}{c}{$QinA$} &       & \multicolumn{3}{c}{$Q$} \\ \cline{2-4}\cline{6-8}\cline{10-12}
    $h$:     & 1     & 10    & 100   &       & 1     & 10    & 100   &       & 1     & 10    & 100 \\ \hline
   $\tau=20$       &       &       &       &       &       &       &       &       &       &       &  \\ 
    N     & A     & A     & A     &       & QinA  & QinA  & QinA  &       & Q     & Q     & Q \\
    D1    & M     & A     & A     &       & QinA  & QinA  & QinA  &       & M     & Q     & Q \\
    D2    & M     & M     & A     &       & QinA  & QinA  & QinA  &       & M     & M     & Q \\
    M1    & QinA  & A     & A     &       & QinA  & QinA  & QinA  &       & Q     & Q     & Q \\
    M2    & A     & A     & A     &       & QinA  & QinA  & QinA  &       & Q     & Q     & Q \\
    Y1    & A     & A     & A     &       & QinA  & QinA  & QinA  &       & Q     & Q     & Q \\
    Y2    & A     & A     & A     &       & QinA  & QinA  & QinA  &       & Q     & Q     & Q \\
          &       &       &       &       &       &       &       &       &       &       &  \\ \hline
   $\tau=15$       &       &       &       &       &       &       &       &       &       &       &  \\
    N     & A     & A     & A     &       & QinA  & QinA  & QinA  &       & Q     & Q     & Q \\
    D1    & QinA  & A     & A     &       & QinA  & QinA  & QinA  &       & Q     & Q     & Q \\
    D2    & M     & M     & A     &       & Bin   & QinA  & QinA  &       & B     & M     & Q \\
    M1    & A     & A     & A     &       & QinA  & QinA  & QinA  &       & Q     & Q     & Q \\
    M2    & A     & A     & A     &       & QinA  & QinA  & QinA  &       & Q     & Q     & Q \\
    Y1    & A     & A     & A     &       & QinA  & QinA  & QinA  &       & Q     & Q     & Q \\
    Y2    & A     & A     & A     &       & QinA  & QinA  & QinA  &       & Q     & Q     & Q \\
          &       &       &       &       &       &       &       &       &       &       &  \\ \hline
    $\tau=25$   &       &       &       &       &       &       &       &       &       &       &  \\
    N     & A     & A     & A     &       & QinA  & QinA  & QinA  &       & Q     & Q     & Q \\
    D1    & QinA  & A     & A     &       & QinA  & QinA  & QinA  &       & Q     & Q     & Q \\
    D2    & QinA  & M     & A     &       & QinA  & QinA  & QinA  &       & B     & M     & Q \\
    M1    & A     & A     & A     &       & QinA  & QinA  & QinA  &       & M     & Q     & Q \\
    M2    & M     & A     & A     &       & QinA  & QinA  & QinA  &       & M     & Q     & Q \\
    Y1    & A     & A     & A     &       & QinA  & QinA  & QinA  &       & Q     & Q     & Q \\
    Y2    & A     & A     & A     &       & QinA  & QinA  & QinA  &       & M     & Q     & Q \\ \hline
    \multicolumn{12}{l}{\parbox[t]{14cm}{\footnotesize {\it Notes:} The table presents the model that minimizes the feasible target  $\sum_{s=1}^{T-1} \frac{1}{n}\sum_{i=1}^n(\bar{\mu}_{\alpha}(\mathcal{W}_{i,T})-\bar{\mu}_{\alpha}(\mathcal{W}_{i,T-s})-(Y_{i,T}-Y_{i,T-s}))^2 \pi(\mathcal{W}_{i,T-s},\mathcal{W}_{i,T+\tau})$, where $\bar{\mu}_{\alpha}(\mathcal{W}_{it}) = X_{it}' \bar{\beta}_\alpha^k$ and $\tau=15, 20, 25$. N indicates no weight; D1 and D2 are $L^\infty$ and $L^2$ norms of daily differences; M1 and M2 are $L^\infty$ and $L^2$ norms of monthly differences; Y1 and Y2 are $L^\infty$ and $L^2$ norms of annual differences.}}    
    \end{tabular}
\end{table}

\clearpage
\begin{table}[htbp]
  \centering
  \fontsize{10}{10}\selectfont  
  \caption{Selecting Knots in Piecewise Yield Function Based on Minimized MSE}
    \begin{tabular}{ccccccccccc}\hline
    \multicolumn{5}{c}{Unbalanced Panel}   &       & \multicolumn{5}{c}{Balanced Panel} \\ \cline{1-5} \cline{7-11}
    \\
    \multicolumn{2}{c}{One-knot Piecewise} &       & \multicolumn{2}{c}{Two-knot Piecewise} &       & \multicolumn{2}{c}{One-knot Piecewise} &       & \multicolumn{2}{c}{Two-knot Piecewise} \\ \cline{1-2} \cline{4-5} \cline{7-8} \cline{10-11} 
    \\
    Knot  & MSE   &       & Knots & MSE   &       & Knot  & MSE   &       & Knots & MSE \\ \hline
    \\
    29    & 0.046905 &       & 24, 26 & 0.046478 &       & 30    & 0.029987 &       & 29, 33 & 0.047226 \\
    28    & 0.046959 &       & 23, 27 & 0.046483 &       & 31    & 0.030061 &       & 30, 32 & 0.047235 \\
    30    & 0.047258 &       & 25, 26 & 0.046488 &       & 29    & 0.030243 &       & 29, 34 & 0.047239 \\
    27    & 0.047331 &       & 24, 27 & 0.046497 &       & 32    & 0.030493 &       & 30, 31 & 0.047239 \\
    26    & 0.047903 &       & 31, 32 & 0.046499 &       & 28    & 0.030736 &       & 29, 32 & 0.047244 \\
    31    & 0.048043 &       & 30, 33 & 0.046499 &       & 33    & 0.031261 &       & 29, 35 & 0.047269 \\
    25    & 0.048573 &       & 22, 27 & 0.046505 &       & 27    & 0.031377 &       & 30, 33 & 0.047282 \\
    32    & 0.049233 &       & 23, 26 & 0.046507 &       & 26    & 0.032090 &       & 29, 31 & 0.047285 \\
    24    & 0.049271 &       & 30, 32 & 0.046536 &       & 34    & 0.032273 &       & 24, 26 & 0.047301 \\
    23    & 0.049950 &       & 25, 27 & 0.046538 &       & 25    & 0.032816 &       & 28, 36 & 0.047301 \\ \hline
    \\
    \multicolumn{11}{l}{\parbox[t]{15cm}{\footnotesize {\it Notes:} For illustration purpose, in each case, only the  smallest ten MSEs and their corresponding knots are presented.}}
    \end{tabular}
  \label{tab:knots_mse}
\end{table}

\begin{table}[htbp]
  \centering
  \fontsize{8}{8}\selectfont
  \caption{Weather Impacts on Corn Yields: Unbalanced Sample}\label{tab:yld_est_unbalanced}
    \begin{tabular}{lcccc} \hline
    & (1) & (2) & (3) & (4) \\ \hline
    Average temperature: April & 0.0110*** &       &       &                 \\
          & [0.0020] &       &       &                 \\
          &       &       &       &  \\
    Average temperature: May & 0.0033 &       &       &                 \\
          & [0.0029] &       &       &                 \\
          &       &       &       &  \\
    Average temperature: June & -0.0093 &       &       &                 \\
          & [0.0057] &       &       &                 \\
          &       &       &       &  \\
    Average temperature: July & -0.0586*** &       &       &                 \\
          & [0.0084] &       &       &                 \\
          &       &       &       &  \\
    Average temperature: August & -0.0305*** &       &       &                 \\
          & [0.0036] &       &       &                 \\
          &       &       &       &  \\
    Average temperature: September & 0.0021 &       &       &                 \\
          & [0.0047] &       &       &                 \\
          &       &       &       &  \\
    GDD (8-32C, in 100C) &       & 0.0802* &       &                 \\
          &       & [0.0316] &       &                 \\
          &       &       &       &  \\
    GDD, squared &       & -0.0018* &       &                 \\
          &       & [0.0007] &       &                 \\
          &       &       &       &  \\
    HDD (34C+), squared root &       & -0.1240*** &       &                 \\
          &       & [0.0123] &       &                 \\
          &       &       &       &  \\
    Degrees accumulated above 0C &       &       & 0.0002*** & 0.0002 \\
          &       &       & [0.0001] & [0.0001]    \\
          &       &       &       &  \\
    Degrees accumulated above 24C &       &       &       & 0.0059**  \\
          &       &       &       & [0.0017]    \\
          &       &       &       &  \\
    Degrees accumulated above 26C &       &       &       & -0.0107*** \\
          &       &       &       & [0.0022]    \\
          &       &       &       &  \\
    Degrees accumulated above 29C &       &       & -0.0056*** &                 \\
          &       &       & [0.0007] &                 \\
          &       &       &       &  \\
    Precipitation & 0.1794*** & 0.1141*** & 0.1055*** & 0.1068*** \\
          & [0.0169] & [0.0189] & [0.0201] & [0.0205]    \\
          &       &       &       &  \\
    Precipitation, squared & -0.0115*** & -0.0087*** & -0.0082*** & -0.0084*** \\
          & [0.0012] & [0.0014] & [0.0015] & [0.0015]    \\
          &       &       &       &  \\
    Signal-to-noise ratio &      0.2529	& 0.3813	& 0.4082	& 0.3935 \\
    Observations &        120,995  &        120,995  &        120,995  &        120,995  \\ \hline
        \multicolumn{5}{l}{{\it Notes:} Standard errors (in brackets) are state-clustered. Significance: * .05, ** .01, *** .001.}
    \end{tabular}
\end{table}

\begin{table}[htbp]
  \centering
  \fontsize{8}{8}\selectfont
  \caption{Weather Impacts on Corn Yields: Balanced Sample}\label{tab:yld_est_balanced}
    \begin{tabular}{lcccc} \hline
    & (1) & (2) & (3) & (4) \\ \hline
    Average temperature: April & 0.0126*** &       &       &                 \\
          & [0.0021] &       &       &                 \\
          &       &       &       &  \\
    Average temperature: May & 0.0008 &       &       &                 \\
          & [0.0028] &       &       &                 \\
          &       &       &       &  \\
    Average temperature: June & 0.0014 &       &       &                 \\
          & [0.0038] &       &       &                 \\
          &       &       &       &  \\
    Average temperature: July & -0.0453*** &       &       &                 \\
          & [0.0082] &       &       &                 \\
          &       &       &       &  \\
    Average temperature: August & -0.0320*** &       &       &                 \\
          & [0.0028] &       &       &                 \\
          &       &       &       &  \\
    Average temperature: September & 0.0075 &       &       &                 \\
          & [0.0050] &       &       &                 \\
          &       &       &       &  \\
    GDD (8-32C, in 100C) &       & 0.1768*** &       &                 \\
          &       & [0.0284] &       &                 \\
          &       &       &       &  \\
    GDD, squared &       & -0.0041*** &       &                 \\
          &       & [0.0007] &       &                 \\
          &       &       &       &  \\
    HDD (34C+), squared root &       & -0.1491*** &       &                 \\
          &       & [0.0101] &       &                 \\
          &       &       &       &  \\
    Degrees accumulated above 0C &       &       & 0.0003*** & 0.0003*** \\
          &       &       & [0.0000] & [0.0000]    \\
          &       &       &       &  \\
    Degrees accumulated above 29C &       &       &       & -0.0055*** \\
          &       &       &       & [0.0007]    \\
          &       &       &       &  \\
    Degrees accumulated above 30C &       &       & -0.0099*** &                 \\
          &       &       & [0.0006] &                 \\
          &       &       &       &  \\
    Degrees accumulated above 33C &       &       &       & -0.0084**  \\
          &       &       &       & [0.0024]    \\
          &       &       &       &  \\
    Precipitation & 0.2245*** & 0.1498*** & 0.1380*** & 0.1347*** \\
          & [0.0293] & [0.0316] & [0.0294] & [0.0293]    \\
          &       &       &       &  \\
    Precipitation, squared & -0.0156*** & -0.0123*** & -0.0117*** & -0.0114*** \\
          & [0.0025] & [0.0026] & [0.0025] & [0.0025]    \\
          &       &       &       &  \\
    Signal-to-noise ratio & 0.3019	& 0.6194	& 0.6225	& 0.5688     \\
    Observations &      44,814  &      44,814  &      44,814  &      44,814  \\ \hline
        \multicolumn{5}{l}{{\it Notes:} Standard errors (in brackets) are state-clustered. Significance: * .05, ** .01, *** .001.}    
    \end{tabular}
\end{table}

\end{document}